\documentclass[letter,12pt]{article}
\pdfoutput=1  
\usepackage{jheppub}
\usepackage{graphicx,amssymb,amsmath,color,scalefnt}
\usepackage{hyperref}
\usepackage{multirow,soul}
\usepackage{booktabs}
\usepackage{cleveref}

\usepackage{tikz}
\usetikzlibrary{arrows,shapes}
\usetikzlibrary{trees}
\usetikzlibrary{matrix,arrows} 				
\usetikzlibrary{positioning}				
\usetikzlibrary{calc,through}				
\usetikzlibrary{decorations.pathreplacing}  
\usepackage{pgffor}							

\usetikzlibrary{decorations.pathmorphing}	
\usetikzlibrary{decorations.markings}
\tikzset{
    vector/.style={decorate, decoration={snake}, draw},
	provector/.style={decorate, decoration={snake,amplitude=2.5pt}, draw},
	antivector/.style={decorate, decoration={snake,amplitude=-2.5pt}, draw},
    fermion/.style={draw=black, postaction={decorate},
        decoration={markings,mark=at position .55 with {\arrow[draw=black]{>}}}},
    fermionbar/.style={draw=black, postaction={decorate},
        decoration={markings,mark=at position .55 with {\arrow[draw=black]{<}}}},
    fermionnoarrow/.style={draw=black},
    gluon/.style={decorate, draw=black,
        decoration={coil,amplitude=4pt, segment length=5pt}},
    scalar/.style={dashed,draw=black, postaction={decorate},
        decoration={markings,mark=at position .55 with {\arrow[draw=black]{>}}}},
    scalarbar/.style={dashed,draw=black, postaction={decorate},
        decoration={markings,mark=at position .55 with {\arrow[draw=black]{<}}}},
    scalarnoarrow/.style={dashed,draw=black},
    electron/.style={draw=black, postaction={decorate},
        decoration={markings,mark=at position .55 with {\arrow[draw=black]{>}}}},
	bigvector/.style={decorate, decoration={snake,amplitude=4pt}, draw},
}

\tikzstyle{block} = [draw, rectangle, 
    minimum height=3em, minimum width=6em]

\usepackage{lscape}

\newcommand{\beq}{\begin{equation}}
\newcommand{\eeq}{\end{equation}}
\def\bea{\begin{eqnarray}}
\def\eea{\end{eqnarray}}
\newcommand{\bei}{\begin{itemize}}
\newcommand{\eei}{\end{itemize}}

\newcommand{\Fig}[1]{Fig.~\ref{#1}}
\newcommand{\Eq}[1]{Eq.~(\ref{#1})}
\newcommand{\Sec}[1]{Section~\ref{#1}}

\def\={\,=\,}
\def\+{\,+\,}
\def\-{\,-\,}
\def\TeV{\ifmmode {\mathrm{Te\kern -0.1em V}}\else
                   \textrm{Te\kern -0.1em V}\fi\,}%
\def\GeV{\ifmmode {\mathrm{Ge\kern -0.1em V}}\else
                   \textrm{Ge\kern -0.1em V}\fi\,}%
\def\MeV{\ifmmode {\mathrm{Me\kern -0.1em V}}\else
                   \textrm{Me\kern -0.1em V}\fi\,}%
\def\keV{\ifmmode {\mathrm{ke\kern -0.1em V}}\else
                   \textrm{ke\kern -0.1em V}\fi\,}%
\def\eV{\ifmmode  {\mathrm{e\kern -0.1em V}}\else
                   \textrm{e\kern -0.1em V}\fi\,}%
\let\tev=\TeV
\let\gev=\GeV
\let\mev=\MeV

\def\iab{\mbox{ab$^{-1}$~}}
\def\ifb{\mbox{fb$^{-1}$~}}

\begin{document}

\title{Higgs, top and electro-weak precision measurements at future $e^+ e^-$ colliders; a combined effective field theory analysis with renormalization mixing}

\author[a]{Sunghoon Jung,} 
\author[a]{Junghwan Lee,} 
\author[b]{Martin Perell\'{o},}
\author[c]{Junping Tian,} 
\author[b]{Marcel Vos}

\affiliation[a]{CTP, Department of Physics and Astronomy, Seoul National University, Seoul 08826, Korea}
\affiliation[b]{IFIC, Universitat de Val\`encia and CSIC, c./ Catedr\'atico Jos\'e Beltr\'an 2, E-46980 Paterna, Spain}
\affiliation[c]{ICEPP, The University of Tokyo, Hongo, Bunkyo-ku, Tokyo, 113-0033, Japan}

\emailAdd{sunghoonj@snu.ac.kr}
\emailAdd{ghe1266@snu.ac.kr}
\emailAdd{martin.perello@ific.uv.es}
\emailAdd{tian@icepp.s.u-tokyo.ac.jp}
\emailAdd{marcel.vos@cern.ch}


\abstract{
This paper presents a combined analysis of the potential of a future electron-positron collider to constrain the Higgs, top and electro-weak (EW) sectors of the Standard Model Effective Field Theory (SMEFT). The leading contributions of operators involving top quarks arise mostly at one-loop suppressed order and can be captured by the renormalization group mixing with Higgs operators. We perform global fits with an extended basis of 29 parameters, including both Higgs and top operators, to the projections for the Higgs, top and electro-weak precision measurements at the International Linear Collider (ILC). The determination of the Higgs boson couplings in the 250~\gev stage of the ILC is initially severely degraded by the additional top-quark degrees of freedom, but can be nearly completely recovered by the inclusion of precise measurements of top-quark EW couplings at the LHC. The physical Higgs couplings are relatively robust, as the top mass is larger than the energy scale of EW processes. The effect of the top operators on the bounds on the Wilson coefficients is much more pronounced and may limit our ability to identify the source of deviations from the Standard Model. Robust global bounds on all Wilson coefficients are only obtained when the 500~\gev stage of the ILC is included.  
}


\maketitle

\section{Introduction}

With the discovery of the Higgs boson, all particles postulated by the Standard Model (SM) of particle physics have been experimentally confirmed. Many attempts have been made to embed the SM in a more complete theory, that incorporates a description of gravity, neutrino masses, or dark matter, to name a few of the most popular targets. Currently, none of these extensions have imposed themselves as the ``standard'' paradigm. 

In the meantime, experiments are leaving no stones unturned looking for hints of new physics. The LHC provides precise measurements of a wealth of SM processes in an unexplored energy regime. Future colliders may extend the energy reach and precision still further. We focus on an electron-positron collider with sufficient energy to produce Higgs bosons and eventually top quarks, and adopt in particular the scenario\footnote{A ``Higgs factory'' can also be implemented as a linear collider based on {\em warm} radio-frequency technology (CLIC~\cite{Charles:2018vfv,Linssen:2012hp}) or a 100~km ring (FCCee~\cite{Abada:2019zxq}, CEPC~\cite{CEPCStudyGroup:2018ghi}).}  of the ILC~\cite{Bambade:2019fyw,Baer:2013cma}. 

Effective-field theory (EFT) or the Standard Model EFT (SMEFT) is a crucial tool in modern high-energy physics, as it provides a relatively model-independent framework to order and interpret the enormous wealth of measurements from experiments at colliders and elsewhere. This approach is particularly needed when one is to extract some properties in a model-independent way. Some of the earliest works along this direction have focused on LHC Higgs physics~\cite{Khachatryan:2016vau,Ellis:2018gqa,Brivio:2016fzo} and top physics~\cite{Brivio:2019ius,Hartland:2019bjb,Durieux:2019rbz,Buckley:2015lku,Buckley:2015nca}\footnote{Recently, differential distributions of EW processes have also been included in SMEFT analyses~\cite{Banerjee:2019twi}.}. EFT studies also provide a global framework to compare scenarios for future colliders in terms of the sensitivity of their precision measurements to new physics. We build on studies for future collider prospects that have been performed in the Higgs/EW~\cite{Barklow:2017suo,Barklow:2017awn} and top sector~\cite{Durieux:2019rbz,Durieux:2018tev}.

In this study, we combine the contributions from the Higgs/EW and top sectors that have been treated separately so far. The top-sector contributions to Higgs+EW precision observables have been ignored as they are one-loop suppressed compared to those of Higgs operators and SM. However, these are leading contributions of the top sector at the initial stages of ILC at 250~\gev, before top quarks can be directly produced.

Therefore, it is important to include top contributions and assess whether the Higgs precision achievable without model-independent top effects can be retained even with them, what capabilities of future colliders are needed, and whether the top sector can be precisely constrained without direct top productions. 

To this end, we extend the previous SMEFT basis of Ref.~\cite{Barklow:2017suo,Barklow:2017awn} to include seven additional operators involving top quarks. We thus include a total of 29 degrees of freedom that affect Higgs production and decay rates, EW precision measurements and high-energy di-boson production, and the EW interactions of the top quark. We include all and only leading contributions of these operators, up to {\em log-enhanced} one-loop contributions compared to the SM ones, as captured by Higgs and top operator mixing through the renormalization group (RG) evolution of Higgs operators due to top operators\footnote{The authors of Ref.~\cite{Durieux:2018ggn,Vryonidou:2018eyv} have developed a similar approach and used it to study the prospects of circular electron-positron colliders. Their studies are compared with ours in \autoref{app:durieux}. }.

We first treat the additional top operators as a {\em threat} to the Higgs fit and evaluate how the extension of the operator basis with top-quark operators affects the determination of the Higgs boson couplings. We investigate how precise measurements of top quark electro-weak couplings at the LHC or future measurements at the ILC at $\sqrt{s}=$ 500~\gev can mitigate this effect. We also consider the new {\em opportunity} that the interplay between top and Higgs physics offers. We discuss to what extent the 250~\gev ``Higgs factory" programme can do top physics without actually producing top quarks by using Higgs/EW precision measurements to set indirect bounds on the top operator coefficients. 

The paper is structured as follows. In \Sec{sec:motiv} and \Sec{sec:theory} we present the motivation and the theory framework, with details of the operator basis, power counting of top effects, and renormalization group effects. \Sec{sec:data} presents the benchmark datasets that are used in the fit. The main results are presented in \Sec{sec:results}. In \Sec{sec:summary}, we summarize the most important findings of the study and discuss their implications.

\section{Motivation and background}
\label{sec:motiv}

The top-sector's contributions to the Higgs+EW observables are the leading top contributions to the initial ``Higgs factory'' stage of future electron-positron colliders. There are various good reasons to expect that top-sector contributions can be important, numerically.

\subsection{Top-Higgs interplay in the SM}

In practice, we expect the most important next-to-lowest effects would come from loop effects induced by top-quark related operators. It is first because the top-EW interactions are only poorly constrained by the LHC and the initial (250 GeV) stage of future Higgs factories does not reach the top quark pair production threshold.

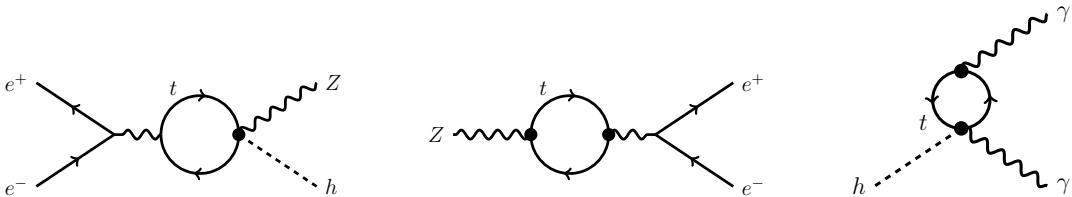
\begin{figure}[h!]
\centering
\resizebox{0.3\columnwidth}{!}{\begin{tikzpicture}[line width=1.5 pt,node distance=1 cm and 1.5 cm]
  \coordinate[] (aux);
  \coordinate[left = 0.9 of aux] (aux2);
  \coordinate[above left = of aux2, label = left: $e^+$] (v1);
  \coordinate[below left = of aux2, label = left: $e^-$] (v2);
  \coordinate[right = of aux] (aux3);
  \coordinate[above right = of aux3, label = right: $Z$] (v3);
  \coordinate[below right = of aux3, label = right: $h$] (v4);  
  \coordinate[right = 0.7 of aux] (label1);
  \coordinate[above = 0.9 of aux, label = right: $t$] (label2);
  \draw[fermion] (aux) arc [start angle=180, end angle=0, radius=0.75cm];
  \draw[fermion] (aux3) arc [start angle=0, end angle=-180, radius=0.75cm];
  \draw[fermion] (aux2) -- (v1);
  \draw[vector] (aux) -- (aux2);
  \draw[fermion] (v2) -- (aux2);
  \draw[vector] (v3) -- (aux3);
  \draw[scalarnoarrow] (v4) -- (aux3);
  \draw[fill=black] (aux3) circle (.1cm);  
\end{tikzpicture}}
\quad \quad
\resizebox{0.3\columnwidth}{!}{\begin{tikzpicture}[line width=1.5 pt,node distance=1 cm and 1.5 cm]
  \coordinate[] (aux);
  \coordinate[left = of aux, label = left: $Z$] (aux2);
  \coordinate[right = of aux] (aux3);
  \coordinate[right = 0.9 of aux3] (aux4);
  \coordinate[above right = of aux4, label = right: $e^+$] (v3);
  \coordinate[below right = of aux4, label = right: $e^-$] (v4);  
  \coordinate[right = 0.7 of aux] (label1);
  \coordinate[above = 0.9 of aux, label = right: $t$] (label2);
  \draw[fermion] (aux) arc [start angle=180, end angle=0, radius=0.75cm];
  \draw[fermion] (aux3) arc [start angle=0, end angle=-180, radius=0.75cm];
  \draw[vector] (aux) -- (aux2);
   \draw[vector] (aux3) -- (aux4);
  \draw[fermion] (aux4) -- (v3);
  \draw[fermion] (v4) -- (aux4);
  \draw[fill=black] (aux) circle (.1cm);
  \draw[fill=black] (aux3) circle (.1cm);
\end{tikzpicture}}
\quad \quad
\resizebox{0.2\columnwidth}{!}{\begin{tikzpicture}[line width=1.5 pt,node distance=1 cm and 1.5 cm]
  \coordinate[] (aux);
  \coordinate[above right = of aux, label = right: $\gamma$] (v1);  
  \coordinate[below = of aux] (aux2);
  \coordinate[below left = of aux2, label = left: $h$] (v3);
  \coordinate[below right = of aux2, label = right: $\gamma$] (v4);  
  \coordinate[left = 0.9 of aux] (label1);
  \coordinate[below = 0.9 of label1, label = right: $t$] (label2);
  \draw[fermion] (aux) arc [start angle=90, end angle=270, radius=0.5cm];
  \draw[fermion] (aux2) arc [start angle=270, end angle=450, radius=0.5cm];
  \draw[vector] (v1) -- (aux);
  \draw[vector] (v4) -- (aux2);
  \draw[scalarnoarrow] (v3) -- (aux2);
    \draw[fill=black] (aux2) circle (.1cm);
  \draw[fill=black] (aux) circle (.1cm);
\end{tikzpicture}}
\caption{Example loop diagrams involving top quarks in $e^+e^-\to Zh$ production (left), $Z\to e^+e^-$ (middle) and $h\to\gamma\gamma$ (right). The solid black dots indicate vertices that are affected by the operators considered in this paper.}
\label{fig:loop_diagrams_higgs_production}
\end{figure}

Figure~\ref{fig:loop_diagrams_higgs_production} shows three example loop processes where top-EW couplings can contribute. The contribution of top-loop diagrams involving poorly constrained operators can easily exceed the experimental accuracies of the rate measurements for those processes.
The first diagram contributes directly to the dominant Higgs production process, and would hence affect the Higgs coupling measurements. The second diagram contributes to the $Z$-pole process and will play a role in electro-weak precision observables\footnote{For an analogy, consider the $W$-mass measurement. The SM computation of $m_W$ receives an uncertainty from top mass uncertainty ($\Delta m_W\sim$ 6~\mev for $\Delta m_t\sim $ 1~\GeV~\cite{Awramik:2003rn}). At a certain precision of $m_W$, further improvements of the experimental measurement will not increase the precision of the comparison, unless one can improve the top-quark mass measurement as well.} (EWPO). The third diagram contributes to the $h\to\gamma\gamma$ branching ratio. 

\begin{figure}[h!]
\centering
\resizebox{0.27\columnwidth}{!}{\begin{tikzpicture}[line width=1.5 pt,node distance=1 cm and 1.5 cm]
    \coordinate[] (aux);
    \coordinate[left = of aux, label = left: $h$] (v1);
    \coordinate[above right =  of aux] (v2);
    \coordinate[below right =  of aux] (v3);
    \coordinate[right = of v2, label = right: $\gamma$] (v4);
    \coordinate[right = of v3, label = right: $\gamma$] (v5);
    \coordinate[right = 0.9cm of aux, label = right: $t$] (aux7);
    \draw[scalarnoarrow] (v1) -- (aux);
    \draw[fermion] (aux) -- (v2);
    \draw[fermion] (v2) -- (v3);
    \draw[fermion] (v3) -- (aux);
    \draw[vector] (v4) -- (v2);
    \draw[vector] (v5) -- (v3);
        \draw[fill=black] (aux) circle (.1cm);
  \draw[fill=black] (v2) circle (.1cm);
  \draw[fill=black] (v3) circle (.1cm);
\end{tikzpicture}}
\quad \quad
\resizebox{0.27\columnwidth}{!}{\begin{tikzpicture}[line width=1.5 pt,node distance=1 cm and 1.5 cm]
    \coordinate[] (aux);
    \coordinate[left = of aux, label = left: $h$] (v1);
    \coordinate[above right =  of aux] (v2);
    \coordinate[below right =  of aux] (v3);
    \coordinate[right = of v2, label = right: $g$] (v4);
    \coordinate[right = of v3, label = right: $g$] (v5);
    \coordinate[right = 0.9cm of aux, label = right: $t$] (aux7);
    \draw[scalarnoarrow] (v1) -- (aux);
    \draw[fermion] (aux) -- (v2);
    \draw[fermion] (v2) -- (v3);
    \draw[fermion] (v3) -- (aux);
    \draw[gluon] (v4) -- (v2);
    \draw[gluon] (v5) -- (v3);
    \draw[fill=black] (aux) circle (.1cm);    
\end{tikzpicture}}
\quad \quad
\resizebox{0.27\columnwidth}{!}{\begin{tikzpicture}[line width=1.5 pt,node distance=1 cm and 1.5 cm]
    \coordinate[] (aux);
    \coordinate[left = of aux, label = left: $h$] (v1);
    \coordinate[above right =  of aux] (v2);
    \coordinate[below right =  of aux] (v3);
    \coordinate[right = of v2, label = right: $\gamma$] (v4);
    \coordinate[right = of v3, label = right: $Z$] (v5);
    \coordinate[right = 0.9cm of aux, label = right: $t$] (aux7);
    \draw[scalarnoarrow] (v1) -- (aux);
    \draw[fermion] (aux) -- (v2);
    \draw[fermion] (v2) -- (v3);
    \draw[fermion] (v3) -- (aux);
    \draw[vector] (v4) -- (v2);
    \draw[vector] (v5) -- (v3);
    \draw[fill=black] (aux) circle (.1cm);
  \draw[fill=black] (v2) circle (.1cm);
  \draw[fill=black] (v3) circle (.1cm);    
\end{tikzpicture}}
\caption{Loop diagrams for Higgs boson interactions with photons (left), gluons (middle), and with a photon and $Z$-boson (right). The solid black dots indicate vertices that are affected by the operators considered in this paper.}
\label{fig:loop_diagrams_higgs}
\end{figure}
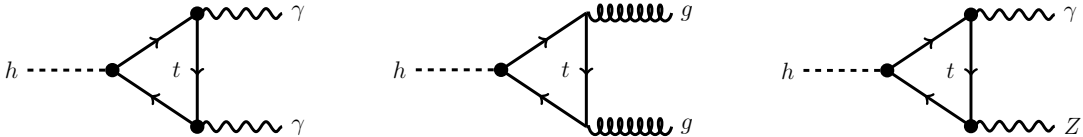

Another reason is that the top quark couples to the SM most strongly. This makes the top-loop the dominant contribution to the loop-induced Higgs decays and productions in the SM, as shown in \Fig{fig:loop_diagrams_higgs}.
In a calculation that explicitly resolves the loop diagrams, the three diagrams depend on various top-EW couplings, including the $t\bar{t}h$, $t\bar{t}\gamma$, $t\bar{t}g$, and $t\bar{t} Z$ vertices\footnote{Indeed, in the so-called {\em resolved} version of the fit in $\kappa$ framework used for early LHC analyses~\cite{Khachatryan:2016vau} the scale factors for the effective Higgs boson coupling to gluons are written as $\kappa_g \sim \kappa_t + \Delta \kappa_g$ and that to photons as $\kappa_\gamma \sim -0.28 \kappa_t + 1.28 \kappa_W + \Delta \kappa_\gamma$~\cite{Dawson:2013bba}, where $\kappa_t$ and $\kappa_W$ are the scale factors that multiply the SM predictions for the Higgs boson couplings to the top quark and the $W$-boson ($\kappa_g = \kappa_\gamma = \kappa_t = \kappa_W = 1$ and $\Delta \kappa_g = \Delta \kappa_\gamma = 0$ in the SM).}. Also, such vertex contributions are \emph{not} loop suppressed compared to those of SM ones. Therefore, Higgs precision measurements can be  sensitive to new physics that affect the top-EW vertices. Higgs factories operating at $\sqrt{s}=$ 250~\gev can probe these vertices indirectly. The inclusion of these loop effects is described in \Sec{sec:yukawa} and a quantitative analysis is presented in \Sec{sec:topyukawa} and \Sec{sec:indirecttopEW}.

\subsection{BSM motivation}

A further entanglement of the Higgs and top sectors may arise in new physics models beyond the SM (BSM). Generally speaking, extensions of the Standard Model are under no obligation to respect the separation between the top and Higgs sectors. Rather, concrete extensions of the SM do activate multiple operators across several sectors. 

A well-known example is found in composite Higgs models~\cite{Agashe:2004rs,Kaplan:1983fs}. The Higgs boson couplings are expected to deviate from the SM predictions as $\kappa \sim 1 - c \times \frac{1 \, \tev}{f}$, where $f$ is the scale associated with the new physics and $c\sim$ 3\% for vector boson couplings and $c\sim$ 3 -- 9\% for fermion couplings~\cite{Dawson:2013bba}. As the top quark is the heaviest SM particle and consequently tends to mix strongly with the composite sector, the top-EW couplings are typically altered sizably. A custodial symmetry is usually invoked to protect the coupling of the $Z$-boson to left-handed bottom quarks, but cannot simultaneously avoid large corrections to the EW couplings of the top quark~\cite{Agashe:2006at,Grojean:2013qca}. Ref.~\cite{Richard:2014upa} indeed collects a number of concrete proposals for composite Higgs models that predict sizeable (up to 10-20\%) deviations from the SM for the top quark couplings to the $Z$-boson. 
Ref.~\cite{Durieux:2018ekg} quantitatively demonstrated that the measurements of Higgs boson and top (and bottom) quark couplings provide complementary handles. Together, precise measurements of the top, Higgs and EW sectors probe the parameter space of typical models up to a scale well beyond the direct reach of colliders. A comprehensive study of the constraints of precision measurements on such models requires a combined analyses of Higgs and top measurements.

\section{Theoretical framework}
\label{sec:theory}

In this section, we present the extended basis of dimension-six operators, including the Higgs operators proposed in Ref.~\cite{Barklow:2017awn} and additional top operators. Then we detail various top-quark effects that we include, with the power counting rules to select which top effects to include with the renormalization group calculation.

\subsection{Higgs/EW operators}
\label{sec:higgsop}

Following \cite{Barklow:2017awn}, we use the 10 `Higgs' operators below to parameterize the lowest-order modifications of Higgs and EW observables.
\bea
 &\mathcal{O}_H=\partial^\mu(\Phi^\dagger\Phi)\partial_\mu(\Phi^\dagger\Phi), ~~~&~~~ \mathcal{O}_T=(\Phi^\dagger \overleftrightarrow{D}^\mu \Phi)(\Phi^\dagger \overleftrightarrow{D}_\mu \Phi), \nonumber\\
 &\mathcal{O}_6=(\Phi^\dagger\Phi)^3, ~~~&~~~ \mathcal{O}_{WB}=\Phi^\dagger t^a \Phi W_{\mu\nu}^a B^{\mu\nu}, \nonumber\\
& \mathcal{O}_{BB}=\Phi^\dagger\Phi B_{\mu\nu}B^{\mu\nu}, ~~~&~~~ \mathcal{O}_{3W}=\epsilon_{abc}W_{\mu\nu}^a W_{\rho}^{b\nu}W^{c\rho\mu}, \nonumber\\
& \mathcal{O}_{HL}=(\Phi^\dagger i\overleftrightarrow{D}^\mu \Phi)(\bar{L}\gamma_\mu L), ~~~&~~~ \mathcal{O}_{HL^\prime}=(\Phi^\dagger t^a i\overleftrightarrow{D}^\mu \Phi)(\bar{L}\gamma_\mu t^a L), \nonumber\\
&\mathcal{O}_{HE}=(\Phi^\dagger i\overleftrightarrow{D}^\mu \Phi)(\bar{e}\gamma_\mu e), ~~~&~~~ \mathcal{O}_{WW}=(\Phi^\dagger \Phi)W_{\mu\nu}^a W^{a\mu\nu},
\eea
with the Lagrangian terms
\bea
\Delta\mathcal{L}_{\rm Higgs} \= &&\frac{c_H}{2v^2}\mathcal{O}_H \+ \frac{c_T}{2v^2}\mathcal{O}_T \- \frac{c_6 \lambda}{v^2}\mathcal{O}_6 \+ \frac{g^2c_{WW}}{m_W^2}\mathcal{O}_{WW} \+ \frac{4gg^\prime c_{WB}}{m_W^2}\mathcal{O}_{WB} \+ \frac{g^{\prime 2}c_{BB}}{m_W^2}\mathcal{O}_{BB} \nonumber \\
&& \+ \frac{g^3c_{3W}}{m_W^2} \mathcal{O}_{3W} \+ \frac{c_{HL}}{v^2}\mathcal{O}_{HL} \+ \frac{4 c_{HL}^\prime}{v^2}\mathcal{O}_{HL^\prime} \+ \frac{c_{HE}}{v^2}\mathcal{O}_{HE},
\label{eq:HiggsLag}\eea
where the Higgs vacuum expectation value $v=246$ GeV and $t^a=\frac{1}{2}\tau^a$ ($\tau^a$ are Pauli matrices). We do not assume oblique~\cite{Peskin:1990zt,Peskin:1991sw} or universal~\cite{Wells:2015uba} corrections and treat all ten operators as independent. We also add Yukawa operators $c_{bH,\, cH,\, \tau H, \, \mu H,}$ to the SM fermions (e.g. $\mathcal{O}_{bH}=(\Phi^\dagger\Phi)(\bar{Q}b \Phi)$) and the collective gluonic operators $c_{gH}$ and ${\cal C}_{W,\, Z}$ to effectively describe the Higgs decay widths and corresponding physical Higgs couplings. We refer to Ref.~\cite{Barklow:2017awn} for a detailed discussion of this operator set. 

We denote $\delta A= \Delta A/A$ as the deviation of the observable $A$ from its SM prediction. We express $\delta A$ in terms of EFT coefficients. We also define the deviation of the {\em physical} Higgs coupling as the deviation of the square-root of the partial width~\cite{Barklow:2017awn,deBlas:2019wgy}
\beq
\delta g(hXX) \, \equiv \, \frac{1}{2} \delta \Gamma(h\to XX),
\label{eq:higgsprecisiondef} \eeq
where $X$ can be any particles to which the Higgs boson decays. For example,
\beq
\delta g(hb\bar{b}) \= \frac{1}{2} \delta \Gamma (h \to b\bar{b}) \= \frac{1}{2}( 2c_{bH} \- 2\delta v \+ \delta Z_h \+ \delta m_h  ),
\eeq
where we explicitly write the dependence on $\delta m_h$ and $\delta v$, and $\delta Z_h=-c_H$ is the correction to the Higgs field strength. This expression is slightly different from that of Ref.~\cite{Barklow:2017awn} where $c_{bH}$ was used to collectively describe $c_{bH},\, \delta v$ and $\delta m_h$ altogether. They are separated to emphasize the dependence on the three parameters, but numerical results remain almost unchanged. We use a similar treatment for $\delta g(h \mu^+ \mu^-),\, \delta g(h \tau^+ \tau^-),\, \delta g(hc\bar{c})$.


The Higgs self-coupling is represented by $\overline{\lambda}$ defined as
\beq
\overline{\lambda} \= \lambda \, \left( 1 \+ \frac{3}{2} c_6 \right),
\label{eq:lambda-bar}
\eeq 
which is a combination of the original Higgs self-coupling $\lambda$ and the operator $c_6$ shifting the Higgs potential. In the Higgs fit at $\sqrt{s}=$ 250~\gev, $\overline{\lambda}$ is constrained by the Higgs mass measurement. Only this combination appears in our global fit until the di-Higgs production $e^+ e^- \rightarrow Zhh$ is available at $\sqrt{s}=$ 500~\gev. Although the Higgs pair production at HL-LHC can provide some handle, the ILC $\sigma(Zhh)$ measurement will eventually provide the most precise and robust determination of $c_6$ independently from $\lambda$~\cite{DiVita:2017vrr}. Instead, for the early stage of the ILC at $\sqrt{s}=$ 250~\gev, it is assumed that the HL-LHC data constrain $c_6$ well enough, so as not to affect the model-independent fit of other parameters~\cite{Durieux:2018ggn}. It was also proposed to use the finite one-loop corrections of $c_6$ to the Higgsstrahlung process $e^+ e^- \to Zh$ at ILC250~\cite{McCullough:2013rea,DiVita:2017eyz} to constrain $c_6$. This contribution is not considered in the power counting we adopt (more on this in \Sec{sec:powercounting}). A more dedicated study could be interesting.

\subsection{Top operators}
\label{sec:topop}

To properly account for the dependence of Higgs precision observables on vertices involving top quarks, we include the following seven `top' operators in our basis. 
\bea
& \mathcal{O}_{tH}=(\Phi^\dagger\Phi)(\bar{Q}t\tilde{\Phi}), ~~~&~~~ \mathcal{O}_{Hq}^{(1)}=(\Phi^\dagger i\overleftrightarrow{D}_\mu \Phi)(\bar{Q}\gamma^\mu Q), \nonumber\\
& \mathcal{O}_{Hq}^{(3)}=(\Phi^\dagger i\overleftrightarrow{D}_\mu^a \Phi)(\bar{Q}\gamma^\mu \tau^a Q), ~~~&~~~ \mathcal{O}_{Ht}=(\Phi^\dagger i\overleftrightarrow{D}_\mu \Phi)(\bar{t} \gamma^\mu t), \nonumber\\
& \mathcal{O}_{Htb}=(\tilde{\Phi}^\dagger i D_\mu \Phi)(\bar{t} \gamma^\mu b), \nonumber \\
& \mathcal{O}_{tW}=(\bar{Q}\sigma^{\mu\nu}t)\tau^a \tilde{\Phi} W_{\mu\nu}^a, ~~~&~~~ \mathcal{O}_{tB}=(\bar{Q}\sigma^{\mu\nu}t)\Tilde{\Phi}B_{\mu\nu},
\eea
with 
\beq
\Delta\mathcal{L}_{\rm top} = -y_t\frac{c_{tH}}{v^2}\mathcal{O}_{tH} + \frac{c_{Hq}^{(1)}}{v^2}\mathcal{O}_{Hq}^{(1)}+ \frac{c_{Hq}^{(3)}}{v^2}\mathcal{O}_{Hq}^{(3)}+ \frac{c_{Ht}}{v^2}\mathcal{O}_{Ht}+ \frac{c_{Htb}}{v^2}\mathcal{O}_{Htb}+ \frac{c_{tW}}{v^2}\mathcal{O}_{tW}+ \frac{c_{tB}}{v^2}\mathcal{O}_{tB},
\label{eq:topLag}\eeq
where $Q$ is the third-generation left-handed quark doublet and $\tilde{\Phi}=i\tau_2 \Phi^\dagger$. 
These are the top operators that affect Higgs and EW observables at one loop. We do not consider four-fermion operators that involve two top quarks. In this work, we calculate these one-loop effects using the Renormalization Group (RG) evolution and mixing of top and Higgs operators. Our power counting rule is discussed in \Sec{sec:powercounting}, top RG effects in \Sec{sec:RGE}, and top effects that are not captured in this way in \Sec{sec:yukawa}.

In several extensions of the Standard Model, the effect of the top operators may be accompanied by bottom-quark effects. We have evaluated the one-loop effects on the global fit and find them to be negligible. This is discussed in detail in \autoref{app:RGEwb}.

\subsection{Top Yukawa coupling} \label{sec:yt}

The operator coefficient $c_{tH}$ shifts the top Yukawa coupling and the top mass at tree level. Including also Higgs operator contributions, we can write the following expression:
\bea
\Delta {\cal L} &=&  -\frac{y_t v}{\sqrt{2}}( 1 + \tfrac{1}{2} c_{tH}) \bar{t} t \- \frac{y_t}{\sqrt{2}}(1+ \tfrac{3}{2} c_{tH} +\tfrac{1}{2} \delta Z_h ) h \bar{t} t \\
&=& - m_t \bar{t} t \- \frac{ m_t}{v} ( 1 + c_{tH} -\frac{1}{2}c_H) h \bar{t} t.
\eea
In the second line, we define the top mass as $m_t = y_t v / \sqrt{2} (1+\tfrac{1}{2} c_{tH})$ and rewrite the Yukawa coupling in terms of the redefined top mass. Thus, we define the deviation of the top-quark Yukawa coupling as
\beq
\delta y_t \, \equiv \,  c_{tH} \- \frac{1}{2} c_H  \- \delta v .
\label{eq:topyukawa} \eeq

The top-quark Yukawa coupling and its tree-level modification can be measured separately from the top mass and the $t\bar{t} h$ cross-section. The latter is available at the LHC and HL-LHC prospects will be used throughout in this work. The ILC can measure the top-quark Yukawa coupling directly at $\sqrt{s}=500$ GeV stage (or, even better, 550 GeV~\cite{Fujii:2019zll}). The top-quark Yukawa coupling also affects Higgs decays $h\to \gamma \gamma,\, Z \gamma,\, gg$ via one-loop diagrams, which are available already at the low-energy running of ILC~\cite{Boselli:2018zxr}. These are finite one-loop effects, but are important because SM contributions are also at one-loop. They are indeed included in our work consistently with the power-counting rule in \Sec{sec:powercounting} and further discussed in \Sec{sec:yukawa}.

The coefficient $c_{tH}$ appears in our global fits only through the top Yukawa coupling as in \Eq{eq:topyukawa}. It does not RG mix with any Higgs operators. The only exception is $c_6$ (as shown in \Eq{eq:c6RG}), but $c_{tH}$'s RG contributions cannot be measured in our global fits since $c_6$ appears only at one energy scale (i.e. from the $Zhh$ production cross section at $\sqrt{s}=$ 500~\gev). Thus, we do not include the RG evolution of $c_6$.

\subsection{Power counting for top-loop contributions} \label{sec:powercounting}

In this work, we consider a subset of the leading effects of top operators on precision measurements of Higgs and EW sectors. Although it is best to include the full leading effects, most of them are loop contributions, and hence not easy to calculate and understand. As a first step toward a complete study, it is better to focus on a consistent subset of important effects. This subsection develops a power counting to select such a subset.

One could hope to include all top contributions at a given {\it absolute order}. The one-loop corrections then unambiguously specify the top contributions from one-loop diagrams. But the fact that not all leading SM processes are generated at the same absolute order makes the issue complicated. Consider the Higgs decays $h \to \gamma \gamma, Z \gamma, gg$ versus the other Higgs processes. In the SM, the former are generated at one-loop while the latter are already present at tree-level. In a numerical analysis, we do include all those processes and treat them equally even though their absolute orders are different. It is because they are {\it leading} contributions to each observable. The underlying logic is that if measurement uncertainties of tree- and loop-induced observables are similar, those leading contributions will receive similar constraints, regardless of their absolute orders. Thus, they will be similarly important. The leading contribution has an important role, irrespective of its absolute order; it provides a new information for the first time, while subleading corrections usually only modify the leading information. 

The same issue applies to top contributions to Higgs physics observables. Although we will be interested in leading top contributions regardless of their absolute orders, not all leading contributions are equally important. If the leading top contribution arises at five-loop level, while the SM process is at tree-level, it is very unlikely that such leading top contribution have an impact, unless the observable is measured with exquisite precision. Here comes the notion of the {\it relative orders} of top contributions compared to the SM. 

In all, {\bf our rule is to include all and only leading top contributions up to the logarithmic one-loop order relative to the SM.} Most leading top effects arise at the one-loop order compared to the SM. Among the full one-loop corrections to SM processes, our rule is to include only logarithmic ones, which can be readily obtained through the RG evolution of Higgs operators induced by top operators. In other words, we include leading top effects up to the logarithmic one-loop compared to the SM.  
Whenever some Higgs operator contributes to some Higgs observables, its RG mixing with top operators accounts for those top operators' logarithmic one-loop corrections to the observables. This is our main method to include the leading top-loop effects in this work.

The logarithmic corrections are not full one-loop effects, but they are a convenient and self-consistent subset. The RG effects respect all the symmetries of the theory, including gauge invariance, so that most non-zero effects in the full one-loop corrections also arise in RG\footnote{Some external momentum dependences of full one-loop effects are not captured by RG calculations. Such dependences can induce useful information on differential and energy-dependent observables.}. Due to the logarithm of scale ratios (e.g. \Eq{eq:RGsol}), the RG contributions are often the dominant contribution among full one-loop effects, although this is not always the case.  

This approach is convenient because the RG evolution is much easier to calculate. It needs to be calculated only once for all operators, independently of the observables, while every new full one-loop effects must be calculated separately (the majority has been done~\cite{Zhang:2012cd,Vryonidou:2018eyv,Durieux:2018ggn}). Moreover, their contribution is easier to understand, e.g. based on operator mixing patterns~\cite{Grojean:2013kd,Jung:2014kxa}. Thus, although full one-loop analysis will have to be eventually done (see, e.g.~\cite{Durieux:2018ggn}), at this early stage we use RG effects as a convenient proxy for the leading top effects.

The power counting rule we propose is particularly useful if all observables are measured with similar errors. All leading contributions will receive similar constraints, regardless of their absolute orders. But if some observables are measured better, this logic may become biased. There is no obvious mathematical answer for which top contributions at which order must be included. At least, the RG parts are theoretically consistent and practically convenient proxies of leading loop effects. 

\medskip

There is a notable exception and subtlety with this power counting rule. It arises again from the fact that not all top contributions are at the same absolute and relative orders. Although SM $h \to \gamma \gamma, Z \gamma$ are induced at one-loop, $c_{WW,WB,BB}$ contribute at the (absolute) tree-level. 
Then, according to the power counting rule, two-loop RG evolutions of $c_{WW,WB,BB}$ and finite one-loop corrections are needed. But we include only one-loop RG evolutions of $c_{WW,WB,BB}$. This may leave some numerical uncertainties.

Some top effects are not captured by the renormalization of Higgs operators.
As discussed in \Sec{sec:yt} and \Sec{sec:yukawa}, there are tree-level shifts of SM parameters due to top operators: $c_{tH}$ shifts the top Yukawa, and $c_{Hq}^{(1)}, c_{Hq}^{(3)}, c_{Ht}$ shift the $Zt\bar{t}$ coupling at the tree-level. We include such relative tree-level effects.

\medskip
The power counting and the corrections that we include are summarized in \autoref{tab:power_counting_summary}. We also assess the numerical impact of our power counting rules in Appendices. First, the top contributions treated differently in our approach and in \cite{Durieux:2018ggn,Vryonidou:2018eyv} are compared in \autoref{app:durieux}; second, the uncertainties in our global fit introduced by the choice of RG scale $Q$ (see next subsection) are estimated in \autoref{app:differentQ}. We can conclude that the differences between different approaches are mostly within ${\cal O}(10)$\% and our results are stable with respect to the variation of $Q$.

\begin{table}[t]
    \centering
    \begin{tabular}{l||c|c|c} 
    \toprule
 & Higgs loop production/decay & other observables  &  top production \\ \hline\hline
 SM  & finite 1-loop & tree-level  &  tree-level \\ \hline
 Higgs operator        &  tree-level from $c_{WW,WB,BB}$ &  tree-level   &   none  \\
                                 & finite 1-loop from other operators & & \\ \hline
top operator         & log 1-loop via $\dot{c}_{WW,WB,BB}$ & log 1-loop via $\dot{c}$ &  tree-level \\ 
                              & log 2-loop via other $\dot{c}$ & & \\ 
                              & finite 1-loop via tree-shift of $y_t, g_{Ztt}$ & & \\ \bottomrule
    \end{tabular}
    \caption{Summary of the \emph{absolute} orders of the contributions that we include. Our power counting is discussed in \Sec{sec:powercounting}; the RG-induced logarithmic loop effects of top operators  in \Sec{sec:RGE}; the finite one-loop effects of top operators in loop-induced Higgs decays in \Sec{sec:yukawa}. The treatment of top observables are from Ref.~\cite{Durieux:2019rbz}, where Higgs loop effects are neglected.}
    \label{tab:power_counting_summary}
\end{table}
%

\subsection{Renormalization group evolution and scale choice}
\label{sec:RGE}

The logarithmic one-loop contributions of top operators to Higgs/EW observables are captured by Higgs and top operator mixing, through the RG evolution of Higgs operators due to top operators. The RG equations of Higgs operators have the following form:
 \beq
 \dot{c}_i \equiv 16\pi^2\frac{dc_{i}}{d\ln{\mu}} \= \gamma_{ij}c_{j},
 \eeq
where $i$ index represents Higgs operators and $j$ top operators. We do not consider Higgs self running between Higgs operators, as they are subleading corrections, while the RG mixing with top operators are leading top contributions to Higgs+EWPO observables. Likewise, we do not consider RG evolutions of top operators, as discussed in the previous subsections.
An approximate solution to the RG equations is given by:
\beq
 c_{i}(Q) \,\simeq\, c_{i}(Q_0)+\frac{1}{16\pi^2}\gamma_{ij}c_{j}(Q_0)\ln{\frac{Q}{Q_0}}.  \label{eq:RGsol}
\eeq 
 
The RG equations of the Higgs operators and $\gamma_{ij}$ are given by~\cite{Alonso:2013hga,Grojean:2013kd,Jenkins:2013zja,Jenkins:2013wua}: (in the MS-bar subtraction scheme with dimensional regularization)
\bea 
 \dot{c}_{H} &=& (12 y_t^2 N_c - 4 g^2 N_c) c_{Hq}^{(3)} - 12 y_t y_b N_c c_{Htb}, \\
 \dot{c}_{T} &=& (4 y_t^2 N_c - \frac{8}{3} g^{\prime 2} Y_h Y_u N_c)c_{Ht} - (4 y_t^2 N_c + \frac{8}{3}g^{\prime 2} Y_h Y_q N_c) c_{Hq}^{(1)} + 4 y_t y_b N_c c_{Htb}, \\
\dot{c}_{WW} &=& \frac{1}{4}(-2 g y_t N_c c_{tW}), \\
 \dot{c}_{BB} &=& \frac{1}{4t_W^2} \left( -4 g^{\prime} y_t (Y_q+Y_u) N_c c_{tB} \right),\\
 \dot{c}_{WB} &=& \frac{1}{8 t_W} \left( 2 g y_t N_c c_{tB} + 4 g^{\prime} y_t (Y_q+Y_u) N_c c_{tW} \right), \\
 \dot{c}_{HL} &=& \frac{1}{2} Y_l g^{\prime 2} \left(\frac{16}{3}Y_q N_c c_{Hq}^{(1)}+\frac{8}{3}Y_u N_c c_{Ht} \right), \\
 \dot{c}_{HL}^{\prime} &=& \frac{2}{3} g^2 N_c c_{Hq}^{(3)},\\
 \dot{c}_{HE} &=& \frac{1}{2} Y_e g^{\prime 2} \left(\frac{16}{3} Y_q N_c c_{Hq}^{(1)}+\frac{8}{3} Y_u N_c c_{Ht} \right), 
\label{eq:RGE24}
\eea 
where $Y_i$ is the hypercharge, $y_{t,b}$ are Yukawa couplings, and $N_c=3$.
These are all relevant RG equations that we include in this work.

\medskip
Before discussing the scale choice, we comment on a few effects we have ignored.
The RG evolution of $c_6$ is irrelevant as $c_6$ appears in only one observable at one energy scale -- $\sigma(Zhh)$ at $\sqrt{s}= 500$ GeV. Thus, $c_6$ and the loop corrections to it cannot be distinguished. Likewise, $c_{bH}$ and similar operators also appear in only one observable. Notably, $c_{tH}$ mixes only with $c_6$ (\Eq{eq:c6RG}) and $c_{bH}$ (\Eq{eq:cbHRG}) among Higgs operators. It is not necessary to include the running of these Higgs operators. $c_{3W}$ does not mix with any top operators. The RG effect of $c_{Htb}$ is small as its RG contributions are proportional to $y_b$. We refer to \autoref{app:RGEwb} for complete expressions including $\dot{c}_6$ and $\dot{c}_{bH}$ and contributions from bottom operators.

Some top-loop effects also arise from the renormalization of SM parameters (as discussed briefly in \Sec{sec:powercounting}). Among SM parameters, only $\dot{y}_t, \dot{y}_b, \dot{\lambda}$ receive top-operator contributions. And only the running of $y_t$ can be relevant to us because $y_t$ can appear at multiple energy scales of $m_t$ and $h\to \gamma\gamma, Z\gamma$, and later $t\bar{t}h$ (although we do not include it). The relevant RG equation is
\beq
\dot{y}_t \= \frac{m_H^2}{v^2} \left(  3 y_t c_{tH} \- y_t ( c_{Hq}^{(1)} + 3 c_{Hq}^{(3)} - c_{Ht} ) \- y_b c_{Htb} \right),
\label{eq:ytrunning}
\eeq
where $m_H^2$ is the quadratic mass parameter in the Higgs potential.

\begin{table}[t]
\begin{center}
\begin{tabular}{l||c|c|c|c|c|c|c|c}
\toprule
 & $G_F$ & EWPO  & $\delta m_{W,Z,h,t}$  &  $\delta \Gamma (h)$ & $W^-W^+$  & $\sigma(\nu \bar{\nu} h)$ & $\sigma(Zh)$ & $\sigma(Zhh)$ \\
\hline 
$Q_{\rm proc}$ [\GeV] &  $m_\mu$  &  $m_Z,m_W$ & $m_{W,Z,h,t}$  & $m_h$ & 250, 500 &  250, 500 & 250, 500 & 500 \\
\bottomrule
\end{tabular}
\caption{The energy scales of observables, $Q_{\rm proc}$. In our main results, EFT top-loop contributions are evaluated at $Q={\rm max}(m_t, Q_{\rm proc})$ in \Eq{eq:Q=mt}, but they are also compared with other choices of $Q$ as explained in text. $\delta m$ and $\delta \Gamma$ may appear in other observables but will still be evaluated at on-shell; examples are discussed in regard of \Eq{eq:obsinobs}. $Q_{\rm proc}$ for top production observables are not needed since top operators do not run in our analysis.} 
\label{tab:scale} 
\end{center}
\end{table}

\medskip
There are two distinct energy scales in the RG calculation. In this paper, $Q_0$ refers to the common scale that we use to express the Wilson coefficients, while $Q$ refers to the renormalization scale for each physical process. In the following we discuss how the choices of the two scales affect the numerical results.

RG effects are evolved down from a common reference scale $Q_0$. This scale can be set to the matching scale of new physics to the EFT, but formally it is arbitrary. In this work, it is just an arbitrary renormalization scale of Higgs and top Wilson coefficients $c_i$ that we use to express theory predictions. The constraints on the Wilson coefficients $c_i(Q_0)$ from a global fit will depend on the choice of $Q_0$. In \autoref{app:RGdep} we show how the covariance matrix obtained at one scale can be evolved to another scale.

The bounds on physical observables (such as physical Higgs coupling precision) must be evaluated at their physical scales and the result must be independent of $Q_0$. 
In \autoref{app:RGdep} we prove that the results for the physical Higgs couplings are indeed independent of the choice of the reference scale $Q_0$.

All RG contributions are evaluated at the renormalization scale $Q$ of each observable. A natural choice is:
\beq
Q \= {\rm max}(m_t, Q_{\rm proc}),
\label{eq:Q=mt}
\eeq
where the top mass $m_t$ is a natural scale below which top-loop effects are suppressed by the heavy top mass. Such decoupling of heavy-particle loop effects must be added by hand in the mass-independent renormalization, e.g. with dimensional regularization and MS-bar subtraction scheme as in this paper. $Q_{\rm proc}$ is the characteristic energy scale of each observable, as collected in \autoref{tab:scale}. If $Q_{\rm proc} < m_t$, operators renormalize only between $m_t$ and $Q_0$ but not between $Q_{\rm proc}$ and $m_t$ because the top quark is heavy and decouples. Of course, the decoupling scale is also somewhat arbitrary. In self-energy diagrams (e.g. second one in \Fig{fig:loop_diagrams_higgs_production}), $2 m_t$ is a more relevant scale at which analyticity produces abrupt changes of loop functions. In the full-matching point of view, $\sqrt{2} m_t$ can be more relevant as the energy dependence of the gauge coupling beta function is highest there~\cite{Weinberg:1980wa}. All these choices are formally equivalent, but numerical differences just reflect the fact that we are terminating at some finite order in perturbative expansion. Therefore, we use the choice in \Eq{eq:Q=mt} as a main one throughout the paper. A comparison with alternative choices, such as $Q={\rm max}(2m_t, Q_{\rm proc})$ and $Q=Q_{\rm proc}$, is presented in \autoref{app:differentQ}.

\medskip
Lastly, observables often involve multiple energy scales. For example,
\beq
\delta \sigma(Zh \to Z b\bar{b})(Q_{\rm proc} = 250) \=  \delta \sigma(Zh)(250) \+ \delta \Gamma(b\bar{b})(m_h) \- \delta \Gamma_{\rm tot}(m_h), \label{eq:obsinobs}
\eeq 
where numbers in parentheses are $Q_{\rm proc}$ values.
Further, $\delta \Gamma$ may depend on, e.g. the $Z$ boson mass $\delta m_Z$ which will always be evaluated at $Q_{\rm proc}=m_Z$.
In the end, the observable $\delta \sigma(Zh \to Zb\bar{b})$ will be written in terms of $c_i(Q_0)$ at $Q_0$ and the covariance matrix of $c_i (Q_0)$ is obtained. The explicit $Q_0$ dependence of the matrix can be used to derive final constraints on the coefficients at some common reference scale $Q_0$.

\subsection{Finite one-loop effects of top couplings}
\label{sec:yukawa}

Several Higgs observables receive effects from top operators that are not logarithmic one-loop. The operator coefficient $c_{tH}$ that shifts the top Yukawa coupling (\Sec{sec:yt}) affects $h\to \gamma \gamma, \, gg, \, Z\gamma$ decays at one-loop (without UV divergences, hence without log-enhancements) and $pp \to t\bar{t}h$ and $e^+e^- \to t\bar{t}h$ production rate at tree-level. In addition to the modifications due to Higgs operators and SM parameter variations, $c_{tH}$ modifies these observables as follows:
\bea
\delta \Gamma ( h \to \gamma \gamma) &=& {\rm Re}\left( \frac{A_{\rm top}}{A_{\rm top} \+ A_W } \right) 2c_{tH} \+ \cdots,  \label{eq:h2rr} \\
\delta \Gamma ( h \to gg) &=& 2 c_{tH} \+ 2c_{gH}  \+ \delta Z_h, \label{eq:h2gg} \\
\delta \sigma ( t\bar{t}h) &=& 2 c_{tH} \+ \cdots, \label{eq:tth}
\eea
where $A_{\rm top, W}$ are SM amplitudes with top and $W$ loops, and $\cdots$ are from the variations other than $c_{tH}$ calculated in \cite{Barklow:2017awn} and \cite{Durieux:2019rbz}. In $\delta \Gamma( h \to gg)$, we write the full expression separating $c_{tH}$ from $c_{gH}$, which were altogether described by a single parameter $c_{gH}$ in \cite{Barklow:2017awn}. In this work, $c_{tH}$ affects various observables differently from $c_{gH}$. Hence, the two operators can be distinguished from each other. Numerical evaluations of these expressions are collected in \autoref{tab:Higgsops} -- \ref{tab:SMparam1}.
The loop diagrams for $h \rightarrow \gamma \gamma$ and $h \rightarrow Z \gamma$ in \Fig{fig:loop_diagrams_higgs} can also be mediated by a $W$-boson. This introduces a dependence of the decay rates on electro-weak couplings. This dependence is accounted for in our fit.

The $h \to Z\gamma$ decay rate is affected by modifications of both the top Yukawa coupling and the $Zt\bar{t}$ coupling. The operator coefficients $c_{Hq(3),\,Hq(1),\, Ht}$ shift the vectorial part of $Zt\bar{t}$ vertex (and $Zb\bar{b}$ similarly) as follow:
\beq
{\cal L} = \frac{g}{c_w} Z^\mu \bar{t} \gamma_\mu t \, (\ell + r) \, \left( 1 + \frac{ - c_{Hq}^{(1)} + c_{Hq}^{(3)} - c_{Ht} }{2(\ell + r)} \right) \, \equiv \, \frac{g}{c_w} Z^\mu \bar{t} \gamma_\mu t \, (\ell + r) \, ( 1+ c_{Ztt}),
\eeq
where $\ell = \frac{1}{2} - \frac{2}{3} s_w^2$ and $r = -\frac{2}{3} s_w^2$. The tree-level shift $(1 + \delta y_t + c_{Ztt})$ modifies the top-loop of $\delta \Gamma (h \to Z \gamma)$ in the same way as $\delta \Gamma(h \to \gamma \gamma)$ in \Eq{eq:h2rr} so that
\beq
\delta \Gamma ( h \to Z \gamma) \= {\rm Re}\left( \frac{A_{\rm top}^{Z\gamma}}{A_{\rm top}^{Z\gamma} \+ A_W^{Z\gamma} } \right) 2(c_{tH}+c_{Ztt}) \+ \cdots.  \label{eq:h2Zr}
\eeq
Numerically, this effect is rather small because the top-loop SM contribution to $h \to Z\gamma$ (denoted by $A_{\rm top}^{Z\gamma}$) is much smaller than the $W$-loop's $A_W^{Z\gamma}$. We comment that $c_{tW}$ seems to also shift the $Zt\bar{t}$ coupling, but this is UV divergent and is actually renormalizing $c_{WW}$ rather than tree-level shifting the coupling.

Lastly, we emphasize again that we include finite one-loop effects in \Eq{eq:h2rr}, (\ref{eq:h2gg}), and (\ref{eq:h2Zr}) because SM contributions are also at one loop so that top effects are at the same level, without an effective loop suppression.

 \subsection{Summary}
 \label{subsec:basis}
 
The basis for our combined EFT fit to Higgs/EW and top physics data has 29 degrees of freedom. These include eight + one Higgs operator coefficients:
\beq
c_{H,\, T,\,WW,\,WB,\,BB,\,HE,\,HL,\,HL^\prime, \, 3W},
\eeq
seven top operator coefficients:
\beq
c_{Ht,\,HQ(1),\,HQ(3),\,tH,\,tB,\,tW,\,Htb},
\eeq
five coefficients for four Yukawa operators and one operator for $h\to gg$:
\beq
c_{bH,\,cH,\,\tau H,\, \mu H,\, gH},
\eeq
four coefficients for non-standard Higgs decays:
\beq
a_{\rm inv},\, a_{\rm oth}, \, {\cal C}_{W,\,Z},
\eeq
and four SM parameters (electroweak gauge couplings, Higgs vacuum expectation value, and Higgs self interaction):
\beq
\delta g, \, \delta g^\prime, \,\delta v, \, \delta \bar{\lambda}.
\eeq
The computed numerical expressions of each observable in terms of operator coefficients are collected in \autoref{app:obs}.

\section{Benchmark datasets}
\label{sec:data}

To study the interplay between Higgs/EW and top measurements and operators we perform fits on several benchmark data sets. The benchmarks include LEP/SLC electro-weak precision measurements, LHC results in Higgs and top physics, and prospects for Higgs/EW and top physics for the high-luminosity phase of the LHC (HL-LHC) and for the ILC runs at $\sqrt{s}=$ 250~\gev and 500~\gev. In this section we provide a brief overview of these scenarios.

The SM parameter values that we use in our fits are from central values of the measurements:
\bea
  m_Z \= 91.1876 \, {\rm GeV}, ~&~ m_W \= 80.385\, {\rm GeV}, ~&~ m_h \= 125.090\, {\rm GeV} \nonumber\\
 m_t \= 173\, {\rm GeV}, ~&~ m_b \= 4.3\, {\rm GeV}, ~&~ m_\mu \= 0.105\, {\rm GeV} \nonumber\\
  s_w^2 \= 0.23152, ~&~ \alpha^{-1} \= 128.9220 ~&~ v \= 246\, {\rm GeV}.
\eea
Here, $m_{t,b}$ come in as $y_{t,b}$ in RG equations, while $m_\mu$ provides the scale $Q$ for $G_F$. 

\subsection{LEP/SLC electro-weak precision observables} 
\label{data:ewpo}
Nine electroweak precision observables are collected in \autoref{tab:ewpo}. The uncertainties are set to the currently available LEP/SLD measurements~\cite{ALEPH:2010aa}. Expected improvements from the LHC and ILC are included for $m_W, m_h, \Gamma_W,$ and $A_\ell$, as in Ref.~\cite{Barklow:2017suo}. This baseline set of  electro-weak precision observables (EWPOs) is included in all benchmark scenarios. Numerical values are given in \autoref{tab:ewpo}. For some fits we consider the Tera-Z precision~\cite{Abada:2019zxq} listed in the same table.
All values are in agreement with those in Ref. \cite{Barklow:2017suo}, except for one. We consider the improvement in $A_\ell$ that can be achieved with a radiative-return analysis at the ILC, which was not considered in Ref.~\cite{Barklow:2017suo}. 
\begin{table}[!h]
    \centering
    \begin{tabular}{l|c|c|c|c|c|c|c|c|c}
    \toprule
observable         & $\alpha^{-1} (m_{Z}^2)$  &  $G_F$             &   $m_{W}$  &  $m_Z$  & $m_h$ & $A_l$ & $\Gamma_l$ & $\Gamma_Z$ & $\Gamma_W$\\
unit               &   -                  & (GeV$^{-2}$) & (\MeV) & (\MeV) & (\MeV) & (\%) & (\MeV) & (\MeV) & (\MeV) \\
\hline
LEP/SLC            & 0.0178 & 0.6$\times 10^{-10}$ & 15 & 2.1 & 240 & 0.13 & 0.086 & 2.3 & 42 \\
+LHC/ILC &  idem  & idem & 5  &  idem & 15 & 0.013 & idem  & idem & 2 \\
+Tera-Z             & 0.00387 & idem & 0.5 & 0.1 & idem & 0.004 & 0.0054 & 0.1 & 1.2 \\ \bottomrule
    \end{tabular}
    \caption{Summary of the uncertainties on electro-weak precision measurements, as included in the fits. The first line (LEP/SLC) lists the current constraints. The second line (LHC/ILC) includes expected improvements at the LHC and the first stage of the ILC. The third line reflects the precision envisaged for a Tera-Z programme at a circular $e^+e^-$ collider. The uncertainties labelled +LHC/ILC form the baseline scenario included in all fits. The value denoted by ``idem" is identical to that in the earlier row.}
    \label{tab:ewpo}
\end{table}

\subsection{Selected LHC Higgs measurements}
\label{data:hllhchiggs}

Three ratios of Higgs branching ratios are included from the LHC. For rare decays, the LHC prospects after the full high-luminosity phase are very competitive~\cite{ATL-PHYS-PUB-2014-016}. As systematic errors largely cancel, these ratios are expected to improve with statistics as $1/\sqrt{N}$. Expectations are taken from the HL-LHC prospects~\cite{Cepeda:2019klc} for 3~\iab: BR($ZZ^*$)/BR($\gamma \gamma$) $\sim$ 2\%, BR($Z\gamma$)/BR($\gamma \gamma$) $\sim$ 20\%, and BR($\mu^+\mu^-$)/BR($\gamma \gamma$) $\sim$ 8\%. This precision exceeds what was assumed in Ref.~\cite{Barklow:2017suo}, that was based on the more conservative ATLAS study~\cite{ATL-PHYS-PUB-2014-016}. This set of results is included in all three benchmark scenarios.

\subsection{LHC top production - run 2}
\label{data:lhctop}

Measurements are included of a variety of processes that are sensitive to top electro-weak couplings, including associated $t\bar{t}+V/h$ production, electro-weak single-top production, and top-decay. The analyses of the LHC run 2 data set are based on 36~\ifb at $\sqrt{s}=$ 13~\tev. The top decays analysis is performed with 20~\ifb at $\sqrt{s}=$ 8~\tev. The constraints are based on the analysis in Ref.\cite{Durieux:2019rbz}. The $t\bar{t}$ and $hh$ production rates are not used, but they are implicitly assumed to  constrain the one-loop effects of $c_{tG}$ and $c_6$, respectively. 

\subsection{HL-LHC top production - S2 scenario}
\label{data:hllhctop}
The same measurements and final states are included as in the run 2 scenario above, where uncertainties are extrapolated to the total integrated luminosity of 3~\iab at 14~\tev envisaged in the high-luminosity stage of the LHC. The statistical uncertainties and experimental systematics are expected to scale as $1/\sqrt{N}$, while the current theory uncertainties on SM predictions are divided by two.

\subsection{ILC Higgs/EW measurements at $\sqrt{s}=$ 250~\gev}
\label{data:ilc250higgs}
This data set includes Higgs and EW observables. The $W^+W^-$ production process is used to constrain Triple Gauge boson Couplings (TGCs). Higgs measurement include the results of the Higgsstrahlung ($e^+e^- \to Zh$) recoil mass analysis (total and differential cross sections, and cross section times branching ratios) and an analysis of the Vector Boson Fusion (VBF) channel ($e^+e^- \to \nu \bar{\nu} h,\, h\to b\bar{b}$). A total integrated luminosity of 2~\iab is shared equally between two beam polarizations ($\pm30, \mp 80$). The uncertainties are given in Ref.~\cite{Barklow:2017suo}.

\subsection{ILC Higgs/EW measurements at $\sqrt{s}=$ 500~\gev}
\label{data:ilc500higgs}
The same observables are used as in the ILC250 scenario, plus seven $\sigma \times BR$ measurements in the VBF channel. The ILC500 stage envisages a total integrated luminosity of 4~\iab. Uncertainties are given in Ref.~\cite{Barklow:2017suo}.

\subsection{ILC top quark pair production at $\sqrt{s}=$ 500~\gev}
\label{data:ilc500top}
This projection includes measurements of a set of optimal observables~\cite{Durieux:2018tev} in $t\bar{t}$ production at $\sqrt{s}=$ 500~\gev (4~\iab with two beam polarizations). The results are presented in Appendix C.2.5 of \cite{Durieux:2019rbz}. Associated $t\bar{t}h$ production is considered separately, with a $13\%$ uncertainty on the cross section at $\sqrt{s}=$ 500~\gev, which can be improved to 6\% at $\sqrt{s}=$ 550~\gev~\cite{Durieux:2019rbz}.

\section{Global-fit analysis}
\label{sec:results}

\subsection{Method} \label{sec:methodology}

We perform a global fit to find the optimal values and projected uncertainties for the 29 degrees of freedom listed in \Sec{subsec:basis}. From the results, we also reconstruct the projected uncertainties on physical Higgs couplings and top Yukawa as the combined errors on $\delta g(hXX)$ (\Eq{eq:higgsprecisiondef}) and $\delta y_t$ (\Eq{eq:topyukawa}), respectively. 

The global fit minimizes the total $\chi^2$ 
\bea
      \chi^2 &=& \sum_{m,n} (\hat{O}_{exp}-\hat{O}_{thy})_m(\sigma^{-2})_{mn}(\hat{O}_{exp}-\hat{O}_{thy})_n \\\nonumber
      &=& \sum_{I,J} c_I(Q_0) \, \mathcal{C}ov^{-1}_{IJ}(Q_0) \, c_J(Q_0),
\label{eq:chi2}
\eea
where $m,n$ specify observables and $\sigma_{mn}$ is a matrix of error correlations among observables. Theory predictions in terms of operator coefficients at $Q_0$ turn the correlation matrix into the covariance matrix of operators at $Q_0$, $\mathcal{C}ov(Q_0)$, with operator indices $I,J$. The fact that the observables can be expressed in terms of the coefficients at any scales implies that the $\mathcal{C}ov(Q_0)$ at one scale $Q_0$ is physically equivalent to the $\mathcal{C}ov(Q_0^\prime)$ at another scale $Q_0^\prime$, connected by the RG evolutions of operators; see \autoref{app:RGdep} for detailed discussions and examples. Our covariance matrices for benchmark scenarios are collected in \autoref{app:cov}.

\subsection{Results for the ILC at $\sqrt{s}=$ 250~\gev}
\label{sec:ILC250}

\begin{figure}[t]
      \centering
      \includegraphics[width=\textwidth]{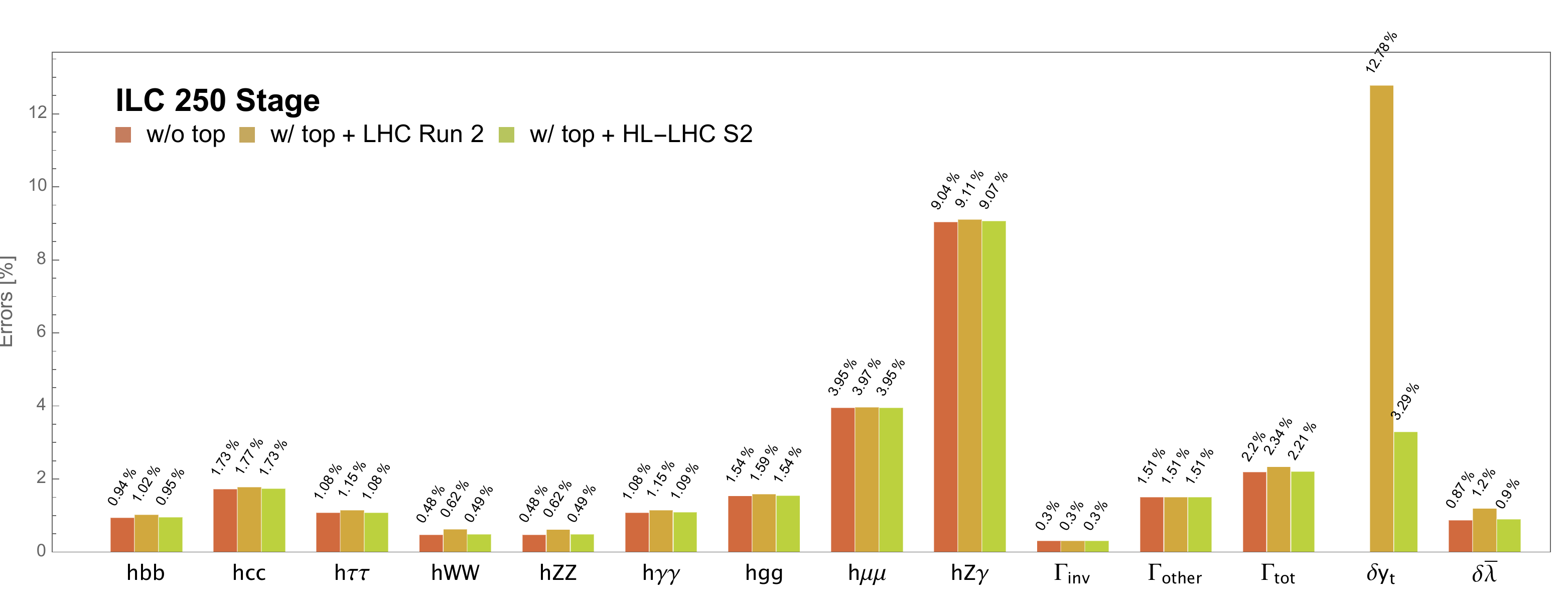}
      \includegraphics[width=\textwidth]{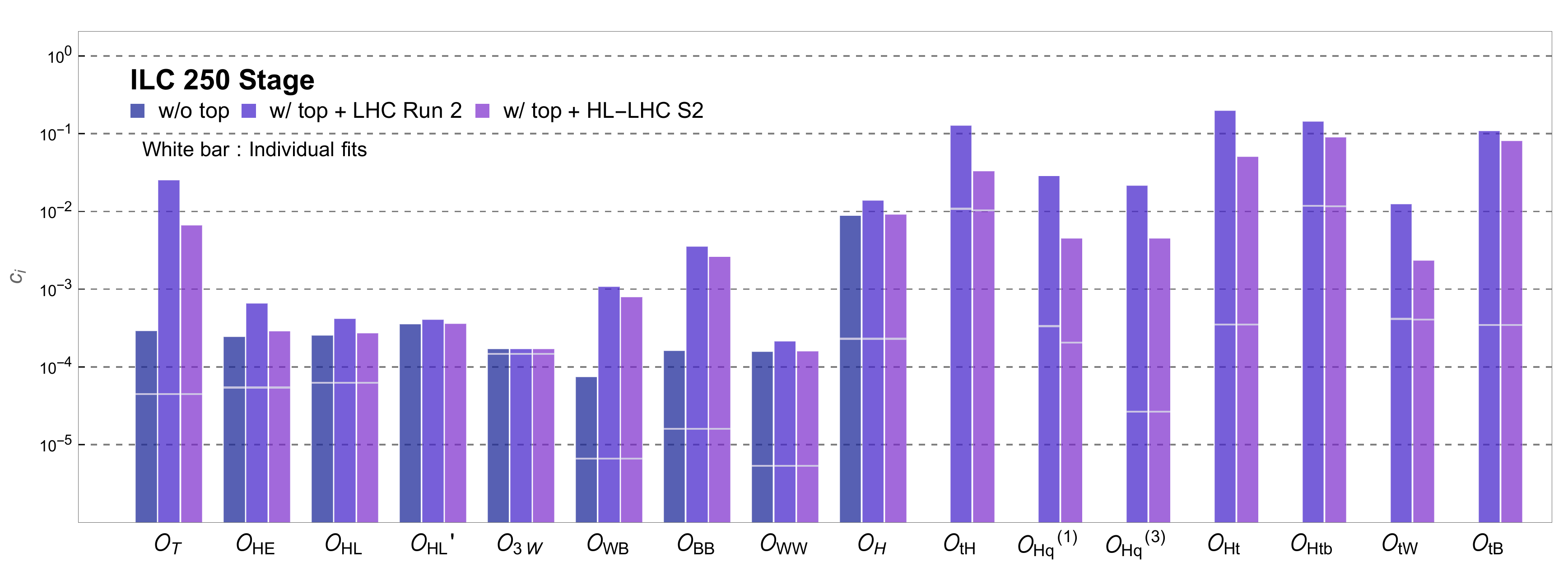}
      \caption{Global fit results for the ILC 250 scenario. The upper panel presents the result in terms of the precision on the physical Higgs couplings (\Eq{eq:higgsprecisiondef}). The lower panel presents the $1\sigma$ bounds on the operator coefficients, renormalized at $Q_0=1$ TeV  with a suppression scale $v$. RG contributions are evaluated at $Q={\rm max}(m_t, Q_{\rm proc})$ as in \Eq{eq:Q=mt} and \autoref{tab:scale}. In both panels, the first column corresponds to a 22-parameter fit without top operators~\cite{Barklow:2017suo}, used as a reference throughout the paper. The second column presents the result that is obtained when the basis is extended with the seven top operator coefficients described in \Sec{sec:topop} and LHC run 2 data are added. The last column repeats the same fit with the expectations of the S2 scenario for the measurement of the top quark electro-weak couplings at HL-LHC. In the bottom panel, white marks are results with only one operator. Results are tabulated in \autoref{tab:HiggsCoup} and \ref{tab:OpConst}.}
      \label{fig:250higgs}
\end{figure}

In this section, we compare several fits on the initial ``Higgs factory" stage of the ILC, which collects 2~\iab at $\sqrt{s}=$ 250~\gev. The benchmark data set for this fit includes the baseline EWPO described in \Sec{data:ewpo}, the measurements of rare Higgs branching ratios at the LHC of \Sec{data:hllhchiggs} and the ILC Higgs/EW data of \Sec{data:ilc250higgs}. The fits on the extended basis include also LHC measurements of the top quark EW couplings, either at the current run 2 precision (\Sec{data:lhctop}) or the projected precision for the full HL-LHC program in scenario S2 (\Sec{data:hllhctop}).

The results of these fits are presented in \Fig{fig:250higgs}. The upper panel presents the expected precision of the physical Higgs couplings (defined as the $1\sigma$ constraint on the square root of the decay width (\Sec{sec:higgsop}); $\delta \bar{\lambda}$ and $\delta y_t$ are defined in  \Eq{eq:lambda-bar} and \Eq{eq:topyukawa}, respectively). The Higgs coupling precision is evaluated at $Q_{\rm proc} =m_h$, and the result is independent of $Q_0$ (see \autoref{app:RGdep}). The lower panel presents the $1\sigma$ bounds on the operator coefficients. The operator coefficients are renormalized at $Q_0=1$~\tev with a suppression scale $v$. 

For each coupling or operator, three results are presented. The result labeled as ``w/o top" is intended for reference: it corresponds to the 22-parameter fit of Ref.~\cite{Barklow:2017suo} without top operators~\footnote{The relevant typos in Ref.~\cite{Barklow:2017suo} are collected in~\autoref{app:typos} and are corrected in the fit presented in this paper.}. 
The second bar, labeled ``w/top + LHC run 2" corresponds to a fit with the complete 29-parameter basis including top operators of \Sec{subsec:basis}. To ensure convergence of the fit, it includes the current LHC run 2 top physics data of \Sec{data:lhctop}; global fits do not converge without top data as too many parameters need to be determined.
The last bar, ``w/top + HL-LHC S2", includes the expected improvements in top physics of the HL-LHC scenario ``S2" of \Sec{data:hllhctop}.

\medskip
Remarkably, the ILC precision on the physical Higgs couplings remains robust against model-independent top effects. The global fits with 22 parameters (without tops) and 29 parameters (with tops) yield similar precision. The most pronounced deterioration is observed in the $hZZ$ and $hWW$ couplings in the ``LHC run 2" scenario, where the precision degrades by 0.48\% and 0.62\%, respectively. The parameter $\bar{\lambda}$ degrades by 40\%. The precision of the Higgs coupling measurements improves again with the inclusion of stronger top production data from HL-LHC, almost to the level without top effects. 

The robustness of the bounds on the physical Higgs couplings is understood as follows. The renormalization scale $Q={\rm max}(m_t, Q_{\rm proc})$ for top-loop effects is $m_t$ for many observables, including Higgs decay widths with $Q_{\rm proc}=m_h < m_t$ and EWPOs with $Q_{\rm proc} = m_Z <m_t$. Thus, those observables depend on common combinations of  Higgs operators and their RG corrections by top operators: 
$\tilde{c}_i(m_t) = c_i + \frac{1}{16\pi^2}\gamma_{ij} c_j(Q_0) \log \frac{m_t}{Q_0}$ (\Eq{eq:RGsol}). 
In the limit where only a single scale $Q$ is relevant, all top effects appear through $\tilde{c}_i(Q)$ so that one can redefine all Higgs operator coefficients $c_i$ by fixed combinations $\tilde{c}_i(Q)$.
Physical Higgs couplings will then depend only on $\tilde{c}_i$ in the same way they depended on $c_i$ in the global fit without top operators. Then the global fit results for physical Higgs couplings with top operators must be equivalent to the one without top operators. If all observables are at similar scales, one expects that the Higgs coupling precision is robust against the extension of the basis with the top operators. In \autoref{app:differentQ}, we present fit results for several different choices of the scale $Q$, corroborating this explanation. In other words, the heaviness of the top quark makes the Higgs coupling precision rather insusceptible to model-independent top effects.

This is remarkable. RG effects are discernible from tree-level effects through the measurements at multiple energy scales because RG effects vary with the energy scale while tree-level effects remain constant. Thus, one may expect that measurements at multiple energy scales are the ones responsible to help achieve high precision in global-fit analyses. However, we just saw a very different conclusion; having many observables with common energy scales actually allows to measure certain combinations of tree-level and RG effects (appearing at the common scale) more precisely. This resulted in the robust Higgs coupling precision at the ILC.

On the other hand, by actually the same reason, the constraints on operator coefficients degrade much more significantly with top effects, as shown in the bottom panel of \Fig{fig:250higgs}.
In the presence of top operators, it is difficult to constrain individual operators that combine into physical Higgs couplings. But many common scales (that made Higgs coupling precision robust) do bother disentangling tree-level effects of Higgs operators and RG effects of top operators. The degeneracies between them are not well resolved by Higgs+EW data at 250~\gev alone. Only with the higher-precision top production data of the HL-LHC S2 scenario, the degeneracies are reduced and operators are better constrained in the global fit. However, the precision does not fully recover to the level of the reference fit at the ILC 250 stage.

Apparently, the two panels of \Fig{fig:250higgs} would seem to lead to different conclusions. While the physical Higgs couplings are affected by 30\% to 40\% at most, the bounds on operator coefficients can degrade by orders of magnitudes. The projection onto the physical Higgs couplings can be useful because they are robustly and semi-directly measured and theoretical predictions of Higgs coupling deviations in various new physics models are available. But these are not full information of new physics effects. The projection onto the operator constraints is still needed to pinpoint the origin of the deviations and to distinguish new physics models.

Lastly, we discuss one feature shown in \Fig{fig:250higgs}. The Higgs couplings that seem to be most sensitive to top effects are $g(hWW), g(hZZ)$ and $\bar{\lambda}$, as shown in the second bars. This is related to the worsening of $c_{T,WB,BB}$ constraints, as shown in the bottom panel. Why are these operators particularly sensitive to top effects? $c_T$ RG mixes most strongly with top operators, in particular with $c_{Ht}$\footnote{$c_{Hq}^{(1)}$ also RG mixes, but it is better constrained than $c_{Ht}$ (\Fig{fig:250higgs}) by $b\bar{b}$ observables at LEP and LHC~\cite{Durieux:2019rbz}.}, and $c_{WB,BB}$ strongly with $c_{tB}$\footnote{$c_{tW}$ is better constrained than $c_{tB}$ by top-decay and single-top measurements at LHC~\cite{Durieux:2019rbz}. It is also why $c_{WW}$ (mixing with $c_{tW}$ but not with $c_{tB}$) is not degraded as much as $c_{WB,\, BB}$.}. These top operators are also not well constrained.

All in all, the ILC precision of the physical Higgs couplings are remarkably robust against the presence of poorly constrained top operators that affect the Higgs measurements through top-loop effects. However, even if the physical couplings of the Higgs boson are well constrained, strong degeneracies or ``blind directions" may be present in the basis of operator coefficients that prevent an unambiguous new physics interpretation of the result.
Therefore, precise measurements of the top-quark EW couplings are important to take full advantage of the ILC Higgs factory stage.

\subsection{The role of beam polarization}
\label{sec:ILCvsCircular}

The possibility of highly polarized beams is one of the distinguishing features of linear colliders. It is instructive to compare the impact of top operators on the model-independent precision measurements with and without beam polarization. The one with polarization represents linear $e^+ e^-$ colliders while the other represents circular $e^+e^-$ colliders. Following Ref.~\cite{Barklow:2017suo}, we compare those results in \Fig{fig:PolHcoup}. The results labeled ``polarized" are for the ILC 250 benchmark scenario of \Fig{fig:250higgs}, consisting of 1~\iab for each of the two beam polarizations; the results labeled ``unpolarized" include Higgs measurements in 5~\iab without beam polarization and the improved Tera-$Z$ estimates for the $Z$-pole EWPO (as described in \Sec{data:ewpo}). The light and dark shadings represent the results with and without top operators. In the 29-parameter fit with top operators, HL-LHC S2 top production is included for both polarized and unpolarized scenarios.

As shown in the figure, the power of beam polarization is not significant in terms of the robustness against top effects. The Higgs coupling precisions are all robust irrespective of the existence of beam polarization, as expected from the discussion in \Sec{sec:ILC250}. The operator constraints are slightly more robust in the polarized scenario, but the difference is not large. As an interesting remark, the polarization effects become more pronounced if we had used $Q=Q_{\rm proc}$, where more various scales are involved; see \autoref{app:differentQ}. The beam polarization effectively doubles the number of independent observables, which allows to better disentangle top and Higgs contributions.

\begin{figure}[t!]
      \centering
      \includegraphics[width=\textwidth]{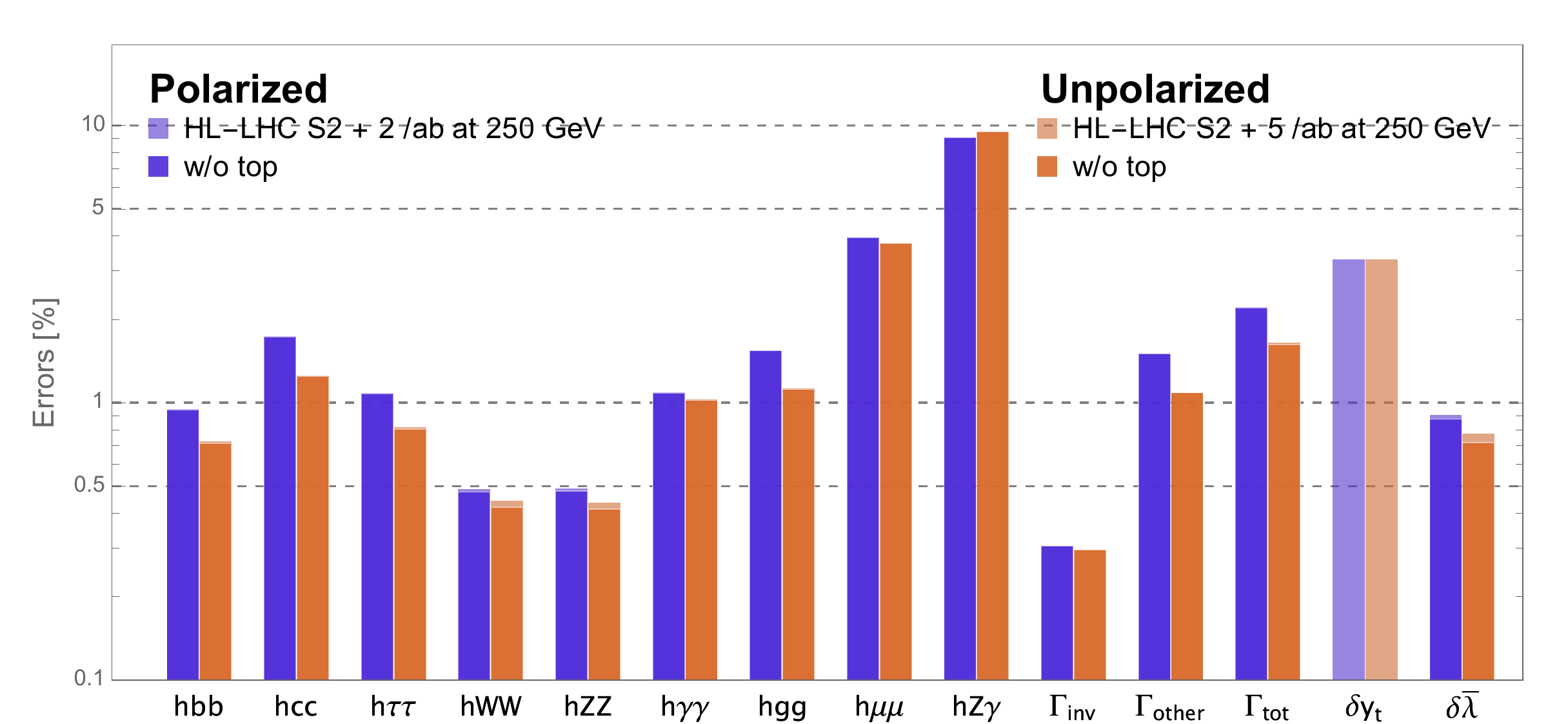}
      \includegraphics[width=\textwidth]{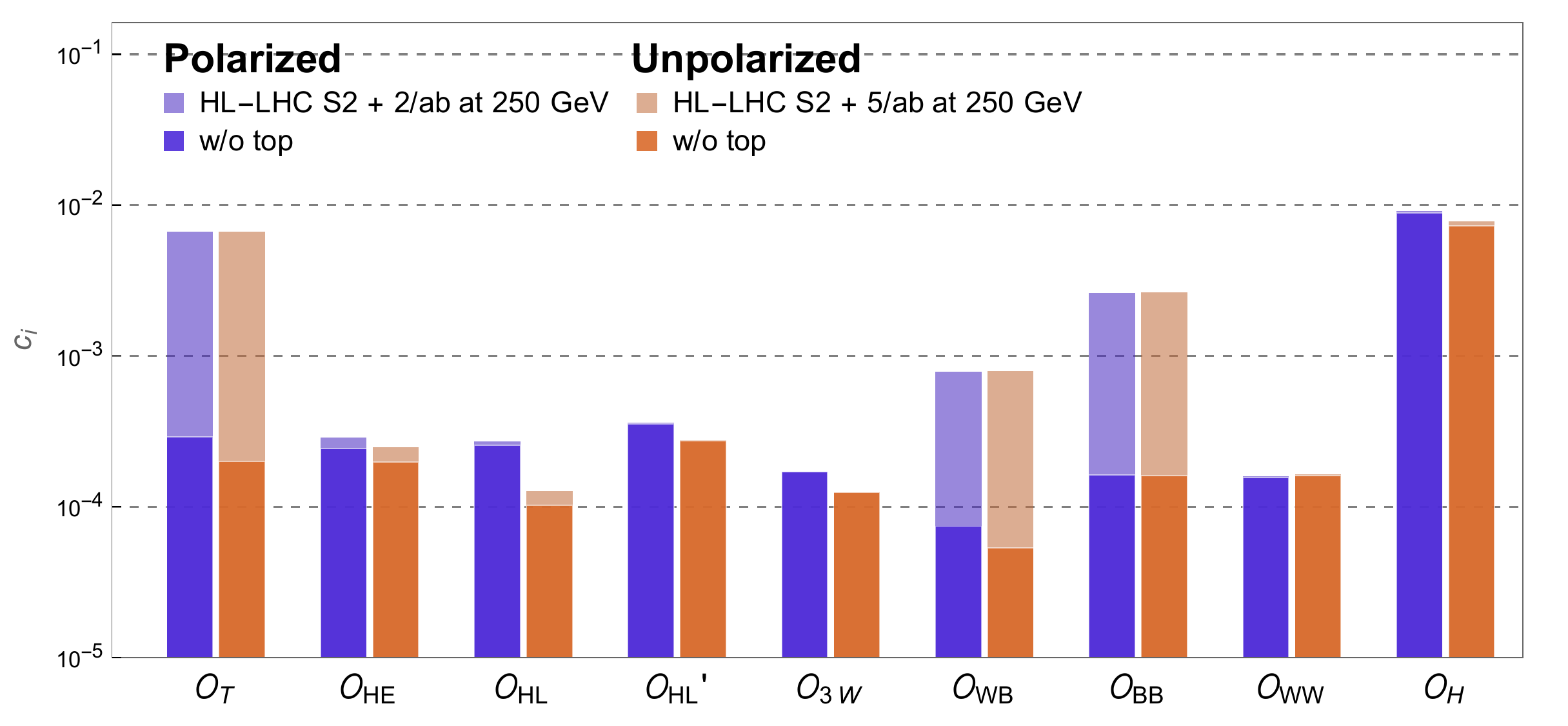}
      \caption{The role of beam polarization at $\sqrt{s}=250$ GeV on the Higgs coupling precision (upper) and Higgs operator constraints (lower) with (light shading) and without (dark shading) top effects. The polarized (blue) dataset includes equal sharing of two opposite beam polarizaitons (1~\iab each) of ILC, and the unpolarized (red) includes a higher luminosity of 5~\iab and enhanced Tera-$Z$ precision for EWPO. HL-LHC S2 top data is added.}
      \label{fig:PolHcoup}
\end{figure}

\clearpage

\subsection{The role of electro-weak measurements}

In \Fig{fig:Comp250} we assess the importance of the EWPO and TGC datasets in the global fit by varying the precisions of EWPO and TGC measurements. The height of the bars indicates the ratio of the bounds obtained with the current LEP/SLD precision and with future precisions of ILC250 (except Tera-$Z$ for unpolarized EWPO). Thus, the bars show improvements of global-fit results obtained from the improvement of EW dataset. The exact dataset values used are collected in \autoref{tab:ewpo}. 
The three bars show different global fits: the first on the reference fit without top operators, and the last two on the fits with top operators including either current LHC run 2 or future HL-LHC S2 bounds on top EW couplings. 

The EW measurements generally become less powerful in the presence of top operators in particular when the top operators are relatively not well constrained at LHC run 2; this can be seen as much smaller heights of the second bars. The degeneracies between Higgs and top effects are one of the bottlenecks for taking advantage of the EW precision measurements. As the degeneracies can be reduced by tighter constraints on the top operators from HL-LHC, the importance of EW/TGC dataset also increases. This means that top data and EW/TGC data are complementary or somewhat orthogonal in resolving those degeneracies; but EWPO and TGC alone are not powerful enough to remove those degeneracies. 
The EW measurements are more important for unpolarized lepton colliders as beam polarization doubles the independent set of observables.
The importance is also more pronounced on operator constraints since EW/TGC dataset can directly constrain some operators and reduce degeneracies between operators.

Among all EW observables, the Higgs mass is the most influential one in global fits. White markers indicate the results that are obtained with $\delta m_h$ fixed to the future LHC/ILC precision of 15 MeV, while all other EWPO continue to vary. These results show that most of the precision improvement from EWPO comes from the improvement of $\delta m_h$. This is in agreement with Ref.~\cite{Lepage:2014fla} that did not even consider top effects, implying that the Higgs mass is most important regardless of model-independent top effects. 
Finally, the further improvement of $\delta m_h$ from the 15 MeV does not bring additional improvements of global fits at ILC 250.

In summary, the improvements of EW/TGC dataset from the current LEP/SLD precision to the future LHC/ILC precision are very important for the global fits, as well as strong bounds on the top-quark electro-weak operators.

\begin{figure}[t]
      \centering
      \includegraphics[width=\textwidth]{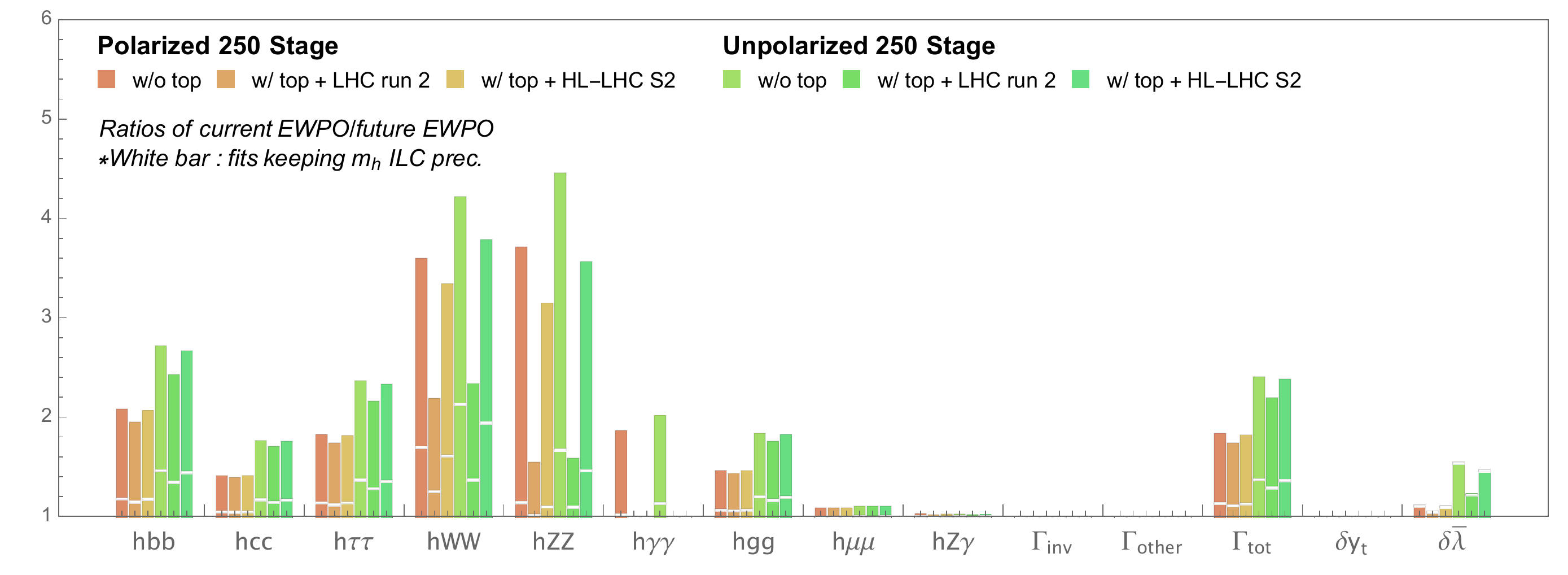}
      \includegraphics[width=\textwidth]{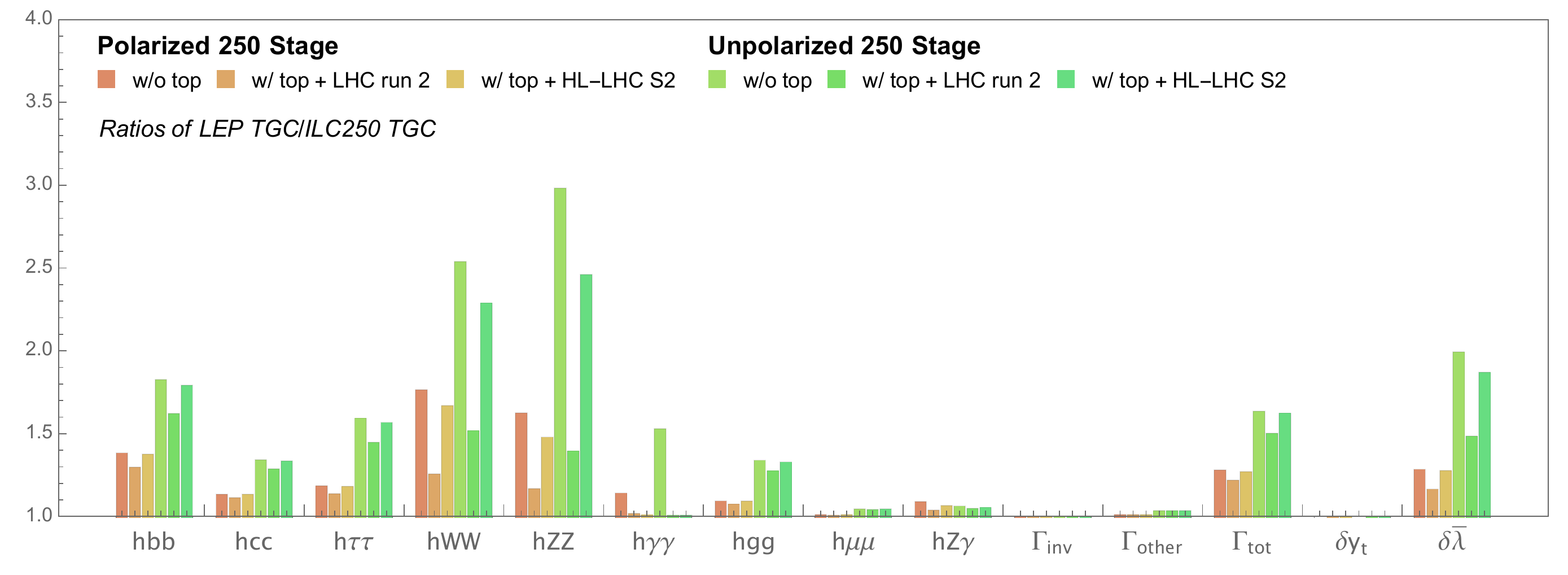}
      \includegraphics[width=\textwidth]{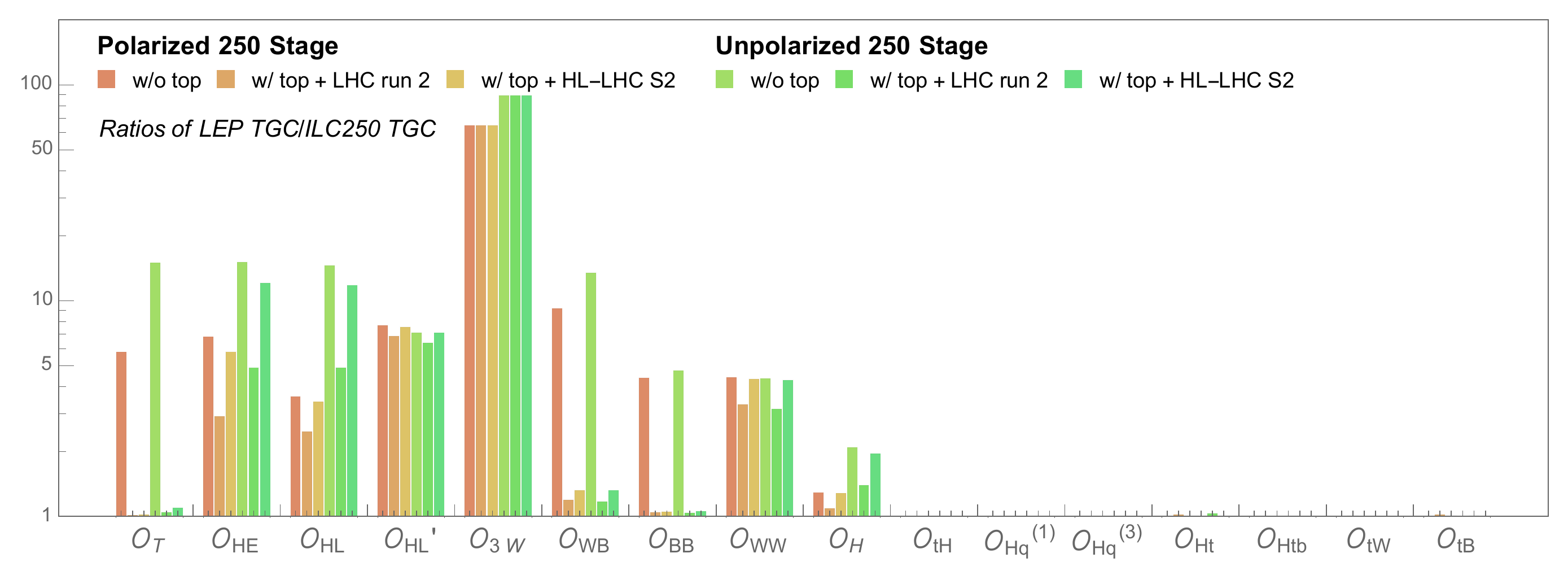}
      \caption{The ratio of Higgs coupling precision obtained by varying EWPO and TGC uncertainties, from current LEP/SLD uncertainties of EWPO (upper panel) and TGC measurements (central panel) to the future uncertainties of ILC250 (Tera-$Z$ for unpolarized EWPO). The white markers indicate the results with the uncertainty $\delta m_h$ fixed to the future LHC/ILC precision of 15 MeV while all other EWPO uncertainties varied. The third panel presents the results of the variation of the TGC precision in the operator basis. The first three columns are with polarized dataset, and the last three are with unpolarized.}
      \label{fig:Comp250}
\end{figure}

\clearpage

\subsection{Results for higher-energy operation -- ILC 500 stage} 
\label{sec:ILC500}

The nominal ILC programme includes 2~\iab at $\sqrt{s}=$ 250~\gev and 4~\iab at $\sqrt{s}=$ 500~\gev. The Higgs measurements at higher energy, with much more abundant VBF production, lead to a significantly improved Higgs fit~\cite{Barklow:2017awn}. The sensitivity of measurements in $e^+e^- \to W^+W^-$ increases with center-of-mass energy. And, finally, the high-energy run opens up top quark pair production, allowing for a precise characterization of the electro-weak interactions of the top quark~\cite{Durieux:2019rbz, Durieux:2018tev, Amjad:2015mma}.

The results of a fit to the combined ILC250+ILC500 scenario are presented in \Fig{fig:500higgs}. The upper and lower panels again correspond to the physical Higgs coupling basis and the effective operator coefficients. For each coupling precision and operator constraint, four results are shown. The first bar corresponds to the 22-parameter reference fit to the two ILC stages of Ref~\cite{Barklow:2017suo}. The second, third and fourth bar are obtained with the full 29-parameter basis. The second bar uses only Higgs and EW observables, the third adds the HL-LHC top constraints in the S2 scenario, and the fourth adds the constraints on top EW operators from the ILC run at $\sqrt{s}=$ 500~\gev.

Notably, global fits in the 29-parameter basis now converge without top production data (second bar). The dataset with two ILC energy stages provides a sufficiently rich set of measurements to constrain also the top operator coefficients. However, the Higgs coupling precision in this case is significantly degraded with respect to the reference fit, and indirect constraints on top operators remain much worse than those from measurement in top production. Higgs coupling precision without top production data is not as robust as in ILC250 since ILC500 now provides a new energy scale so that the degeneracies with top effects become more important. 
Adding ILC 500 top production data (fourth bar) can almost fully recover the Higgs coupling precision. HL-LHC top data is also useful, but not quite sufficient to restore the precision to that of the reference fit without top operators.

A more remarkable impact is found in the basis of operator coefficients. In particular, the coefficients $c_{Ht,\, tB}$ and $c_{T, \, WB, \,BB}$, which mix strongly with each other, are strongly affected. The mixing consequently affects the precision of $\lambda$ and $g(hZ\gamma)$ too. Only in the last scenario, including ILC 500 top data, the degeneracies between Higgs and top operators are fully resolved and all operators and couplings are constrained as well as in the reference fit without top operators.

All in all, ILC 500 is capable of precise and the model-independent test of the SMEFT with top effects. The precision of the top electro-weak coupling measurements at the HL-LHC and in $e^+e^- \rightarrow t\bar{t}$ is essential for model-independent Higgs coupling precision.

\begin{figure}[t]
      \centering
      \includegraphics[width=\textwidth]{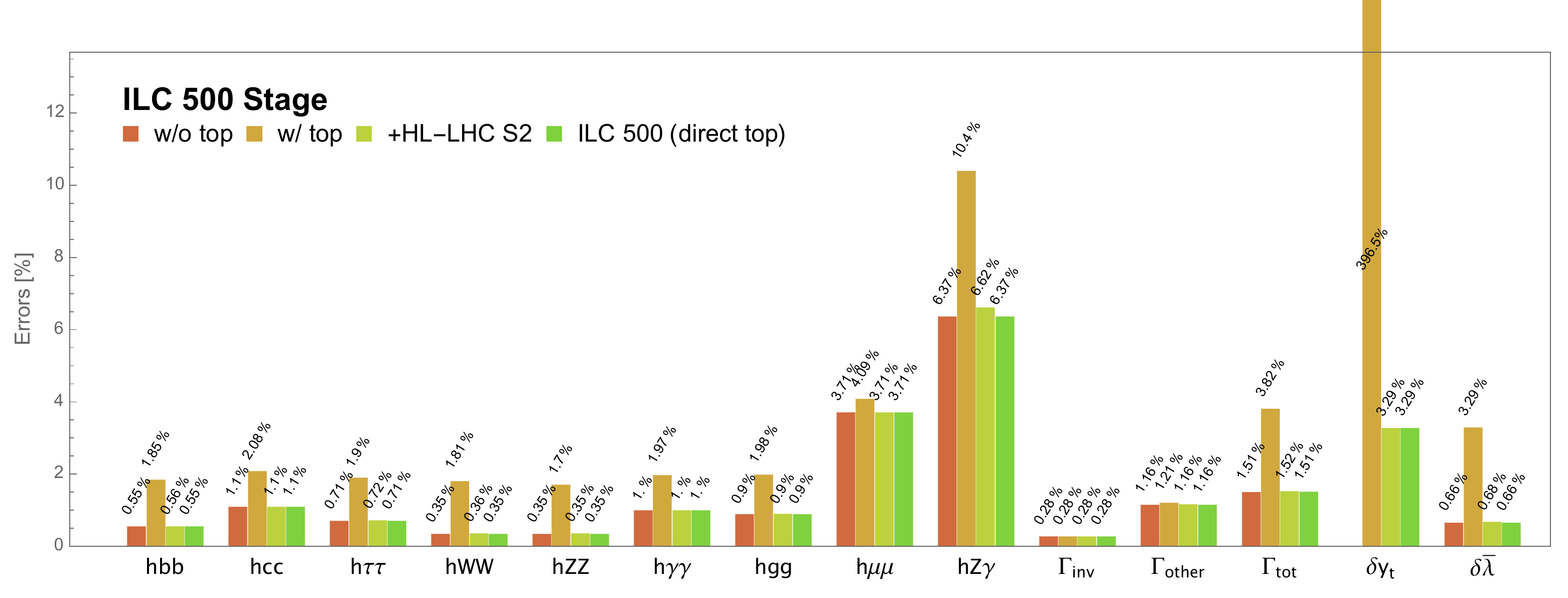}
      \includegraphics[width=\textwidth]{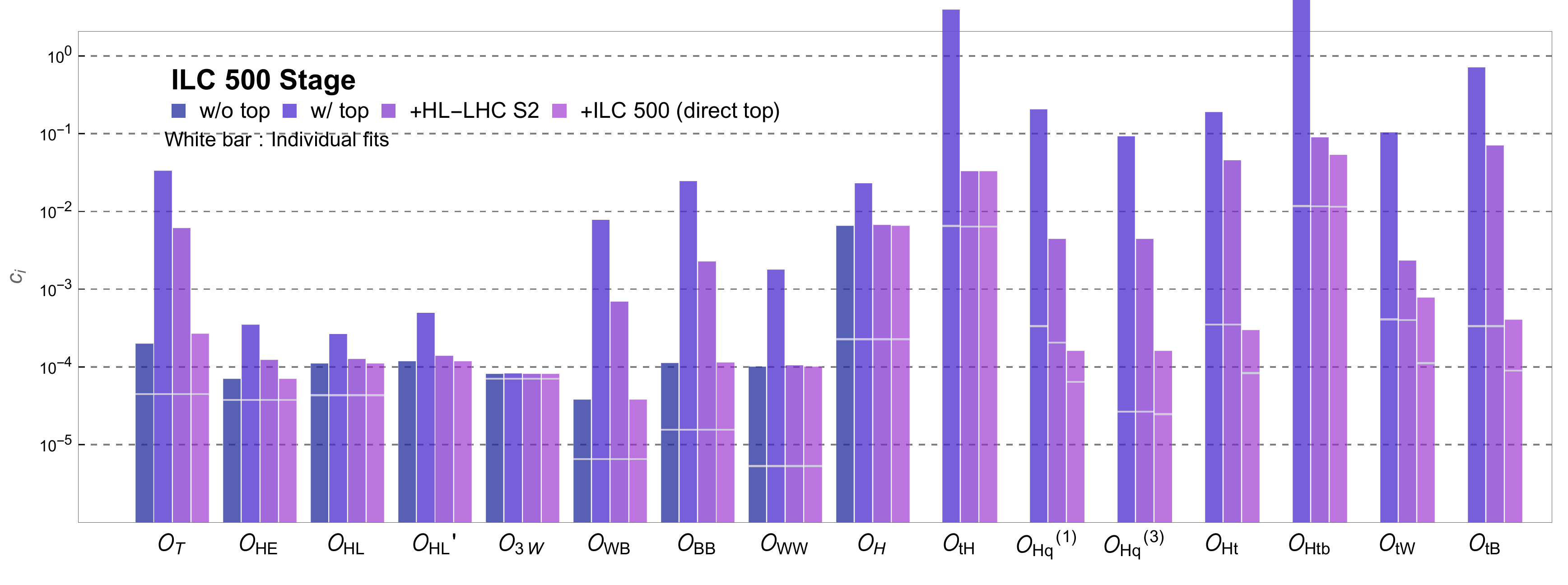}
      \caption{Global-fit results for the ILC250+ILC500 scenario. The upper panel presents the result in terms of the precision on the physical Higgs couplings. The lower panel presents the $1\sigma$ bounds on the operator coefficients, renormalized at $Q_0=1$~\tev with a suppression scale $v$. In both panels, the first column corresponds to a 22-parameter fit without top operators, that is used as a reference throughout the paper. The second column presents the result that is obtained when the basis is extended with the seven top operator coefficients described in \Sec{sec:topop}. In the third column, LHC run 2 top data are added. In the fourth column, ILC top measurements at $\sqrt{s}=$ 500~\gev are added. In the bottom panel, white marks are results with only one operator. Results are tabulated in \autoref{tab:HiggsCoup} and \ref{tab:OpConst}.}
      \label{fig:500higgs}
\end{figure}

\clearpage

\subsection{The Higgs self-coupling}
\label{sec:higgsself}

The triple coupling or self-coupling of the Higgs boson is one of the key objectives of high-energy physics in the next decades~\cite{DiMicco:2019ngk}. A robust extraction of this parameter is an important consideration in the design of the Higgs factory and its operating scenarios~\cite{deBlas:2019wgy,deBlas:2019rxi,Barklow:2017awn,DiVita:2017vrr}. 

The Higgs self-coupling $\lambda$ and the operator coefficient $c_6$ are expected to be measured in di-Higgs boson production at the LHC before the ILC turns on~\cite{DiMicco:2019ngk}. However, a model-independent extraction from LHC data is difficult. One of the most challenging aspects is that the $gg \to hh$ process receives contributions from several diagrams. In addition to the diagram with a triple-Higgs-boson vertex, box and loop diagrams involving $t\bar{t}h$ vertices yield sizeable contributions to the total rate. This results in a strong dependence on several operator coefficients, among which $c_{tH}$~\cite{Azatov:2015oxa}.
\begin{table}[t]
\begin{center}
\begin{tabular}{c|l|c }
  \toprule
   & & ~$\left< (\delta\sigma(Zhh))^2 \right>^{1/2}$ \\
  \hline\hline
  \multirow{2}{*}{w/o top op.} & ~ILC 250 & 2.66\% \\
  & ~ILC 250\,+\,500  & 2.13\% \\
  \hline
  \multirow{3}{*}{w/top op.}
  & ~ILC 250\,+\,500 (Higgs/EW) & 8.06\% \\
  & ~+\,HL-LHC top& 2.62\% \\
  & ~+\,ILC 500 top & 2.13\% \\ 
  \bottomrule
  \end{tabular}
  \caption{Total uncertainties of the $\sigma(Zhh)$ contributed from EFT operators, evaluated at $Q=500$ GeV. Measurement uncertainties are estimated to be 16\%~\cite{Duerig:2016dvi,Bambade:2019fyw} but not shown. These results are independent on $Q_0$ (\autoref{app:RGdep}). The first two rows correspond to the fit on the 22-parameter basis without top operators, for ILC Higgs/EW data at $\sqrt{s}=$ 250~\gev and for the complete programme at 250 + 500~\gev. The three last rows correspond to the 29-parameter fit on the complete basis with top operators. The first of them includes Higgs/EW data only, while the second and third add top physics measurements at the HL-LHC and ILC500, respectively.}
\label{tab:Zhhtable}
\end{center} 
\end{table}

In our EFT fit, as discussed at the end of \Sec{sec:powercounting}, the Higgs self-coupling $\lambda$ and $c_6$ appear together in the $\bar{\lambda}$ parameter (\Eq{eq:lambda-bar}), which is constrained by measurement of the Higgs mass. 
Figure~\ref{fig:500higgs} shows that the reference result $\delta \bar{\lambda} = $ 0.66\% of the 22-parameter fit at ILC500 is degraded to 3.29\% when the basis is extended with top operators. The precision recovers to 0.68\% after inclusion of HL-LHC top data in the S2 scenario, and fully recovers with the top measurements in $e^+e^- \to t\bar{t}$ production at ILC500. 

Even if the $\bar{\lambda}$ coefficient is tightly bounded, we cannot constrain $\lambda$ and $c_6$ individually with the 250~\gev data alone (remember that finite loop effects from the Higgs operator on the $Zh$ cross section are not included in the fit). The  coefficients $c_6$ is precisely measured by $\sigma(Zhh)$ at ILC500. This measurement separates the two parameters, since $\lambda$ and $c_6$ enter in different combinations in that process. Here, we evaluate the effect of Higgs and top operator coefficients on the extraction of $c_6$.

\autoref{tab:Zhhtable} shows the total uncertainty on the cross-section $\delta \sigma(Zhh)$ from EFT coefficients. This EFT error contribution in the model-independent extraction of $c_6$ must be kept small - by precise measurements of single-Higgs and top production rates - to convincingly attribute a measured deviation in the double Higgs production cross section to a shift in the triple Higgs coupling. The errors are evaluated at $Q=500$ GeV, and the results are independent of the choice $Q_0$.

With only Higgs+EW data, the EFT uncertainty contribution is 8.06\%, better than the expected measurement error of 16\%~\cite{Duerig:2016dvi,Bambade:2019fyw}. This means that the measurement of $c_6$ from  $\sigma(e^+e^- \to Zhh)$ is robust, even in the presence of top operators. Addition of the HL-LHC top production data improves the uncertainty to 2.6\%. Finally, $e^+e^- \to t\bar{t}$ at $\sqrt{s}=$ 500~\gev fully recovers the constraint 2.13\% of the reference fit without top operators. 

An interesting aside is that, even though the final errors are 2.13\% both with or without top operators, individual EFT contributions to the total error are quite different. In the fit on the extended basis, $c_T$ becomes the dominant source of EFT uncertainties while $c_H$ was the dominant contribution before top operators were added. 

In summary, the extraction of the Higgs boson self-coupling from the di-Higgs production rate at the ILC at $\sqrt{s}=$ 500~\gev is robust against the impact of other Higgs operators and top operators. With the inclusion of HL-LHC and ILC500 results on top electro-weak couplings, the effect of those operators on the extraction is reduced to well below the expected measurement uncertainty.

\subsection{Indirect bounds on top-EW couplings}
\label{sec:indirecttopEW}

\begin{figure}[t!]
      \centering
      \includegraphics[width=\textwidth]{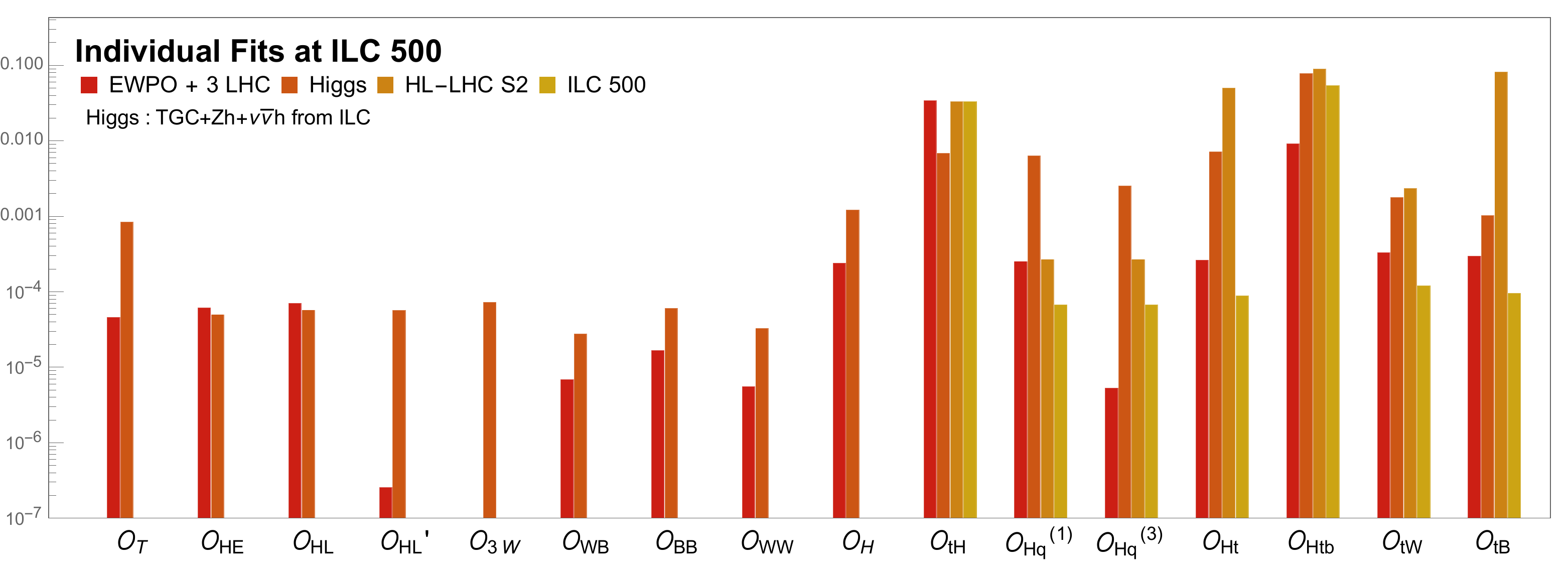}
      \caption{The individual constraints on the EFT coefficients by each set of measurements at ILC 500 and HL-LHC stages. The first column (EWPO+3 LHC) shows the individual bounds by 9 EWPOs and 3 branching ratios measurements from LHC. The second column (Higgs) is from the TGC measurements and Higgs production processes ($Zh\+\nu\bar{\nu}h$) from ILC. The third and fourth ones represent those from direct top data of HL-LHC and ILC 500-top respectively. }
\label{fig:IndFitTopEW}
\end{figure}

The effect of the top operator coefficients on the Higgs and EW observables offers a way to probe these operators during the first ``Higgs factory" stage of the ILC project at $\sqrt{s}=$ 250 GeV, when the top quark pair production process is not yet accessible. The indirect bound from the Higgs/EW fit is therefore the first top quark physics of the new project. 

The individual bounds on all operator coefficients from a fit to the ILC500 projection are presented in \Fig{fig:IndFitTopEW}. The measurements are grouped in two broad categories: electro-weak precision measurements and Higgs coupling measurements. These indirect determinations are compared to the projections for HL-LHC top physics results in the S2 scenario and to the bounds expected from ILC top measurements at $\sqrt{s}=$ 500~\gev.

The Higgs/EW precision data at $\sqrt{s}=$ 250~\gev are sensitive to the top electro-weak couplings through loop effects. The complete set of relations is found in \autoref{tab:Topops1}. The Higgs decay widths $\Gamma(h\rightarrow \gamma \gamma)$ and $\Gamma(h\rightarrow Z\gamma)$ are particularly sensitive to the dipole operator coefficients $c_{tW}$ and $c_{tB}$ that affect the $Zt\bar{t}$ and $\gamma t\bar{t}$ vertices and to the top-quark Yukawa coupling and hence $C_{tH}$. Figure~\ref{fig:250higgs} and \ref{fig:500higgs} show that the ILC precision can measure these widths to ${\cal O}(1)\%$ precision. In \Fig{fig:IndFitTopEW} the individual bounds from Higgs measurements are indicated as the second, orange bar. The indirect bounds from Higgs observables are indeed found to be quite powerful for a number of operators. The individual bounds on $c_{Ht}$, $c_{tH}$ and, especially $c_{tB}$ are expected to improve the bounds from the HL-LHC measurements by one to two orders of magnitude.

Also the electro-weak precision measurements, indicated with the first, red bar in \Fig{fig:IndFitTopEW}, offer very good sensitivity. Renormalization-group mixing leads to a dependence of the electro-weak precision observables on top operators. Most individual bounds on top operator coefficients exceed
the S2 projection for the HL-LHC, and in several cases ($c_{Hq}^{(3)}$, $c_{Ht}$, $c_{tB}$) by more than two orders of magnitude. In some cases, these individual bounds are competitive even when compared to the limits from direct $e^+e^- \rightarrow t\bar{t}$ production in the 500~\gev run. 

The Higgs and electro-weak measurements at $\sqrt{s}=$ 250~\gev therefore provide a powerful indirect handle on the top electro-weak couplings and can improve the legacy bounds from the LHC top physics programme. If these couplings receive corrections from new physics, the Higgs/EW measurements may well be the first place where significant deviations from the SM show up. 
We find, however, that these individual bounds are generally not robust and an unambiguous determination of the coefficients in a global fit requires the inclusion of $e^+e^- \rightarrow t\bar{t}$ data at higher energy. The bound on the dipole operator $c_{tB}$ is the only case, where the addition of the Higgs/EW data improves the global limit over the result of a fit to HL-LHC top physics data. For this operator a weak sensitivity of rare associated production processes ($pp \rightarrow t\bar{t}\gamma$, $pp \rightarrow t\bar{t}Z$, $pp \to tZq$) at the HL-LHC coincides with a relatively strong sensitivity of the Higgs/EW observables.

\subsection{The top-quark Yukawa coupling}
\label{sec:topyukawa}

The top-quark Yukawa coupling is an important parameter of the SM and merits a dedicated discussion. 

The indirect sensitivity to $c_{tH}$ stems from loop-induced Higgs boson decay widths $\Gamma(h\rightarrow \gamma \gamma)$, $\Gamma(h\rightarrow Z\gamma)$ and $\Gamma(h \rightarrow gg)$ that involve top-quark loops. No other observables in \autoref{tab:Topops1} depend on $c_{tH}$. In \Fig{fig:IndFitTopEW} the indirect bound on $c_{tH}$ from the ILC Higgs data is much more stringent than the direct measurement in $pp \rightarrow t\bar{t}H$ at the HL-LHC. This finding confirms the good indirect sensitivity to the Yukawa coupling found by Ref.~\cite{Boselli:2018zxr}.

However, as for the other top operators, the excellent individual bounds do not translate into a robust measurement of the top-quark Yukawa coupling. The top-quark Yukawa contributions to the Higgs decay widths are entangled with other contributions. The degeneracy between the different operator coefficients cannot be lifted with Higgs and electro-weak data alone. Indeed, the global bound on the top Yukawa in a model-independent fit is entirely dominated by the direct measurements in $t\bar{t}H$ production. 

Before proceeding to discuss the direct measurements, we explore possible improvements in the indirect constraints on top Yukawa. First, $\Gamma(h \rightarrow gg)$ was identified as the most promising indirect constraint in Ref.~\cite{Boselli:2018zxr}. In addition to $c_{tH}$, $\Gamma(h\to gg)$ receives contributions from $c_{tG}$ and $hgg$ contact operators. Although the former is likely well constrained by the measurements of the top quark pair production cross section at hadron colliders with a global limit -0.4 $< c_{tG}/\Lambda^2 < $ 0.4~\cite{Hartland:2019bjb}, the degeneracy with the $hgg$ operator is very hard to lift~\cite{Azatov:2016xik}. On the other hand, for $\Gamma(h \to \gamma \gamma)$ and $\Gamma(h\to Z \gamma)$, $c_{tH}$ must be separated from $c_{tW,tB}$ as well as $c_{WB,BB}$. These operators can perhaps be better constrained if differential analyses of $t\bar{t}X$ production at the LHC can improve the bounds significantly over the precision envisaged S2 scenario.

The direct measurements of the top quark Yukawa coupling have been discussed in some detail in Ref.~\cite{Fujii:2019zll}. The HL-LHC measurements of the $t\bar{t}H$ cross section yield a precision of 3.29\% in our fit\footnote{We note here that, while the extraction of the top Yukawa coupling from the $t\bar{t}h$ cross section is the most robust handle on this coupling, it is by no means immune to degradation due to the presence of other operators, as demonstrated by recent global fits~\cite{Hartland:2019bjb}. Also the claim of Ref.~\cite{Plehn:2015cta} of a 1\% precision on the Yukawa coupling from the ratio of the $t\bar{t}H$ and $t\bar{t}Z$ rates must be carefully assessed in a global environment.}. 
The precision can be further improved with $t\bar{t}h$ measurements from ILC runs at high energy. The ILC must be operated at $\sqrt{s}=$ 550~\gev to match the 3\% precision of the HL-LHC in the direct extraction from the $e^+e^- \to t\bar{t}h$ channel and confirm a possible deviation from the SM. 
The operation of a linear collider at 1~\tev can improve the constraint by another factor two. The analysis in Ref.~\cite{Durieux:2019rbz} shows that the extraction from the $t\bar{t}h$ rate at the ILC is robust when data at 1~\tev are added to constrain the coefficients of $e^+e^-t\bar{t}$ operators in a global fit. The combination of all these measurements will allow to measure top-quark Yukawa coupling with approximately 1\% precision.

Finally, we note an interesting direction to constrain $c_{tH}$. The operator RG mixes with $c_6$. But we did not need to include the running of $c_6$ because $c_6$ enters in only one observable $\sigma(Zhh)$ at $\sqrt{s}=500$ GeV. If finite loop corrections of $c_6$ to $\sigma(Zh)$ at $\sqrt{s}=250$ GeV~\cite{McCullough:2013rea} can be considered in the future, multiple energy scales may yield a new way to constrain the $c_{tH}$'s RG contributions. The HL-LHC Higgs pair production~\cite{DiMicco:2019ngk}  may even be used too. A more dedicated study is postponed to a future work~\cite{TianPeskinLambda}.

\medskip
In summary, we confirm the good indirect sensitivity of the 250~\gev run to the top-quark Yukawa coupling that was signalled by Ref.~\cite{Boselli:2018zxr}. The ILC Higgs programme at 250~\gev can provide individual, single-parameter bounds on $c_{tH}$ that are more stringent than those from the HL-LHC. Hence, the ILC initial stage may observe strong deviations from the Standard Model predictions in the $H\rightarrow \gamma \gamma $ and $H \rightarrow gg$ decay rates, should the top-quark Yukawa coupling be affected by physics beyond the Standard Model. However, we find that the indirect extraction of $c_{tH}$ from the Higgs branching ratios is not robust in a global fit. This means that a deviation from the SM cannot unambiguously be attributed to $c_{tH}$; the effect could be caused by another operator. To pinpoint the new physics effect, a measurement of the associated production rate of a Higgs boson with a top quark pair at higher energy remains necessary.

\section{Summary}
\label{sec:summary}

In this paper, we have presented a global SMEFT fit for the ILC by combining contributions from Higgs/EW and top sectors. In addition to the 22 parameters of Higgs operators and SM parameters which are complete at the lowest order~\cite{Barklow:2017suo,deBlas:2019rxi}, this requires seven top-quark operators. They contribute to the Higgs/EW precision observables at the one-loop suppressed order compared to those of Higgs operators (except for tree-level shifts of $t\bar{t}h$ and $t\bar{t}Z$ couplings), but they are leading contributions of top operators. The 29-parameter fit then provides a complete and model-independent description of leading Higgs+EW+top effects on precision observables at future lepton colliders.

The loop calculations of top effects necessarily introduce new uncertainties. The top-loop contributions were included through the renormalization mixing with Higgs operators. Although good proxies of full one-loop effects, the RG contributions are only a subset of the full one-loop results\footnote{
While including full one-loop effects for all the relevant observables is beyond the scope of this paper, we recognize that the important effect from the left-over finite contributions would be mainly coming from $c_{tB}$ term in $\Gamma(h\to\gamma\gamma)$, provided the prospected constraints on top operators from HL-LHC; see details in Appendix~\ref{app:durieux2}.}.
The impact of RG effects depend on the renormalization scale $Q$ to which the Higgs operators will run from a common scale $Q_0$ where all the operators are defined. $Q={\rm max}(m_t,Q_{\rm proc})$ was used as the default choice, but the variations with respect to  $Q=Q_{\rm proc}$ and ${\rm max}(2m_t,Q_{\rm proc})$ reflect the errors from even higher-order quantum corrections.

With this formalism, we have evaluated the prospects of the ILC in the extended 29-parameter basis, in comparison with those of the previous 22-parameter fit of Ref.~\cite{Barklow:2017awn}. We have found that the Higgs fit at ILC 250~\gev is strongly affected by the inclusion of top effects. This result, and those of Ref.~\cite{Vryonidou:2018eyv,Durieux:2018ggn},  show that results obtained with today's state of the art SMEFT fits must be interpreted with care, as they may not hold in a more complete fit\footnote{As a familiar example, one can consider the well-known $\kappa$-framework to interpret Higgs boson coupling measurements. Clearly, the limitations of the $\kappa$ framework affect the conclusions of a benchmark analysis in important ways: due to the assumptions inherent in the fit, it fails to acknowledge the role of EWPO and beam polarization, to name just two examples. The importance of these aspects becomes apparent only in a more advanced EFT approach.}.

To mitigate the effect of the additional degrees of freedom, their contribution must be disentangled by including precise measurements of the top quark electro-weak couplings. The precision of the HL-LHC top physics program, as envisaged in the S2 scenario of Ref.~\cite{Durieux:2019rbz}, is sufficient to restore the precision of the physical Higgs boson couplings. But it still leaves several of the underlying EFT operator coefficients relatively poorly constrained. This implies that, even if the Higgs couplings can be precisely determined and their deviations from SM predictions can be well detected, it becomes harder to pinpoint the source of the deviations, should they be observed. The projection of the EFT fit results on the physical Higgs couplings offers a set of bounds with an intuitive {\em physical} interpretation, but can obscure potentially very important information. It is therefore important to present both sets of results -- Higgs coupling precisions and operator constraints -- to fully characterize the analyzing power of the data.

We also assess whether the Higgs/EW precision data in the initial stage at $\sqrt{s}=$ 250~\gev can place bounds on top operators before top quark pairs are produced at the ILC.
These indirect bounds are very very competitive if only a single top operator is considered in a fit. For example, the $h \rightarrow gg$ and $h \rightarrow \gamma \gamma$ decay rates yield indirect sensitivity on the top coupling at sub-\% precision, well beyond what can be achieved in associated $t\bar{t}h$ productions in all planned lepton or hadron colliders. However, these individual bounds turn out to be not robust in the presence of other operators with non-zero coefficients. A robust determination of top operators requires the inclusion of $e^+e^- \rightarrow t\bar{t}$ and $e^+e^- \to t\bar{t} h$ data at higher center-of-mass energy.

In the complete ILC programme, with a second energy stage at 500 GeV, a precise characterization of the $e^+e^- \rightarrow t\bar{t}$ process provides very precise constraints on the top operators. A combined fit on Higgs, EW and top quark precision data then over-constrains the EFT fit, yielding robust bounds on all 29 operators and the unambiguous identification of the origin of any deviations from the SM. 

\acknowledgments{
This work builds on the effort of the LHC collaborations and the ILC and CLIC simulation studies. We would like to acknowledge the work of our colleagues in this place. We owe special thanks to Gauthier Durieux, Michael Peskin and Cen Zhang for their feedback on the project and write-up. 
SJ and JL are supported by Grant Korea NRF-2019R1C1C1010050, 2017R1D1A1B03030820, 2015R1A4A1042542, and SJ also by POSCO Science Fellowship. MP and MV are supported by the Spanish national program for particle physics, projects FPA2015-65652-C4-3-R (MINECO/FEDER) and PGC2018-094856-B-100, and PROMETEO grant 2018/060 of the Generalitat Valenciana. MP is supported by the “Severo Ochoa” Grant SEV-2014-0398-05, reference BES-2015-072974.
JT was supported by the Japan Society for the Promotion of Science (JSPS) under Grants-in-Aid for Science Research 15H02083.
}

\newpage
\appendix

\section{Typos and corrections in Ref.\cite{Barklow:2017awn}}
\label{app:typos}

Ref.~\cite{Barklow:2017awn, Barklow:2017suo} and the distributed C++ code, on which our study is based, contain several mild typos and mistakes. We collect them in this appendix using the same notation, and we make the corrected Mathematica code available upon request.
\begin{enumerate}
 \item Typos in the paper \cite{Barklow:2017awn}:
 \begin{enumerate}
  \item  Some expressions for the TGCs written in the Section 3 and Appendix of Ref.~\cite{Barklow:2017awn} should be corrected in a right form. The amplitudes of $e_{R,L}^-e_{L,R}^+ \to W_L^+ W_L^-$ are 
 \bea
 \mathcal{A}_R &=& e^2 \kappa_A \- g_R g_Z \kappa_Z, \\
 \mathcal{A}_L &=& e^2 \kappa_A \- g_L g_Z \kappa_Z - \frac{g_W^2}{2}.
 \eea
 The $\kappa_V$ denote the deviation of the triple gauge couplings between $W^+W^-$ and $V(=Z,A)$. The $g_W$ is the W boson coupling to leptons(electron and neutrino) that is introduced only to the amplitude for $e_L^-e_R^+$ through the neutrino exchange diagram. In the Ref.~\cite{Barklow:2017awn}, the sign of the terms for $g_L(g_R)g_Z\kappa_Z$ is written as plus but the negative one is correct. The fits were done with the correct sign but one term for $\delta e$ in $\delta g_{Z,eff}$ was missed. By the definition of the effective TGC (\cite{Barklow:2017awn}), 
 \beq
 \delta g_{Z,eff} = \frac{1}{g c_w^2} \left(2 \Delta \mathcal{A}_L - \Delta \mathcal{A}_R \right), 
 \eeq
 the $\delta g_{Z,eff}$ has one more term which was missed and we find 
 \beq
 \delta g_{Z,eff} \= \delta g_Z \+ \frac{1}{c_w^2}((c_w^2-s_w^2)\delta g_L \+ s_w^2 \delta g_R \- 2 \delta g_W \+ 2 s_w^2 \delta e).
 \eeq
  This can be checked using the fact that the TGCs are independent of $c_{WB,BB,WW}$ because the $\Delta \mathcal{A}_{R,L}$ are also independent of them.
 \end{enumerate}
\item Typos in the code distributed with \cite{Barklow:2017awn}:
\begin{enumerate}
  \item\label{item:typo1} $\delta g_R \= -c_w^2 \delta g + (1+c_w^2) \delta g^\prime - \frac{1}{2s_w^2} c_{HE} - \frac{1}{2}c_w^2(8c_{WW}) + c_w^2(8c_{WB}) + \frac{1}{2}\frac{s_w^2}{c_w^2}(1+c_w^2)(8c_{BB}) $ : the sign of the $(8c_{WW})$ was reversed.
  
  \item \resizebox{0.95\hsize}{!}{ $b_L \= \frac{1}{(1-2s_w^2)} \left[ c_w^2(1-2s_w^2\frac{m_Z^2}{s})(8c_{WW})+2s_w^2(1-2s_w^2)\frac{m_Z^2}{s}(8c_{WB})-\frac{s_w^4}{c_w^2}(1-2c_w^2\frac{m_Z^2}{s})(8c_{BB}) \right]$} \\ : $-\frac{1}{c_w^2}(1-2c_w^2\frac{m_Z^2}{s})(8c_{BB})$ was used instead of the above $(8c_{BB})$ term.
  
  \item $k_h \= -\frac{E_Z m_h^2}{2k^2\sqrt{s}}-\frac{E_Z^2/m_Z^2}{(2+E_Z^2/m_Z^2)}\frac{m_h^2}{E_Z\sqrt{s}}$ in the $Zh$ measurement. \\
  : the first term was inserted as $ -\frac{E_Z m_Z^2}{2k^2\sqrt{s}}$. The Z boson mass was used for the Higgs mass.
  
  \item $\delta \sigma(R) \= 2\delta g_R + 1.40 \eta_Z + 1.02\eta_{ZZ}+18.6\delta Z_Z - 28.7 \delta Z_{AZ} + 0.56\eta_h + \cdots $ for the $Zhh$ cross section. \\
  : $2\delta g_L$ was used instead of the first term, $2\delta g_R$ in the estimate of the uncertainty of $\sigma(Zhh)$.

  \item In the fits for ``$+Zh$" columns in \autoref{tab:typoOp}, the $a_L,\,a_R$ parameters were inserted in the place of $b_L,\,b_R$ as error inputs of the Zh measurement but in ILC 250,500 fits the error inputs were used correctly.
  
  \item\label{item:typof} When the total cross section of $e^-e^+ \to Zh$ is prepared for the its error estimate in the fits for 250+350 GeV in the fourth column of \autoref{tab:typoHiggs}, only left-handed one was inserted, that is right-handed one was missing.
\end{enumerate}
\end{enumerate}

By correcting the typos, we re-obtain the Table 2 in Ref.~\cite{Barklow:2017awn} and Table 3 in \cite{Barklow:2017suo} in \autoref{tab:typoOp} and \ref{tab:typoHiggs}, respectively. 
The impact of the typos on the Higgs precision is rather mild while the impact on some operators is significant. Especially, the constraint on $c_{HE}$ decreases significantly from the ``+LHC'' column as shown in \autoref{tab:typoOp} mainly due to \autoref{item:typo1} and the difference reduces at the ILC 500 and ILC 250+500 fits where the influence of the LHC Higgs measurements becomes weak relatively. We also found that the missing right-handed $\sigma(Zh)$ part (\autoref{item:typof}) in the 250+350~\gev fit reduces the error of the cross section to the half of the original value as shown in the ``+1.5/ab 350~\gev" column of \autoref{tab:typoHiggs}.

\begin{table}[h] 
\begin{tabular}{c|cccccccccc}
  \toprule
      & \multicolumn{2}{c}{prec. EW} & \multicolumn{2}{c}{+\,$WW$} & \multicolumn{2}{c}{+\,LHC} & \multicolumn{2}{c}{+\,$Zh$} & \multicolumn{2}{c}{ILC\,250}\\
  \cmidrule(lr){2-3}\cmidrule(lr){4-5}\cmidrule(lr){6-7}\cmidrule(lr){8-9}\cmidrule(lr){10-11}
  & rep. & corr. & rep. & corr. & rep. & corr. & rep. & corr. & rep. & corr.\\
  \midrule   
  $c_{T}$ & 0.011 & 0.011 & 0.051 & 0.051 & 0.051 & 0.051 & 0.048 & 0.049 & 0.052 & 0.050 \\
  $c_{HE}$ & 0.043 & 0.043 & 0.026 & 0.026 & 0.085 & 0.026 & 0.047 & 0.024 & 0.055 & 0.025 \\
  $c_{HL}$ & 0.042 & 0.042 & 0.035 & 0.035 & 0.035 & 0.035 & 0.032 & 0.03 & 0.039 & 0.030 \\
  $c_{HL}^\prime$ & . & . & 0.028 & 0.028 & 0.028 & 0.028 & 0.028 & 0.027 & 0.047 & 0.039 \\
  $8\,c_{WB}$ & . & . & 0.078 & 0.078 & 0.080 & 0.080 & 0.076 & 0.078 & 0.090 & 0.091 \\
  $8\,c_{BB}$ & . & . & . & . & 0.20 & 0.20 & 0.16 & 0.18 & 0.11 & 0.17 \\
  $8\,c_{WW}$ &  . & . & . & . & 0.21 & 0.21 & 0.13 & 0.17 & 0.13 & 0.19 \\
  $8\,c_{H}$ & . & . & . & . & . & . & 1.12 & 1.12 & 1.20 & 0.99 \\
\end{tabular}
\begin{tabular}{c|cccccccccccc}
   \toprule
   & \multicolumn{2}{c}{prec. EW} & \multicolumn{2}{c}{+\,$WW$} & \multicolumn{2}{c}{+\,LHC} & \multicolumn{2}{c}{+\,$Zh$} & \multicolumn{2}{c}{ILC\,500} & \multicolumn{2}{c}{ILC\,250+500}\\
   \cmidrule(lr){2-3}\cmidrule(lr){4-5}\cmidrule(lr){6-7}\cmidrule(lr){8-9}\cmidrule(lr){10-11}\cmidrule(lr){12-13}
   
   & rep. & corr. & rep. & corr. & rep. & corr. & rep. & corr. & rep. & corr. & rep. & corr.\\
   \midrule  
   $c_{T}$ & 0.011 & 0.011 & 0.046 & 0.046 & 0.047 & 0.047 & 0.041 & 0.041 & 0.037 & 0.041 & 0.030 & 0.036 \\
   $c_{HE}$ & 0.043 & 0.043 & 0.015 & 0.015 & 0.077 & 0.015 & 0.040 & 0.014 & 0.01 & 0.01 & 0.009 & 0.008 \\
   $c_{HL}$ & 0.042 & 0.042 & 0.030 & 0.030 & 0.030 & 0.030 & 0.027 & 0.018 & 0.016 & 0.015 & 0.013 & 0.012 \\
   $c_{HL}^\prime$ & . & . & 0.027 & 0.027 & 0.028 & 0.028 & 0.026 & 0.018 & 0.014 & 0.015 & 0.011 & 0.012 \\
   $8\,c_{WB}$ & . & . & 0.070 & 0.070 & 0.072 & 0.071 & 0.067 & 0.066 & 0.052 & 0.069 & 0.041 & 0.059 \\
   $8\,c_{BB}$ & . & . & . & . & 0.20 & 0.20 & 0.15 & 0.16 & 0.088 & 0.16 & 0.062 & 0.12 \\
   $8\,c_{WW}$ &  . & . & . & . & 0.21 & 0.21 & 0.11 & 0.14 & 0.044 & 0.14 & 0.039 & 0.10 \\
   $8\,c_{H}$ & . & . & . & . & . & . & 4.8 & 2.2 & 1.2 & 1.2 & 0.65 & 0.68 \\
   \bottomrule
\end{tabular}
\caption{The corrected 1 $\sigma$ constraints on the EFT coefficients in $\%$, which can be compared with Table 2 of Ref.\cite{Barklow:2017awn}. The ``rep." columns present the numbers we reproduce for the results of \cite{Barklow:2017awn} with typos; they are same. The ``corr." columns give the corrected results.}
\label{tab:typoOp} \end{table}
\begin{table}[h] 
\resizebox{0.98\hsize}{!}{$
\begin{tabular}{l|cccccccccc}
  \toprule
  & \multicolumn{2}{c}{2/ab w. pol.} & \multicolumn{2}{c}{2/ab 350\,GeV} & \multicolumn{2}{c}{5/ab no. pol.} & \multicolumn{2}{c}{+\,1.5/ab 350\,GeV} & \multicolumn{2}{c}{full ILC} \\
  \cmidrule(lr){2-3}\cmidrule(lr){4-5}\cmidrule(lr){6-7}\cmidrule(lr){8-9}\cmidrule(lr){10-11}
  
  & rep. & corr. & rep. & corr. & rep. & corr. & rep. & corr. & rep. & corr. \\
  \midrule
  $g(hb\bar{b})$ & 1.04 & 0.98 & 1.08 & 1.04 & 0.98 & 0.92 & 0.66 & 0.66 & 0.55 & 0.56 \\
  $g(hc\bar{c})$ & 1.79 & 1.75 & 2.27 & 2.25 & 1.42 & 1.38 & 1.15 & 1.15 & 1.09 & 1.10 \\
  $g(hgg)$ & 1.60 & 1.56 & 1.65 & 1.63 & 1.31 & 1.26 & 0.99 & 0.99 & 0.89 & 0.90 \\
  $g(hWW)$ & 0.65 & 0.54 & 0.56 & 0.50 & 0.80 & 0.73 & 0.42 & 0.42 & 0.34 & 0.36 \\
  $g(h\tau\bar{\tau})$ & 1.17 & 1.11 & 1.35 & 1.31 & 1.06 & 1.00 & 0.75 & 0.75 & 0.71 & 0.73 \\
  $g(hZZ)$ & 0.66 & 0.55 & 0.57 & 0.51 & 0.80 & 0.73 & 0.42 & 0.42 & 0.34 & 0.37 \\
  $g(h\gamma\gamma)$ & 1.21 & 1.12 & 1.15 & 1.11 & 1.26 & 1.19 & 1.04 & 1.04 & 1.01 & 1.01 \\
  $g(h\mu\mu)$ & 5.53 & 5.51 & 5.71 & 5.70 & 5.10 & 5.08 & 4.87 & 4.87 & 4.95 & 4.95 \\
  \hline
  $g(hb\bar{b})/g(hWW)$ & 0.82 & 0.82 & 0.90 & 0.90 & 0.58 & 0.58 & 0.51 & 0.51 & 0.43 & 0.43 \\
  $g(hWW)/g(hZZ)$ & 0.07 & 0.07 & 0.06 & 0.07 & 0.07 & 0.07 & 0.06 & 0.07 0& 0.05 & 0.06 \\
  \hline
  $\Gamma_h$ & 2.38 & 2.25 & 2.50 & 2.38 & 2.11 & 2.01 & 1.49 & 1.49 & 1.50 & 1.54 \\
  $\sigma(e^-e^+ \to Zh)$ & 0.70 & 0.70 & 0.77 & 0.75 & 0.50 & 0.49 & 0.22 & 0.44 & 0.61 & 0.61 \\
  \hline
  $BR(h\to inv)$ & 0.30 & 0.30 & 0.56 & 0.56 & 0.30 & 0.30 & 0.27 & 0.27 & 0.28 & 0.28 \\
  $BR(h\to other)$ & 1.50 & 1.51 & 1.63 & 1.60 & 1.09 & 1.09 & 0.94 & 0.94 & 1.15 & 1.16 \\
 \bottomrule
 \end{tabular}$}
\caption{The corrected Higgs coupling precision in $\%$, which can be compared with Table 3 of Ref.\cite{Barklow:2017suo}. The ``rep." columns present the numbers we reproduce for the results of \cite{Barklow:2017awn} with typos; they are same. The ``corr." columns give the corrected results.}
\label{tab:typoHiggs} \end{table}

\clearpage

\section{Comparison with Ref.\cite{Durieux:2018ggn,Vryonidou:2018eyv}}
\label{app:durieux}

Both our work and Ref.~\cite{Durieux:2018ggn, Vryonidou:2018eyv} perform global-fit analyses for the combined Higgs, EW and top precision at future electron-positron colliders. They have notable differences, which are discussed and compared in this appendix.

Above all, we focus on the projection of linear lepton colliders (using ILC inputs) while \cite{Durieux:2018ggn, Vryonidou:2018eyv} on circular colliders; some of the linear versus circular colliders are also presented in \Sec{sec:ILCvsCircular}. In addition to well-known differences of linear versus circular colliders, we use upgraded direct-top constraints from the 10-parameter fit at the HL-LHC S2 stage~\cite{Durieux:2019rbz}, which yields ${\cal O}(1)$-factor stronger constraints on top operators than the HL-LHC inputs used in \cite{Durieux:2018ggn, Vryonidou:2018eyv}. The upgraded results were not yet available at the time Durieux et al. performed their study.

In the rest of this appendix, starting from different operator choices and assumptions, we discuss important differences in computing the contributions of top operators.

\subsection{Operator choice}\label{app:durieux1}

Although a quite general set of Higgs and top operators were used in both works, different assumptions were made to reduce the number of independent operators in global-fit analyses. They do not induce significant differences in numerical results, but it is worth collecting them here.

On Higgs operators, Ref.~\cite{Durieux:2018ggn,Vryonidou:2018eyv} insisted the universality of the theory and perfect EWPO, while we do not. For the universality introduced in \cite{Wells:2015uba}, they replaced the light fermion operators like ${\cal O}_{HE,HL}$ and ${\cal O}_{HL}'$ with ${\cal O}_W$ and ${\cal O}_B$ using equations of motion. In addition, for the perfect EWPO, they fixed the values of $c_{HWB,HD}$ required to make the oblique parameters, $S$ and $T$, vanish. These Wilson coefficients are replaced by top operator contributions to adjust the oblique parameters to be zero. However, we include $c_{HE,HL,HL^\prime}$ and our counterparts of $c_{HWB,HD}$ ($c_{WB,T}$ as in \Eq{eq:opconversion}) and let them vary freely in the global fit by including the EWPO data. 

As for top operators, Ref.~\cite{Durieux:2018ggn,Vryonidou:2018eyv} removed one combination of ${\cal O}_{Hq}^{(1)}$ and ${\cal O}_{Hq}^{(3)}$ by assuming that the measurement of the $Z\bar{b}b$ coupling is perfect. Moreover, they ignored ${\cal O}_{Htb}$ since its contributions are suppressed by the bottom Yukawa coupling; one exception could be its contribution to $\dot{c}_{bH}$ in \Eq{eq:cbHRG}, but this running is not relevant to our work. On the other hand, we have included all of them, ${\cal O}_{Hq}^{(1)}, {\cal O}_{Hq}^{(3)}$ and ${\cal O}_{Htb}$, as well as the $Z\bar{b}b$ measurement through the work of Ref.~\cite{Durieux:2019rbz}. 

Another difference is on the treatment of the $hgg$ vertex. Although they included both ${\cal O}_{tG} = (\bar{Q}\sigma^{\mu\nu}T^A t)\tilde{\Phi}G_{\mu\nu}^A + h.c.$ and ${\cal O}_{GG} = (\Phi^\dagger \Phi) G_{\mu \nu}^A G^{A \mu \nu}$ which contribute to the $hgg$ vertex, we use $c_{gH}$ to describe all those effects collectively as the $hgg$ contact interaction, as discussed in \Sec{sec:higgsop} and in Ref.~\cite{Barklow:2017awn}. ${\cal O}_{tG}$ will indeed be well constrained by LHC $t\bar{t}$ measurements, as was also assumed in \cite{Durieux:2019rbz} from which we take our direct-top data; see also \Sec{sec:topyukawa}.

\subsection{Finite versus log effects of top quarks} \label{app:durieux2}

Another important difference is the way to compute the contributions of top operators. Our work accounts for the logarithmic terms of top-loop effects, computed by the RG evolutions of Higgs operators. Ref.~\cite{Durieux:2018ggn,Vryonidou:2018eyv} computed one-loop diagrams by top operators instead, which then includes log-terms (captured also in our RG calculation) as well as non-log finite terms of the loop diagrams. On the other hand, our power counting rule allows us to include two-loop top effects in $\delta \Gamma(h \to \gamma \gamma, Z \gamma)$ (see \Sec{sec:powercounting}) which were not added in \cite{Durieux:2018ggn,Vryonidou:2018eyv}. We assess the numerical impact of finite terms and higher-order terms in Appendix~\ref{app:durieux2} and \ref{app:durieux3}, respectively.

\medskip
The log and finite terms of top-loop effects on several Higgs observables can be extracted from Table 14 of \cite{Vryonidou:2018eyv}. The finite terms are obtained from the results with RG scale $\mu_{\rm EFT}=m_h$, where all log terms with $\log \mu_{\rm EFT} / m_h$ vanish. The log terms then can be obtained from the subtraction of two results with $\mu_{\rm EFT} = m_h$ and 1 TeV since finite terms are independent on the RG scales. The extracted results are tabulated in the second and third lines in \autoref{tab:logvsfinite}.

Also listed in the Table are our results of log terms. For the comparison, our calculation must be corrected to account for on-shell mass renormalization and ignored higher-order terms. The former affects $\delta \Gamma(h \to b\bar{b}, \, l \bar{l})$ and the latter affects loop-induced $\delta \Gamma(h \to \gamma \gamma, \, Z\gamma)$ as
\begin{subequations}\label{eq:SS}
\begin{align}
 \delta\Gamma(h\to b\bar{b}) &\= -\frac{1}{2}c_H + 2 c_{bH} + \cdots, \label{eq:Sbb}\\
 \delta\Gamma(h\to l\bar{l}) &\= -\frac{1}{2}c_H + 2 c_{lH} + \cdots, \label{eq:Sll}\\
 \delta\Gamma(h\to \gamma\gamma) &\= 528 \,\delta Z_{A}, \label{eq:Sgg}\\
 \delta\Gamma(h\to Z\gamma) &\= 290 \,\delta Z_{AZ}, \label{eq:SZg}
\end{align}
\end{subequations}
where $\cdots$ denotes the variation of SM parameters $\delta v$ and $\delta \bar{\lambda}$, not relevant to the comparison of top contributions in this appendix. Another piece to correct is to add the running of Yukawa operators $c_{bH}$ and $c_{lH}$, according to \Eq{eq:cbHRG} and 
\beq
\dot{c}_{l H} \= 2 N_c y_t c_H \- 4 N_c (y_t^2+ y_b^2) c_{Hq}^{(3)} \+ 4 N_c y_t y_b c_{Htb}.
\eeq
The corrected log terms are shown in the first line of \autoref{tab:logvsfinite}. Our corrected log terms and those of \cite{Vryonidou:2018eyv} agree well, as it should be.

\begin{table}[t]
\begin{center}
\begin{tabular}{ c l||c c c c c c c }
 \toprule
 channel & & $\mathcal{O}_{Ht}$ & $\mathcal{O}_{Hq}^{(1)}$ & $\mathcal{O}_{Hq}^{(3)}$ & $\mathcal{O}_{Htb}$ & $\mathcal{O}_{tW}$ & $\mathcal{O}_{tB}$ & $\mathcal{O}_{tH}$\\
 \midrule
 $h \to bb$ & ~~Our log(\ref{eq:SS}) & 0 & 0.04 & 1.91 & -7.90 & -0.62 & 0 & 0.48 \\
 & ~~log-term(\cite{Vryonidou:2018eyv}) & 0 & 0.04 & 2.08 & -7.05 & -0.62 & 0 & 0.47 \\
 & ~~finite-term(\cite{Vryonidou:2018eyv}) & 0 & 0.04 & -0.18 & -1.13 & -0.28 & 0 & -0.18 \\ 
 \hline
 $h \to ll$ & ~~Our log(\ref{eq:SS}) & 0 & 0 & 0.94 & -0.03 & 0 & 0 & 0.95\\
 & ~~log-term(\cite{Vryonidou:2018eyv}) & 0 & 0 & 0.94 & -0.03 & 0 & 0 & 0.95 \\
 & ~~finite-term(\cite{Vryonidou:2018eyv}) & 0 & 0 & -0.04 & -0.00 & 0 & 0 & -0.27 \\
 \hline
 $h \to \gamma\gamma$ & ~~Our log(\ref{eq:SS}) & 0 & 0 & 0 & 0 & 200.9 & 366.2 & 0 \\
 & ~~log-term(\cite{Vryonidou:2018eyv}) & 0 & 0 & 0 & 0 & 187.9 & 350.8 & 0 \\
 & ~~finite-term(\cite{Vryonidou:2018eyv}) & 0 & 0 & 0 & 0 & -73.3 & -136.8 & 3.45* \\
 \hline
 $h \to Z\gamma$ & ~~Our log(\ref{eq:SS}) & 0 & 0 & 0 & 0 & 119.3 & -20.9 & 0\\
 & ~~log-term(\cite{Vryonidou:2018eyv})  & 0 & 0 & 0 & 0 & 117.1 & -16.7 & 0\\
 & ~~finite-term(\cite{Vryonidou:2018eyv}) & 1.77* & 1.80* & -1.74* & 0 & -45.8 & 6.97 & 0.72* \\
 \bottomrule
\end{tabular}
\caption{The comparison of log versus finite terms of top-loop contributions in the deviations of the Higgs decay widths. The coefficient of each top operator contribution is shown in \% with $1/({\rm 1 \, TeV})^2$ instead of $1/v^2$ normalization. The first line ``Our log'' shows the RG-running contributions of Higgs operators (induced by top operators) calculated in this work and corrected as in \Eq{eq:SS} to be compared with \cite{Vryonidou:2018eyv}. The second and third lines show the log and finite terms computed in \cite{Vryonidou:2018eyv}. The finite terms are not added in our work. The impact of finite terms is numerically explored in \Fig{fig:finiteterms}. The starred(*) numbers are not relevant to the comparison (but shown for completeness) as they are from the tree-level shifts of $Z\bar{t}t$ and top Yukawa couplings, without log counterparts. $Q=m_h$.}
\label{tab:logvsfinite}
\end{center}
\end{table}

Remarkably, \autoref{tab:logvsfinite} shows that finite terms are usually smaller than or, at most, the same order as the log terms. All non-zero entries of finite terms are also generated by log terms (the starred entries of finite terms are included in our work too, as they are generated by tree-level shift of top couplings). Thus, it is reasonable to expect that RG results are good proxies of loop effects. 

Finite terms could induce ${\cal O}(1)$ uncertainties in our global-fit results without finite terms. Figure~\ref{fig:finiteterms} explores this by adding all finite terms extracted in \autoref{tab:logvsfinite} to our analyses, albeit not full effects of finite terms. First of all, the Higgs coupling precision is worsened by ${\cal O}(1)$ at the ILC 250, but becomes robust against finite terms at the ILC 500. On the other hand, operator constraints are robust initially at the ILC 250 but become affected by ${\cal O}(1)$ at the ILC 500. Notably, adding finite terms could improve some operator constraints (while usually worsening the Higgs coupling precision). As stronger dataset become available, Higgs couplings may be better measured semi-directly while operator constraints may become sensitive enough to feel extra finite terms. Lastly, also marked in \Fig{fig:finiteterms} are the results with only finite terms proportional to $c_{tB}$ added to $\delta \Gamma(h\to \gamma \gamma)$. The $c_{tB}$ was expected to be influential as it is relatively weakly constrained while RG-mixing strongly with $c_{WB,BB}$ (\Sec{sec:ILC250}). Indeed, $c_{tB}$ finite terms seem to cause major impact among finite-term effects.

\begin{figure}[t]
      \centering
      \includegraphics[width=\textwidth]{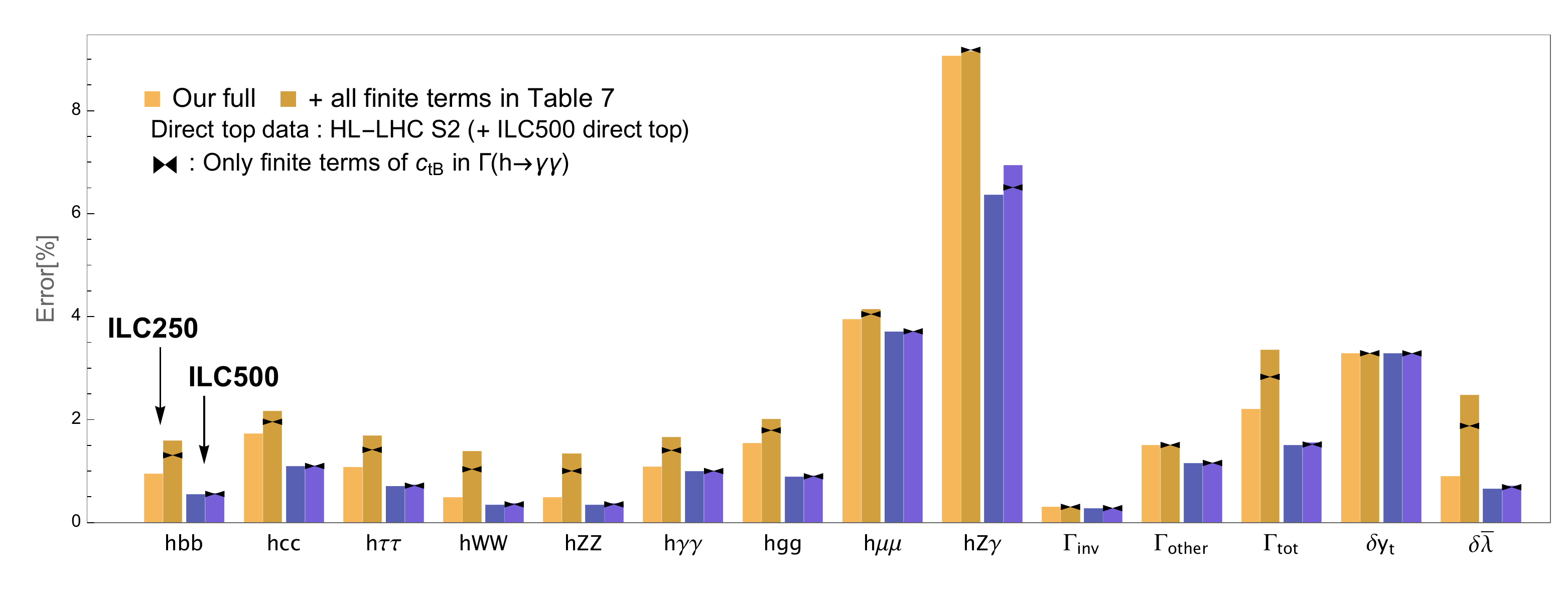}
      \includegraphics[width=\textwidth]{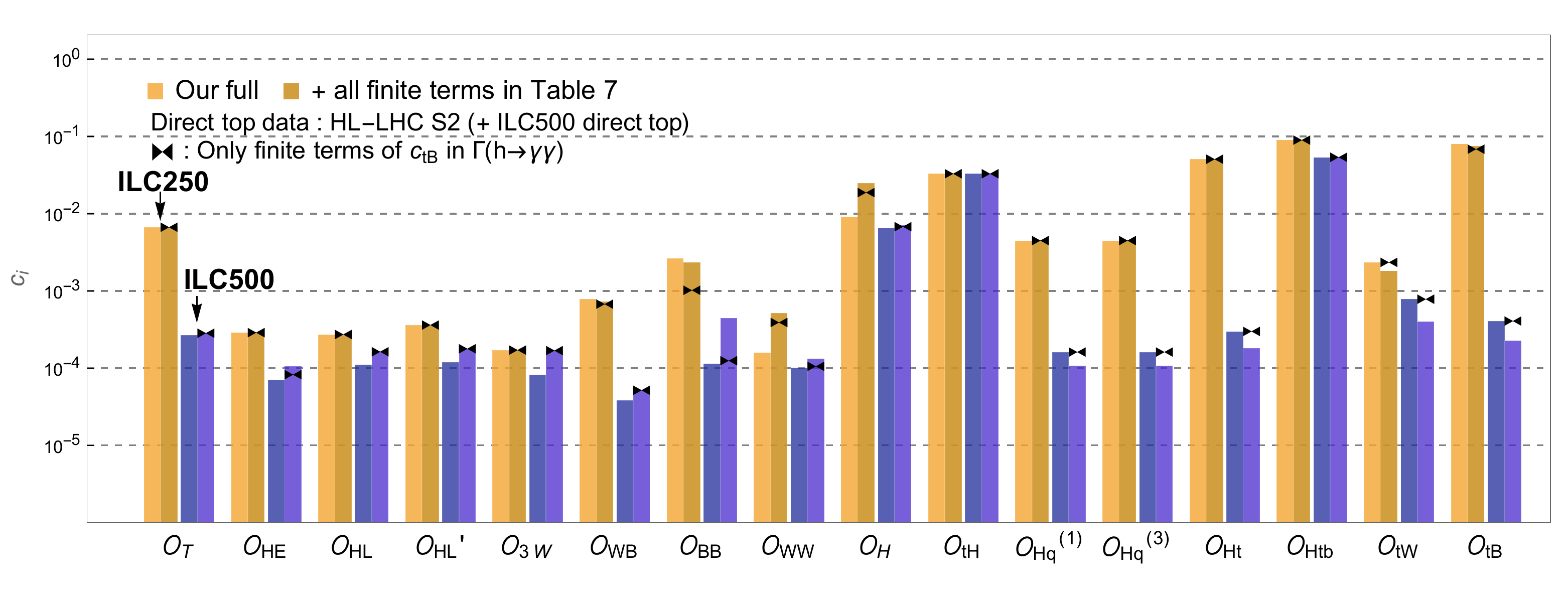}
      \caption{The effects of finite terms of top-loop diagrams on the Higgs coupling precision (upper panel) and operator constraints (lower) at the ILC 250 (yellow) and 500 stage (blue). For each case, two bars are shown: the first bar is just our full result that does not include the finite terms, while the second bar includes all the finite terms in \autoref{tab:logvsfinite}. Also marked in the second bar is the result obtained with only finite terms proportional to $c_{tB}$ in $\delta \Gamma(h \to \gamma \gamma)$.}
      \label{fig:finiteterms}
\end{figure}
%

\subsection{Higher-order effects of top quarks} \label{app:durieux3}

\begin{table}[t]
\begin{center}
\begin{tabular}{ c l||c c c c c c c }
 \toprule
 channel & & $\mathcal{O}_{Ht}$ & $\mathcal{O}_{Hq}^{(1)}$ & $\mathcal{O}_{Hq}^{(3)}$ & $\mathcal{O}_{Htb}$ & $\mathcal{O}_{tW}$ & $\mathcal{O}_{tB}$ & $\mathcal{O}_{tH}$\\
  \midrule
 $h \to \gamma\gamma$ &~~Our full & 0 & 0 & 1.05 & -0.03 & 28.07 & 51.24 & -0.57 \\
 & ~~Without higher-order & 0 & 0 & 0 & 0 & 28.02 & 51.05 & -0.57 \\
 \hline
 $h \to Z\gamma$ &~~Our full & -0.71 & 0.14 & 2.26 & -0.07 & 16.62 & -2.79 & -0.11 \\
 & ~~Without higher-order & -0.29 & -0.29 & 0.29 & 0 & 16.64 & -2.93 & -0.11 \\
 \bottomrule
\end{tabular}
\caption{The higher-order (two-loop) contributions of top quarks in the loop-induced Higgs decay widths. The coefficient of each top operator contribution is shown in absolute value with $1/v^2$ normalization. Although two-loop originated, they are only one-loop suppressed compared to the SM contributions. The first line ``Our full'' shows the full results of this work including such higher-order effects while the second line without them. The higher-order effects were ignored in \cite{Durieux:2018ggn,Vryonidou:2018eyv}. The impact of higher-order effects is numerically explored in \Fig{fig:higherorder}. $Q = {\rm max}(m_t, Q_{\rm proc})$.} 
\label{tab:higherorder}
\end{center}
\end{table}

For $\delta \Gamma( h\to \gamma \gamma, \, Z\gamma)$, Higgs operators contribute either through the tree-level renormalization of SM expressions (hence, $c_{H,T}$ at absolute one-loop) or through the tree-level contact interaction of $h\gamma \gamma$ and $h Z \gamma$ (hence, $c_{WW,WB,BB}$ at absolute tree-level). Since those widths are one-loop induced in the SM, our power counting insists on including absolute two-loop top-quark corrections. On the other hand, Ref.~\cite{Durieux:2018ggn,Vryonidou:2018eyv} considered only absolute one-loop top corrections. To estimate their impact, the higher-order terms of top operators are extracted from our calculation and tabulated in \autoref{tab:higherorder}.

\autoref{tab:higherorder} shows that largest higher-order effects come from $c_{Ht,Hq1,Hq3}$ which renormalize $c_H$ and $c_T$. Although these two-loop effects are smaller than the one-loop effects of $c_{tW,tB}$ renormalizing the tree-level contact interactions induced by $c_{WW,WB,BB}$\footnote{Small higher-order effects proportional to $c_{tW,tB}$ are from the renormalization of SM parameters such as $\delta e$ and $\delta m_W$. As an aside, $c_{Htb}$ terms are small proportional to $y_b$, and ${\cal O}_{tH}$ does not RG mix in our work.}, these ${\cal O}(0.1-1)$ terms are not small compared to most entries of top contributions in \autoref{tab:Topops1} (actually larger than most entries). This is consistent with an underlying logic of our power counting rule that the relative order matters. Moreover, higher-order effects can be leading contributions, e.g. for ${\cal O}_{Hq}^{(3)}$ and ${\cal O}_{Htb}$ in $\delta \Gamma( h\to \gamma \gamma, \, Z\gamma)$.

Figure~\ref{fig:higherorder} shows the impact of higher-order terms by comparing the global-fit results with and without higher-order terms in \autoref{tab:higherorder}. The results with higher-order terms are our full global-fit results. Compared to \Fig{fig:finiteterms}, higher-order effects are somewhat smaller in general (Higgs coupling precision in particular), partly because only two observables are modified by higher-order effects. But still, as stronger dataset becomes available, $g(hZ\gamma)$ precision and several operator constraints become sensitive to higher-order terms and change by ${\cal O}(1)$.

\begin{figure}[t]
      \centering
       \includegraphics[width=\textwidth]{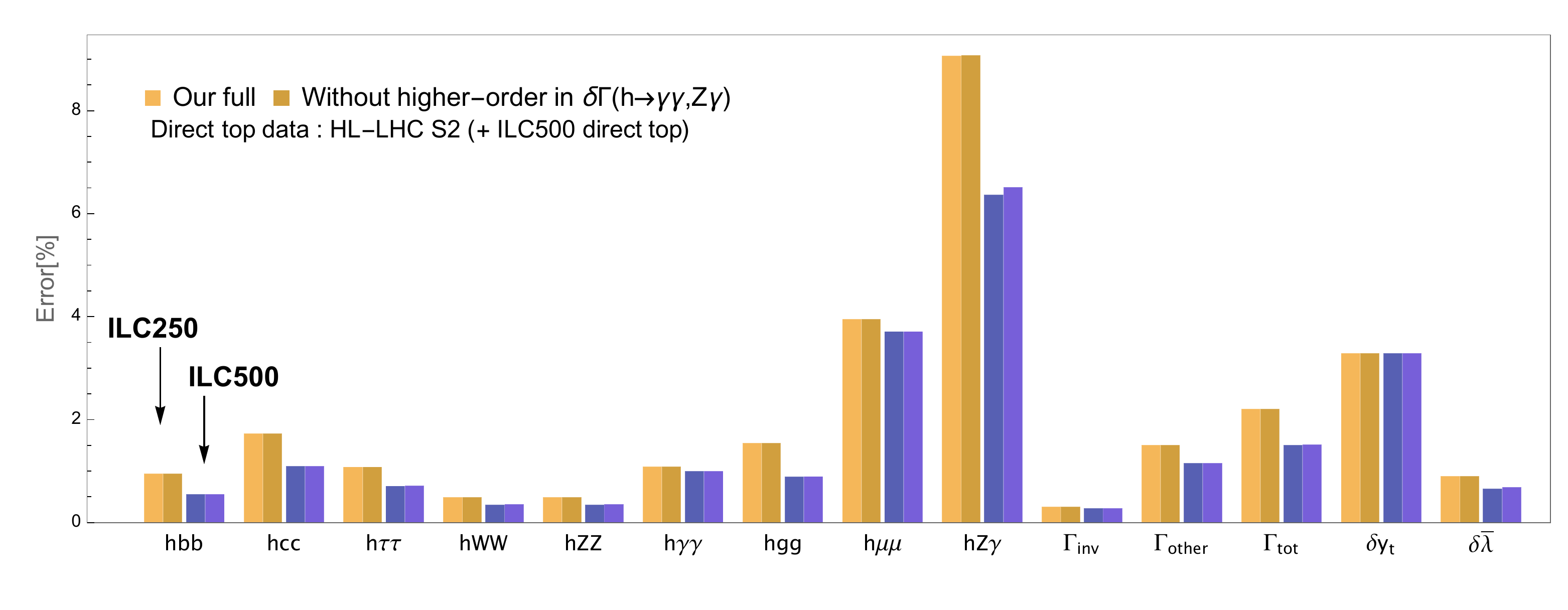}
       \includegraphics[width=\textwidth]{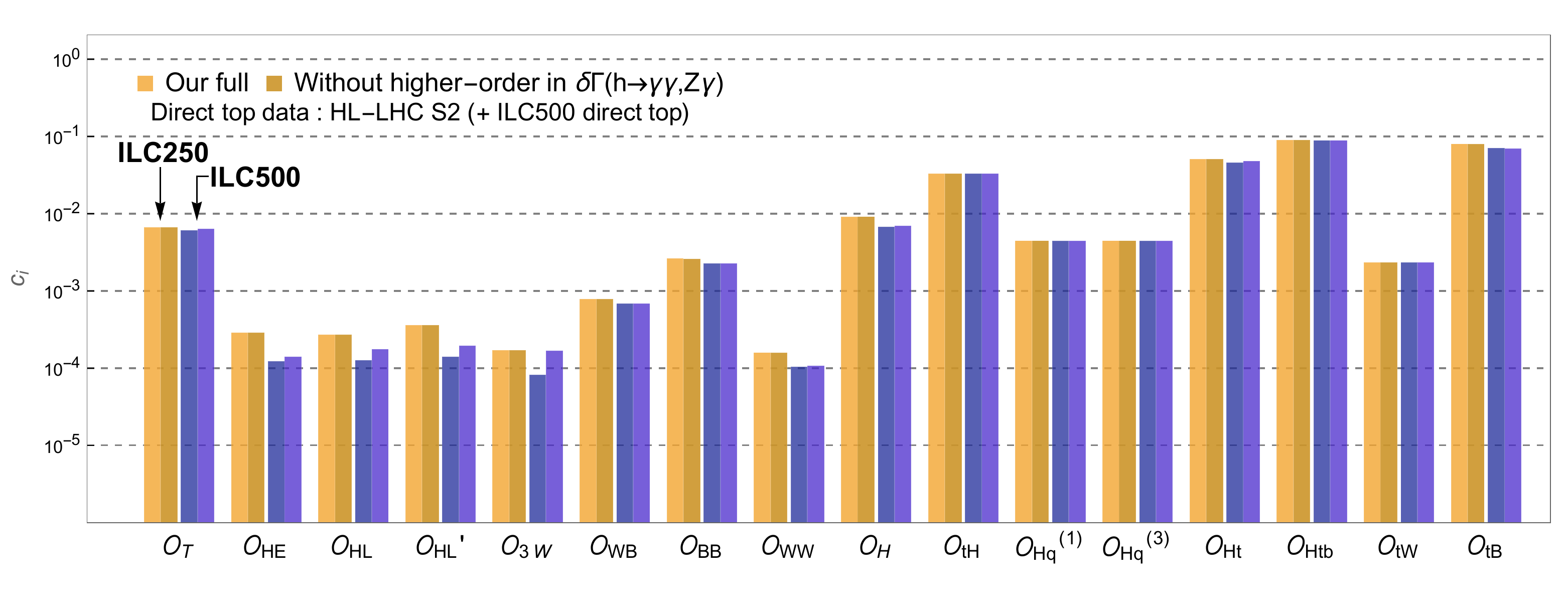}
       \caption{The effects of higher-order (two-loop) top contributions on the Higgs coupling precision (upper panel) and operator constraints (lower) at the ILC 250 (yellow) and 500 stage (blue). For each case, two bars are shown: the first bar is just our full result that includes the higher-order terms, while the second bar does not. Such top  contributions are tabulated in \autoref{tab:higherorder}.}
      \label{fig:higherorder}
\end{figure}

\clearpage

\section{Results with different renormalization scale $Q$}
\label{app:differentQ}

Loop calculations necessarily involve the uncertainty due to the choice of the RG scale $Q$. We have chosen $Q={\rm max}(m_t, Q_{\rm proc})$ in the main text, but also introduced other choices $Q = {\rm max}(2 m_t, Q_{\rm proc})$ and $Q= Q_{\rm proc}$ in \Sec{sec:RGE}. In \Fig{fig:Qvariation}, we compare global-fit results with those three choices of $Q$. 

First of all, the variations of the Higgs coupling precision and operator constraints are $\sim {\cal O}(10\%)$ at the ILC 250 with LHC Run 2. But stronger direct-top constraints from HL-LHC S2 make this variation from top effects smaller. One can expect that at the ILC 500, even though there will be more multiple scales, stronger direct-top data may not significantly increase the variation. These are the uncertainties in our results due to the choice of $Q$. In principle, this size of uncertainty is inevitable in perturbative calculations which can only be reduced by carrying out even higher-order calculations. 

More interestingly, Higgs coupling precision is most robust against the addition of top-quark effects when $Q={\rm max}(2m_t, Q_{\rm proc})$ is used. This is already discussed in \Sec{sec:ILC250} as the main reason why Higgs coupling precision at the ILC can be robust against top effects; the more number of common scales in the fit observables, the better can Higgs couplings be semi-directly measured. 

On the other hand, the very many common scales may harm operator constraints since top RG effects can only be discerned from multiple energy scales. Figure~\ref{fig:Qvariation} indeed shows that operator constraints vary somewhat more than Higgs coupling precision. But notably, operator constraints also become most robust when $Q={\rm max}(2m_t, Q_{\rm proc})$, which provides more numbers of common energy scales $2m_t$ so that RG effects are expected to be harder to discern. This is partly because $2m_t$ is closer to $Q_0 = 1$ TeV from which we RG evolve operators, so that RG effects with $Q= {\rm max}(2m_t, Q_{\rm proc})$ are smaller than with other choices of $Q$. 

Lastly,  \Fig{fig:polQproc} (compared with \Fig{fig:PolHcoup}) shows that the beam polarization becomes more important with $Q=Q_{\rm proc}$ than with other choices. As discussed in \Sec{sec:ILCvsCircular}, it is because more various energy scales are involved so that the beam polarization can efficiently double the number of independent observables, disentangling the Higgs and top operators. Also, finite terms and higher-order effects may not be described by a small number of $\tilde{c}_i$ combinations (introduced in \Sec{sec:ILC250}), which then make the beam polarization more useful.

\begin{figure}[h]
      \centering
      \includegraphics[width=\textwidth]{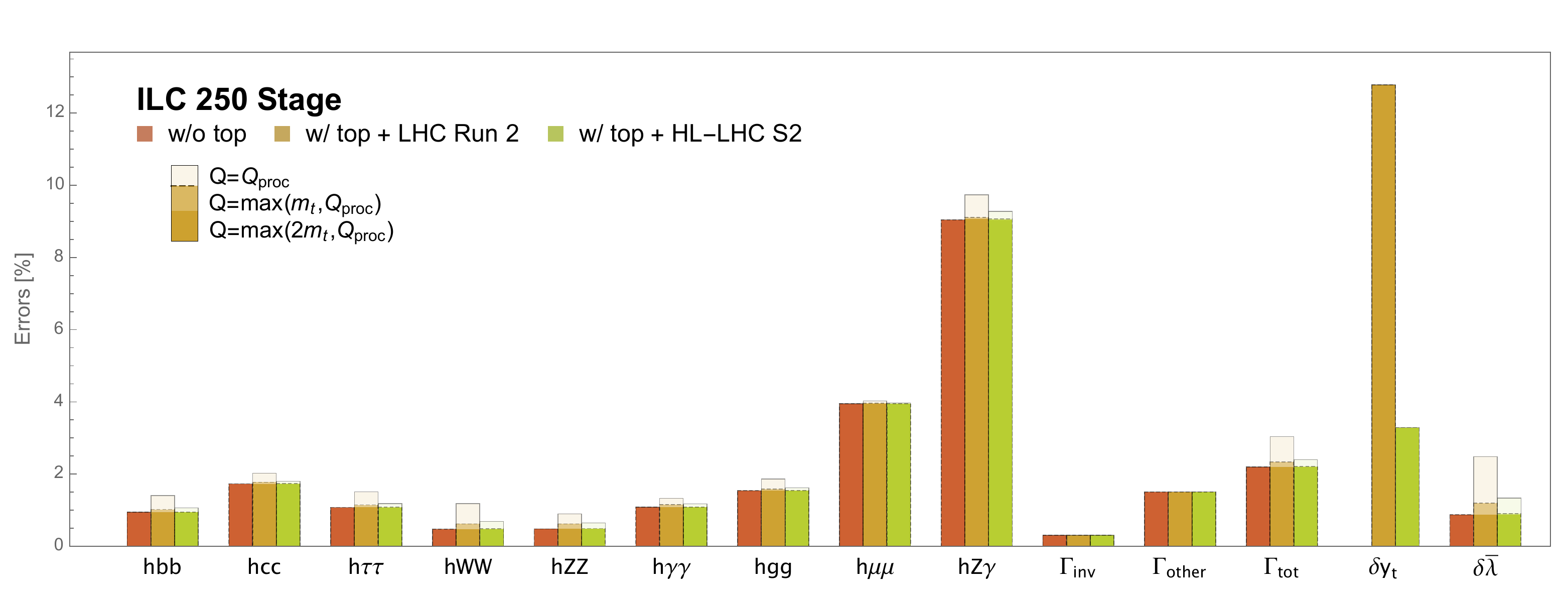}
      \includegraphics[width=\textwidth]{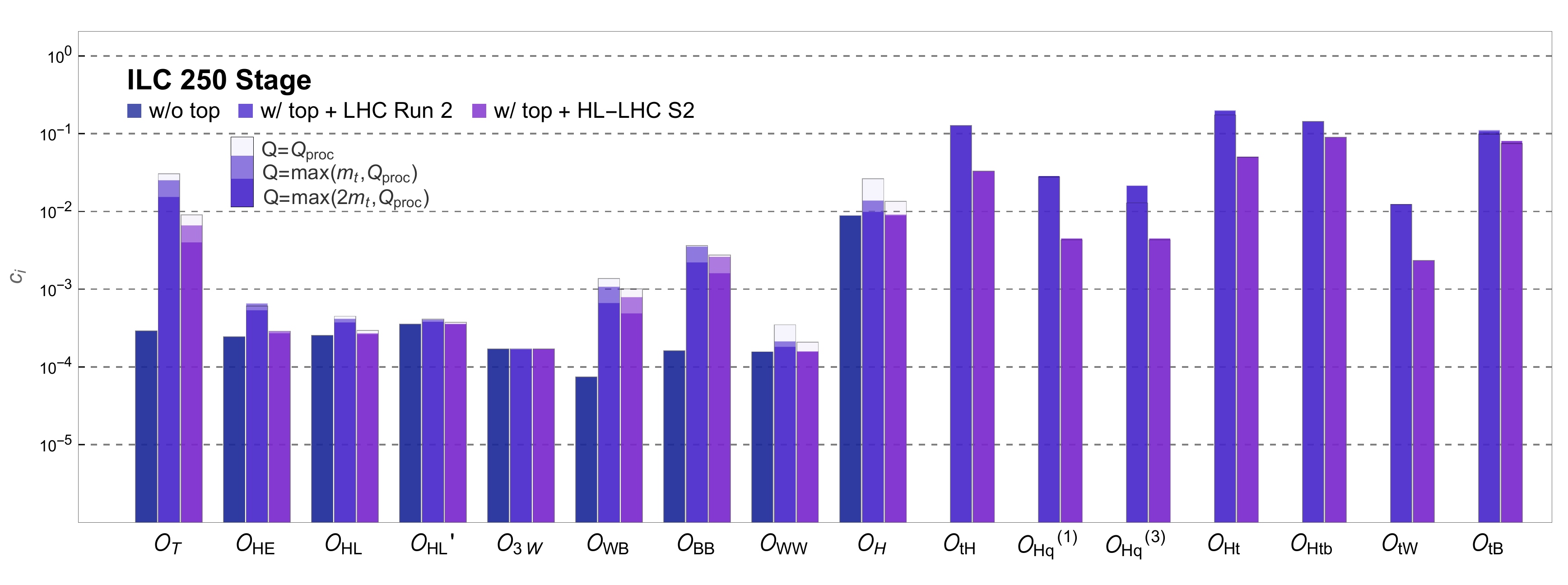}
      \caption{Global-fit results with various choice of $Q=Q_{\rm proc}, \, {\rm max}(m_t,Q_{\rm proc})$ and ${\rm max}(2m_t,Q_{\rm proc})$ at the ILC 250 stage.}
\label{fig:Qvariation}
\end{figure}
\begin{figure}[h]
      \centering
      \includegraphics[width=\textwidth]{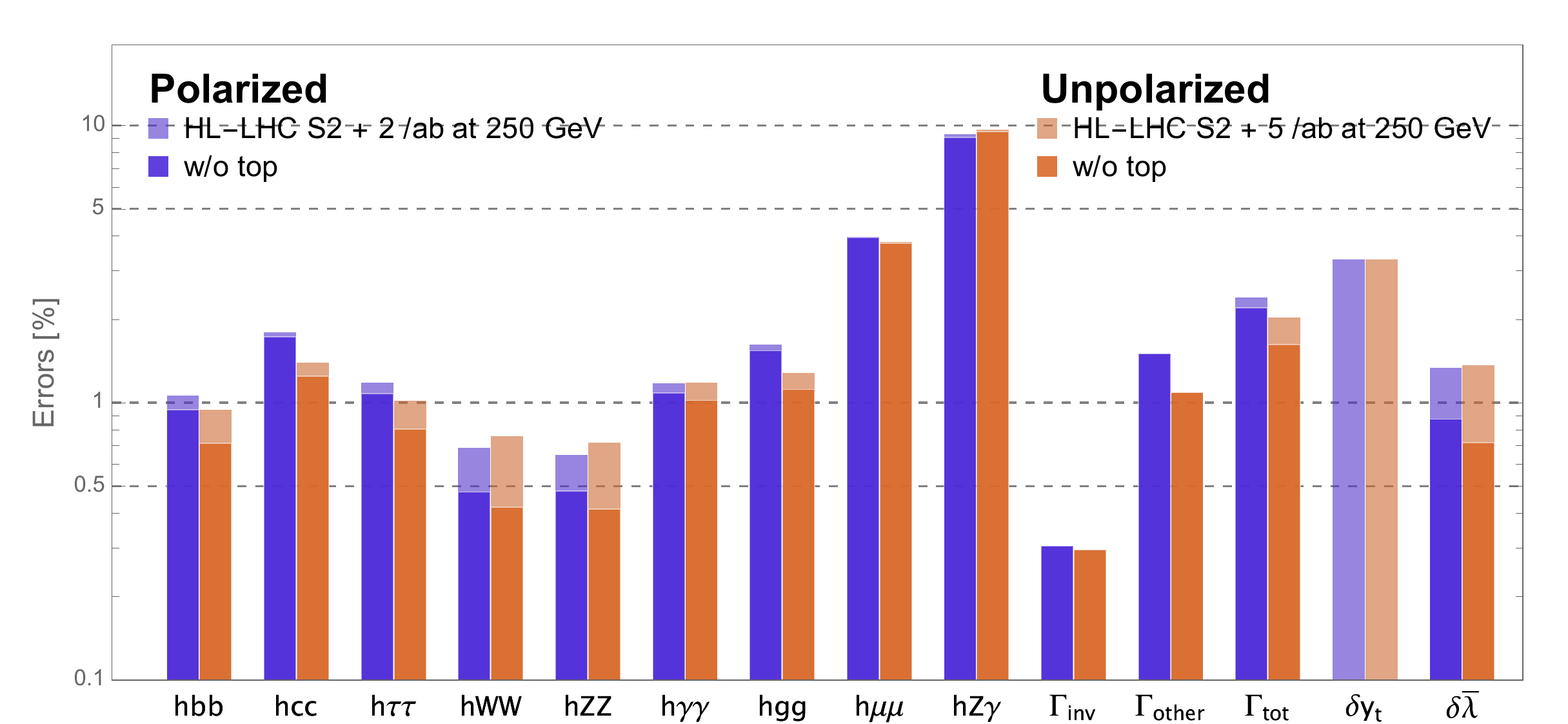}
      \includegraphics[width=\textwidth]{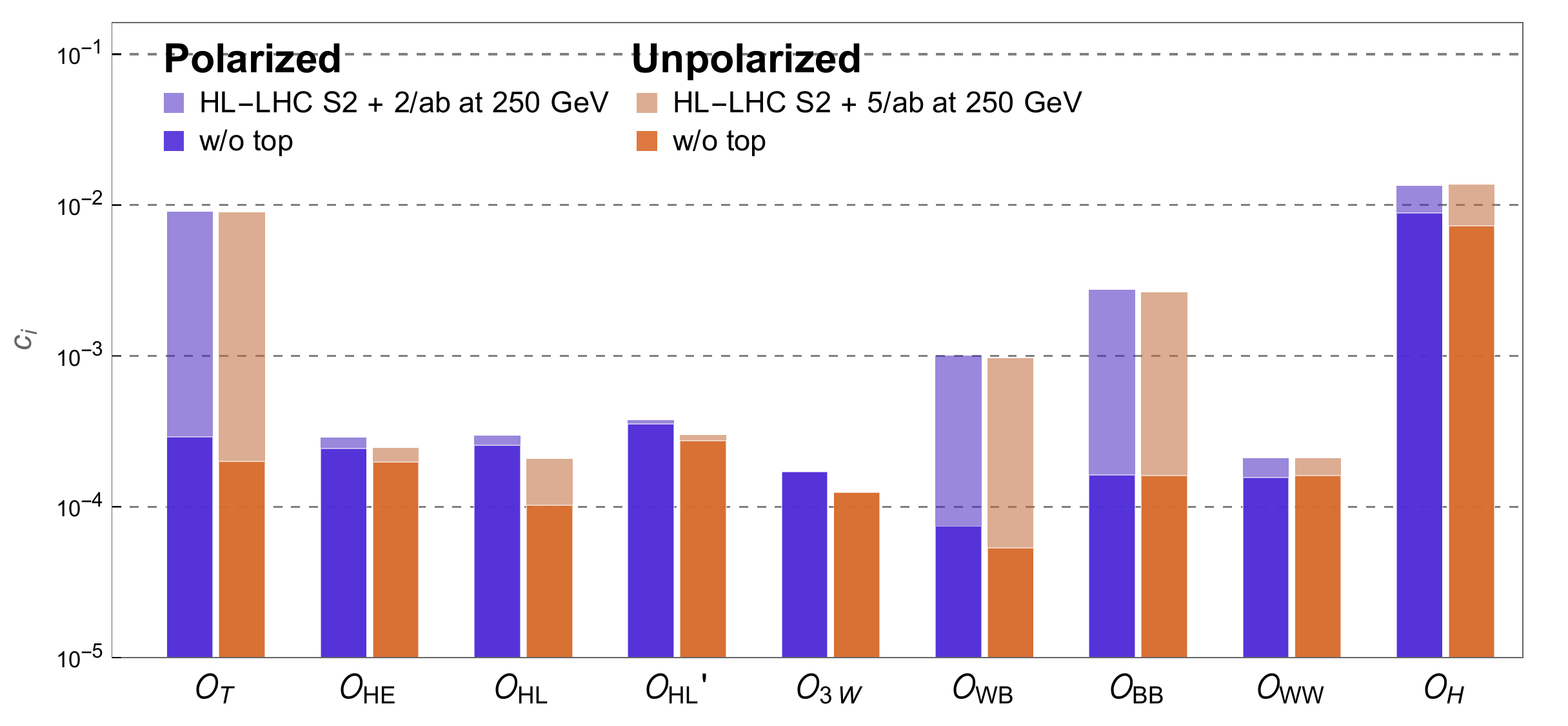}
      \caption{Same as \Fig{fig:PolHcoup} but with $Q=Q_{\rm proc}$. Not only the variations due to top effects become larger, but also the impact of beam polarization becomes more visible.}
\label{fig:polQproc}
\end{figure}

\clearpage

\section{The $Q_0$ scale dependence}
\label{app:RGdep}

As introduced in \Sec{sec:RGE}, the $Q_0$ is a reference scale (different from the RG scale $Q$) at which we renormalize operators and express their constraints. This is arbitrary in the viewpoint of numerical analyses although this can be related to the new physics scale or the matching scale of the SMEFT. Numerically, our final results must not depend on the choice of $Q_0$; unlike the case of $Q$, no inevitable numerical uncertainties are generated. This must imply one property of the covariance matrix: it should change with $Q_0$ in such a way to compensate the RG evolutions of Higgs operators. We detail this in Appendix~\ref{app:RGdep1} and \ref{app:RGdep2}, and apply this to understand the connection of $c_T(Q_0)$ and oblique parameters in Appendix~\ref{app:RGdep3}.

\subsection{The running of the covariance matrix}
\label{app:RGdep1}

The measurements involved in our global fits are performed at various energy scales. Theory predictions are written in terms of the Wilson coefficients renormalized at $Q_0$. The fit results then represent the constraints on the operator coefficients at $Q_0$. If we change the $Q_0$ to $Q_0^\prime$, the constraints change as shown in \Fig{fig:Running}.

As discussed in \Sec{sec:methodology}, the covariance matrix $\mathcal{C}ov_{IJ}(Q_0)$ encodes the RG evolutions of operators from $Q_0$ to the RG scales of observables. It also changes with $Q_0$ in such a way to leave the total $\chi^2$ of the global fit unchanged 
\beq
      \chi^2 = \sum_{I,J} c_I(Q_0) \, \mathcal{C}ov^{-1}_{IJ}(Q_0) \, c_J(Q_0) = \sum_{I,J} c_I(Q_0^\prime) \, \mathcal{C}ov^{-1}_{IJ}(Q_0^\prime) \, c_J(Q_0^\prime).
\label{eq:chi22}
\eeq
The Wilson coefficients $c_I(Q_0)$ RG evolve as 
\begin{equation}
      c_I(Q_0^\prime)=\left({\begin{array}{c} c_i(Q_0^\prime) \\ c_j(Q_0^\prime) \end{array}}\right) \= \left({\begin{array}{cc} 1_{ii^\prime} & \gamma_{ij^\prime} \log(Q_0^\prime/Q_0) \\ 0 & 1_{jj^\prime} \end{array}}\right) \left({\begin{array}{c} c_{i^\prime}(Q_0) \\ c_{j^\prime}(Q_0) \end{array}}\right) 
      \,\equiv\, \Gamma_{IJ}(Q_0^\prime/Q_0) \,c_J(Q_0),
\label{eq:RGrelation}
\end{equation}
where the upper(lower) components $c_i(c_j)$ denote the coefficients of Higgs(top) operators and $I,J$ denote the indices $i,j$ collectively. The block diagonal form of $\Gamma_{IJ}$ is from our scheme where only the RG evolutions of Higgs operators by top operators are considered. Thus, the covariance matrix must evolve as
\begin{equation}
      \mathcal{C}ov(Q_0^\prime) = \Gamma(Q_0^\prime/Q_0) ~\mathcal{C}ov(Q_0)~ \Gamma(Q_0^\prime/Q_0)^T.
      \label{eq:CovTrans}
\end{equation}  
Again, this evolution ensures that global-fit results obtained with different $Q_0$ are physically equivalent (same $\chi^2$), and one result can be obtained from the other through the evolution of the covariance matrix.

\medskip
As an aside, we mention properties of the evolution of covariance matrix. Note that $\Gamma^{-1}(Q/Q^\prime) = \Gamma(Q^\prime/Q)$ and $\Gamma(Q^{\prime \prime} / Q^\prime) \Gamma(Q^\prime / Q) = \Gamma(Q^{\prime \prime}/Q)$ make the evolution similar to (RG-)logarithmic. The diagonal component of the covariance matrix $\mathcal{C}ov_{II}$ is an error of a corresponding Wilson coefficient, and the off-diagonal element $\mathcal{C}ov_{IJ}$ is the error correlation between $c_I$ and $c_J$. These components evolve as
\begin{eqnarray}
      \mathcal{C}ov_{II}(Q_0^\prime) = \mathcal{C}ov_{II}(Q_0)
      + \sum_{J} 2\gamma_{IJ} \log(Q_0^\prime/Q_0) \mathcal{C}ov_{IJ}(Q_0) \\ \nonumber
      + \sum_{J} (\gamma_{IJ} \log(Q_0^\prime/Q_0))^2 \mathcal{C}ov_{JJ}(Q_0).
\end{eqnarray}  
For coefficients that do not RG evolve by top operators, the equation reduces to $\mathcal{C}ov_{II}(Q_0) = \mathcal{C}ov_{II}(Q_0^\prime)$. Thus their constraints do not change with $Q_0$. Examples include the constraints on top operators and $c_{3W}$ as shown in \Fig{fig:Running}.

\begin{figure}[t]
      \centering
      \includegraphics[width=0.8\textwidth]{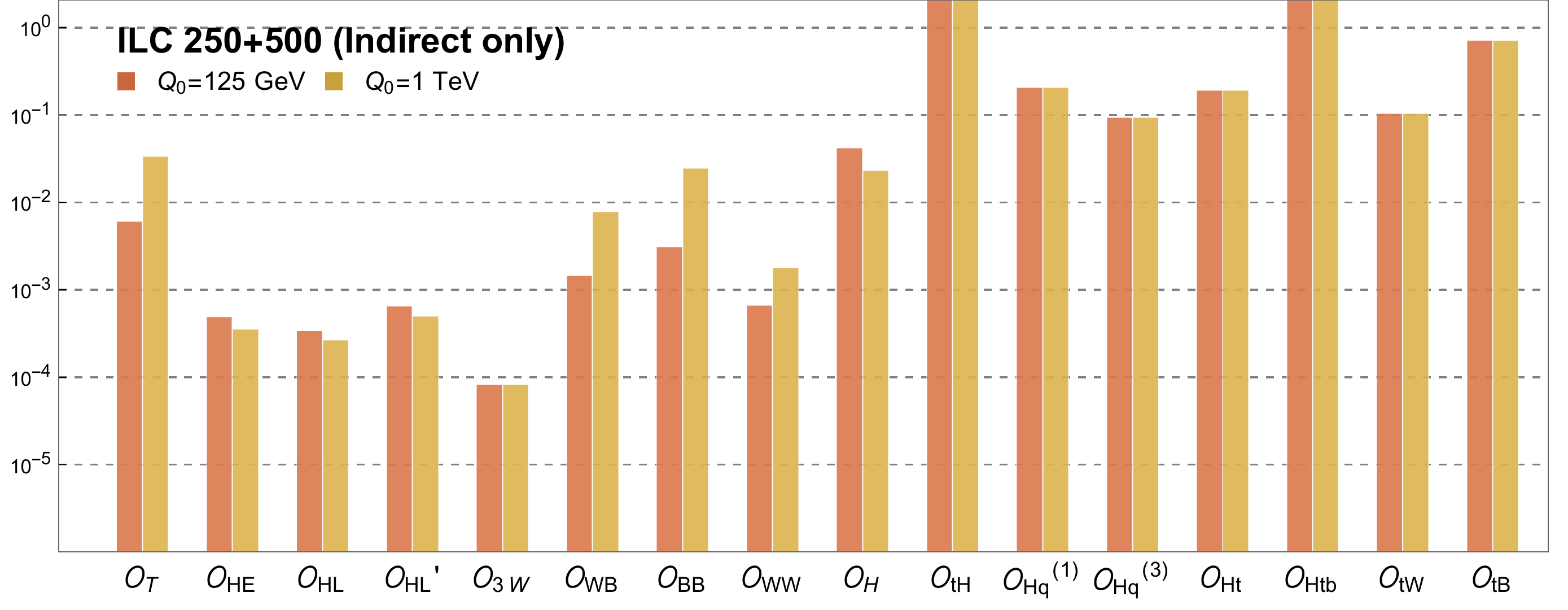}
      \caption{The comparison of global fits performed with $Q_0=m_h$ and 1 TeV. The two result are equivalent because one of them can be obtained by the other via the RG evolution of the covariance matrix. It also shows that ${\cal O}_T$ runs most quickly due to top operators.}
\label{fig:Running}
\end{figure}
%

\subsection{The $Q_0$ independence of the uncertainty of observables} 
\label{app:RGdep2}

The evolution of the covariance matrix also ensures that the (physical) results defined at certain $Q_{\rm proc}$ (such as Higgs coupling precision and oblique parameters) are independent on $Q_0$.
Start from the theory prediction of the deviations of the observable $\delta {\cal O}$ defined at $Q$
  \begin{equation}
      \delta \mathcal{O}(Q) = \sum_{I}a_I(Q)\,c_I(Q)= \sum_{I,J}a_I(Q)\Gamma_{IJ}(Q/Q_0)\,c_J(Q_0) = \sum_{J}a_J(Q_0)\,c_J(Q_0),
  \end{equation} 
where $a_I$ is the constant multiplied in front of each Wilson coefficient $c_I$, i.e. numbers collected in \autoref{tab:Higgsops} and \ref{tab:Topops1}. This determines the rule for the transformation of $a_I$ with the energy scale.
Given the covariance matrix from a global fit, one can calculate the error of the observable $\mathcal{O}$ contributed from EFT operators as
  \begin{eqnarray}
    \sigma_\mathcal{O}^2 = \left< \delta \mathcal{O}^2 \right> &=& \sum_{I,J} a_I(Q) \mathcal{C}ov_{IJ}(Q) a_J(Q) \\\nonumber
    &=& \sum_{I,J,K,L} a_I(Q)\Gamma_{IK}(Q/Q_0) \, \mathcal{C}ov_{KL}(Q_0) \,\Gamma_{JL}(Q/Q_0) a_J(Q) \\\nonumber
    &=& \sum_{I,J} a_I(Q_0) \mathcal{C}ov_{IJ}(Q_0) a_J(Q_0).
  \label{eq:scaleindp}    
  \end{eqnarray}
The second line follows from the matrix relation in \Eq{eq:CovTrans}. This proves that the error of a physical observable contributed from EFT coefficients is independent on our choice of $Q_0$.

\subsection{The interpretation of the $Q_0$ dependence - Oblique parameters}
\label{app:RGdep3}

We turn to discuss one interesting $Q_0$ dependence, which provides another example of how to think of $Q_0$.
Figure~\ref{fig:Running} shows that the constraint on $c_T$ varies significantly with $Q_0$, because $c_T$ receives a large anomalous dimension from top operators. As a result, its constraint is much weaker at higher $Q_0$ scales. Since $c_T$ is closely related to the $T$ parameter that signals the breaking of the custodial symmetry, does this mean that the custodial symmetry can be broken seriously at high energy? Not necessarily. 

\begin{table}[t]
  \begin{center}
  \begin{tabular}{  p{6.8cm}| c | c }
     \toprule
      & $\alpha S$ & $\alpha T$ \\
     \hline\hline
     ~ILC 250+500 without top operators  & ~0.0189\,\%~ & ~0.0175\,\%~ \\
     ~With top operators (indirect-only) & ~0.0268\,\%~ & ~0.0177\,\%~ \\
     ~+\,HL-LHC S2 & ~0.0201\,\%~ & ~0.0175\,\%~ \\
     ~+\,ILC 500 direct-top & ~0.0189\,\%~ & ~0.0175\,\%~ \\
     \bottomrule
  \end{tabular}
  \caption{Global-fit constraints on the oblique parameters, $\alpha \, S$ and $\alpha \, T$ in \Eq{eq:ST}. They are defined at $Q_{\rm proc} = m_Z$, hence top-quark effects are evaluated at $Q={\rm max}(m_t, m_Z)=m_t$. Any choice of $Q_0$ yields the same results.}
  \label{tab:ST}
 \end{center}
\end{table} 

\medskip
We detail these by explicitly checking that our global fit successfully impose the small breaking of the custodial symmetry at low energy $\sim m_Z$ (irrespective of $Q_0$) even though we do not directly use small $T$ as an input data.
The expressions of oblique parameters can be read from the expressions of well-measured electroweak precision observables, $Z$-pole $s_*$ and $m_W^2/m_Z^2$,~\cite{Barklow:2017awn}
\begin{eqnarray} 
  m_W^2/m_Z^2 &=& c_0^2 \+ \frac{c_0^2}{c_0^2-s_0^2} \left(c_0^2\,c_T - 2 s_0^2(c_{HL}^\prime+8c_{WB}) \right), \nonumber\\
  s_*^2 &=& s_0^2 \+ \frac{s_0^2}{c_0^2-s_0^2} \left(c_{HL}^\prime+8c_{WB}-c_0^2 c_T \right) \- \frac{1}{2}c_{HE} \-s_0^2(c_{HL}-c_{HE}),
\end{eqnarray}
as (generalized from \cite{Barklow:2017awn})
\begin{eqnarray}
  \alpha\,S &=& 4s_0^2 \,(8c_{WB}+c_{HL}^\prime) \+ 4 \left(-\frac{1}{2}+s_0^2 \right) c_{HE} \- 4s_0^2c_{HL}, \nonumber\\
  \alpha\,T &=& c_T \- \frac{c_0^2-s_0^2}{c_0^2} c_{HE} \- 2\frac{s_0^2}{c_0^2} c_{HL}.
  \label{eq:ST}
\end{eqnarray}
These are $Z$-pole observables, hence defined at $Q_{\rm proc} = m_Z$.
The expressions are consistent with the oblique parameters, $\hat{S}$ and $\hat{T}$, defined in the universal limit~\cite{Wells:2015uba}
\begin{eqnarray}
      \hat{S} &\equiv& \frac{\alpha}{4 s_0^2} S \= g^2 ( \frac{1}{g g^\prime} C_{HWB} \+ \frac{1}{4} C_{HJW} \+ \frac{1}{4}C_{HJB} ), \\
      \hat{T} &\equiv& \alpha T \= -\frac{1}{2} C_{HD} \+ \frac{g^{\prime\,2}}{2}C_{HJB},
\end{eqnarray}
because our operators correspond to the bosonic currents in the universal theory
\begin{equation}
      c_{HL}^\prime \,\to\, \frac{g^2}{4} C_{HJW}, \quad c_{HL} \,\to\, \frac{g^{\prime\,2}}{2} Y_l\, C_{HJB}, \quad c_{HE} \,\to\, \frac{g^{\prime\,2}}{2} Y_e\, C_{HJB},
\end{equation}
where $Y_{l,\,e}$ are the hypercharges of the $L$ and $e$, respectively.

Using these, we numerically evaluate the global-fit constraints on $\alpha S$ and $\alpha T$ in \autoref{tab:ST}. Whether top operators are added or not, they are indeed constrained to be small for all benchmark scenarios. Top-quark effects on oblique parameters are particularly small once direct-top constraints are included. The oblique parameters are defined at $Q_{\rm proc} =m_Z$, while top-quark effects are decoupled below $m_t$ so that $Q={\rm max}(m_t, m_Z) = m_t$ is used. Thus, (we checked that) these results remain true and same for any value of $Q_0$, again because the constraints on physical observables are independent on $Q_0$. 
The $Q_0$-independence implies that the custodial symmetry is well preserved at low energy regardless of $c_T$ constraints at high $Q_0$. 

What about the custodial symmetry at high energy scales (where $c_T$ is weakly constrained)? 
First of all, $c_T$ (at high $Q_0$) does not necessarily correspond to the $T$ parameter. \Eq{eq:ST} shows that there can be other contributions to EW precision observables because our operator set is not assumed to be oblique or universal. At high energy scales, the meaning of $T$ becomes more subtle too. One main reason is that the $S,T,U$ description is valid at low energy, as this is the Taylor expansion of self-energy corrections in powers of energy scale~\cite{Peskin:1990zt}. Also, the RG evolution itself generates the non-universal effects which cannot be absorbed into the parameters of universal theories~\cite{Wells:2015cre}. It is expected that the non-universal effects interfere with the universal ones, making the meaning of $T$ ambiguous. 
Therefore, the poor constraint on $c_T$ at high $Q_0$ does not necessarily mean the large breaking of the custodial symmetry at those high scales, nor at low scales.

\section{RG evolution with bottom operators and their impact}
\label{app:RGEwb}

Our main discussion focuses on the Higgs and top operators. But many new physics models accompany bottom operators together with top operators. The following 3 `bottom' operators are relevant to us (in addition to the bottom Yukawa operator ${\cal O}_{bH}$ that is already added)
\bea
\mathcal{O}_{Hb} &=& (\Phi^\dagger i \overleftrightarrow{D}_\mu \Phi)(\bar{b} \gamma^\mu b), \nonumber \label{eq:bottomop}\\
\mathcal{O}_{bW} &=& (\bar{Q}\tau^a\sigma^{\mu\nu}b) \Phi W_{\mu\nu}^a,\\
\mathcal{O}_{bB} &=& (\bar{Q}\sigma^{\mu\nu}b)\Phi B_{\mu\nu}, \nonumber
\eea
with the Lagrangian
\beq
\Delta {\cal L}_{\rm bottom} \= \frac{c_{Hb}}{v^2} \, \mathcal{O}_{Hb} \+ \frac{c_{bW}}{v^2} \, \mathcal{O}_{bW} \+ \frac{c_{bB}}{v^2} \, \mathcal{O}_{bB}.
\eeq
These operators are relevant in direct-top and -bottom observables at tree-level as well as Higgs + EW observables at one-loop. The tree-level effects on direct-top constraints are already taken into account in \cite{Durieux:2019rbz}, from which we take our direct-top constraints.

The RG equations of dimension-6 operators are given in~\cite{Alonso:2013hga,Grojean:2013kd,Jenkins:2013zja,Jenkins:2013wua}. Their Warsaw-basis operators can be converted to ours following the rule
\bea
c_H = -2c_{H\square}+\frac{1}{2}c_{HD}, ~&~ c_T = -\frac{1}{2} c_{HD}, ~&~ -\frac{\lambda}{v^2}c_6 = \frac{1}{v^2} c_{HH}, \nonumber \\
\frac{g^2}{m_W^2} c_{WW} = \frac{1}{v^2} c_{HW}, ~&~ \frac{g^{\prime 2}}{m_W^2} c_{BB} = \frac{1}{v^2} c_{HB}, ~&~ \frac{2gg^{\prime}}{m_W^2} c_{WB} = \frac{1}{v^2}c_{HWB}, \nonumber \\
\frac{1}{v^2}c_{HL} = \frac{1}{v^2}c_{Hl}^{(1)}, ~&~ \frac{1}{v^2}c_{HL}^\prime = \frac{1}{v^2}c_{Hl}^{(3)}, ~&~ \frac{1}{v^2}c_{HE} = \frac{1}{v^2} c_{He},
\label{eq:opconversion}
\eea
where operators on the right-hand side are the operators in \cite{Alonso:2013hga} and on the left-hand side are ours; to avoid confusion, we use $c_{HH}$ instead of $c_H$ on the right-hand side. We keep only top contributions to Higgs operators. Keeping terms of order ${\cal O}(g,\,g^\prime,\,y_t,\,y_b)^2$, we already showed all the needed top-operator contributions in \Sec{sec:RGE}. 
In addition, for completeness, we show the RG equations of $c_6$ and $c_{bH}$ as well as additional bottom contributions to Higgs operators
\bea
 \dot{c}_{6} &=& -\frac{1}{\lambda} \left[(8 \lambda y_t N_c - 8 y_t^3 N_c) c_{tH} + (-16 \lambda y_t^2 N_c + \frac{16}{3} \lambda g^2) c_{Hq}^{(3)} + 16 \lambda y_t y_b N_c c_{Htb} \right], \label{eq:c6RG}\\
 \dot{c}_{bH} &=& y_t (2N_c-3) c_{tH} \+ \left( (-4N_c+6)y_t^2 -2 g_1^2 + \tfrac{4}{3} g_2^2 N_c + 12 \lambda \right) c_{Hq}^{(3)} \nonumber \\
&&\+ (-2 g_1^2 + 4 \lambda) c_{Hq}^{(1)} \+ (-2 y_t^3 + 3 g_2^2 y_t - 4 \lambda y_t)/y_b c_{Htb} \+ (-6g_2 y_t) c_{tW}. \label{eq:cbHRG}
\eea
The RG effect on $c_{bH}$ does not vanish in the limit $y_b \to 0$ but is rather enhanced by $y_t/y_b$.

The bottom-operator RG contributions are (in addition to the top contributions in \Sec{sec:RGE})
\bea
 \dot{c}_{H} &=& 0, \\ 
 \dot{c}_{T} &=&  - (4 N_c y_b^2 \+ \frac{8}{3} g^{\prime 2} Y_h Y_d N_c)\,c_{Hb}, \\ 
 \dot{c}_{WW} &=& -\frac{1}{2}g N_c\,( y_b\, c_{bW}), \\
 \dot{c}_{BB} &=& -\frac{1}{t_W^2} g^{\prime} N_c ( y_b (Y_q\+Y_d) \,c_{bB} ),\\
 \dot{c}_{WB} &=& \frac{1}{8 t_W} ( - 2 g N_c y_b \,c_{bB} \- 4 g^{\prime} (Y_q\+Y_d) y_b N_c \,c_{bW} ), \\ 
 \dot{c}_{HL} &=& \frac{1}{2} Y_l g^{\prime 2}(\frac{8}{3}Y_d N_c\, c_{Hb}), \\
 \dot{c}_{HL}^{\prime} &=& 0,\\ 
 \dot{c}_{HE} &=& \frac{1}{2} Y_e g^{\prime 2} (\frac{8}{3} Y_d N_c \,c_{Hb}).
\eea

Figure~\ref{fig:Bottomfit} compares the global-fit results with and without the bottom operators. The bottom operators are well constrained by HL-LHC data, which include direct-bottom productions from $Z\to b\bar{b}$ and single-top productions~\cite{Durieux:2019rbz}. Their RG effects are also often suppressed by small $y_b$. After all, the impact of bottom operators are almost invisible. We can safely ignore bottom operators assuming that direct-top and bottom data can constrain them well.

\begin{figure}[t]
      \centering
      \includegraphics[width=0.85\textwidth]{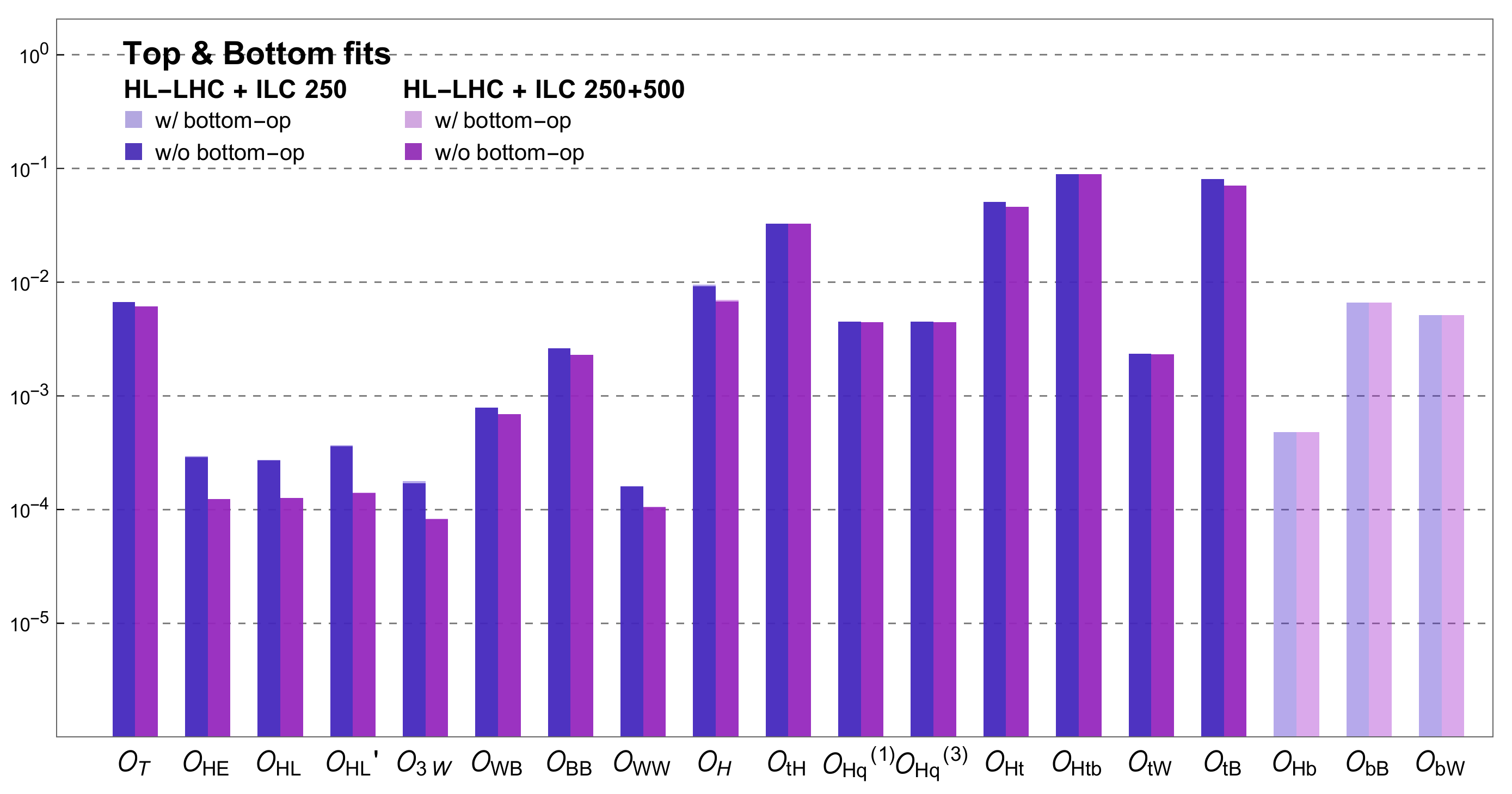}
      \caption{The global-fit $1\sigma$ constraints with (dark) and without (light) the bottom operators in \Eq{eq:bottomop} added to our work. The changes due to bottom operators are almost invisible.}
\label{fig:Bottomfit}
\end{figure}
%

\section{Observables in terms of operators}
\label{app:obs}

\begin{table}[t]
\begin{center}
\resizebox{0.97\hsize}{!}{$
\begin{tabular}{ |c||p{1.3cm}|p{1.4cm}|p{1.3cm}|p{1.3cm}|p{1.3cm}|p{1.3cm}|p{1.3cm}|p{1.3cm}|p{1.3cm}|  }
 \hline
 \multicolumn{10}{|c|}{Higgs Operators} \\
\hline
 Observables &  $c_{T}$ & $c_{HE}$ & $c_{HL}$ & $c_{HL}^\prime$ & $c_{3W}$ & $c_{WB}$ & $c_{BB}$ & $c_{WW}$ & $c_H$  \\
 \hline
 $\delta\alpha^{-1}$ & 0 & 0 & 0 & 0 & 0 & 3.704 & -1.852 & -1.852 & 0 \\
 $\delta G_F$ & 0 & 0 & 0 & 2.0 & 0 & 0 & 0 & 0 & 0 \\
 $\delta A_l$ & 0 & 14.37 & 12.39 & 12.39 & 0 & -53.23 & -22.95 & 76.18 & 0 \\
 $\delta m_W$ & 0 & 0 & 0 & 0 & 0 & 0 & 0 & 4.0 & 0\\
 $\delta m_Z$ & -0.5 & 0 & 0 & 0 & 0 & 1.852 & 0.279 & 3.074 & 0 \\
 $\delta m_h$ & 0 & 0 & 0 & 0 & 0 & 0 & 0 & 0 & -0.5 \\
 $\delta \Gamma_l$ & -0.5 & -1.84 & 2.14 & 2.14 & 0 &  4.97 & 0.58 & 10.06 & 0\\
 $\delta \Gamma_{W}$ & 0 & 0 & 0 & 0 & 0 & 0 & 0 & 12 & 0\\
 $\delta \Gamma_{Z}$ & -0.5 & 0 & 0 & 0 & 0 & 5.03 & 0.61 & 9.97 & 0 \\
 \hline
 \hline
 $\delta g_{Z\,eff} $ & 0 & -0.651 & 1.301 & -1.301 & 0 & 0 & 0 & 0 & 0 \\
 $\delta \kappa_{A\,eff} $ & 0 & 1.160 & 1 & -1 & 0 & 0 & 0 & 0 & 0\\
 $\delta \lambda_{A\,eff}$ & 0 & 0 & 0 & 0 & -2.526 & 0 & 0 & 0 & 0 \\
 \hline
 \hline
 $a_L(250)$ & -1.706 & 0 & 14.0 & 14.0 & 0 & 0.7618 & -0.9244 & 12.71 & 0.0655 \\
 $a_R(250)$ & -1.706 & -16.23 & 0 & 0 & 0 & 8.762 & 2.525 & 1.264 & 0.0655 \\
 $b_L(250)$ & 0 & 0 & 0 & 0 & 0 & 0.4928 & -0.827 & 10.74 & 0 \\ 
 $b_R(250)$ & 0 & 0 & 0 & 0 & 0 & 7.428 & 2.164 & 0.8179 & 0 \\
 $a_L(500)$ & -1.65 & 0 & 55.99 & 55.99 & 0 & 0.556 & -0.955 & 12.37 & -0.412  \\
 $a_R(500)$ & -1.65 & -64.93 & 0 & 0 & 0 & 8.556 & 2.494 & 0.923 & -0.412 \\
 $b_L(500)$ & 0 & 0 & 0 & 0 & 0 & 0.123 & -0.986 & 11.27 & 0 \\ 
 $b_R(500)$ & 0 & 0 & 0 & 0 & 0 & 7.857 & 2.349 & 0.205 & 0\\
 \hline
 \hline
 $\delta \sigma (\nu \bar{\nu} h,250)$ & 0 & 0 & -0.37 & -4.03 & 0 & 0 & 0 & 15.84 & 0.85 \\ 
 $\delta \sigma (\nu \bar{\nu} h,500)$ & 0 & 0 & -0.19 & -6.61 & 0 & 0 & 0 & 17.48 & -0.4 \\
 \hline
 \hline
 $\delta\Gamma(h \to WW^*)$ & 0 & 0 & 0 & 0 & 0 & 0 & 0 & -24.08 & -7.8 \\
 $\delta\Gamma(h \to ZZ^*)$ & 2.31 & 0 & 0 & 0 & 0 & -14.07 & -2.293 & -21.43 & -8.8 \\
 $\delta\Gamma(h \to \gamma\gamma)$ & 0 & 0 & 0 & 0 & 0 & -1963 & 981.6 & 976.4 & -3.1 \\
 $\delta\Gamma(h \to Z\gamma)$ & 3.25 & 0 & 0 & 0 & 0 & -695.8 & -294.2 & 966.6 & -5.8 \\
 \hline
 \multicolumn{10}{|l|}{$\delta \Gamma(h \to f\bar{f}) = 1 - 1.5\,c_H + 2\,c_{fH}\quad (f=b,c,\mu,\tau), \quad \delta \Gamma(h \to gg) = 1 - 1.5\,c_H + 2\,c_{gH} $}\\
 \hline \hline
 $\delta\sigma(Zhh,L)$ & -10.67 & 0 & 112.0 & 112.0 & 0 & 22.17 & -15.14 & 240.4 & -2.19 \\
 $\delta\sigma(Zhh,R)$ & -10.67 & -129.8 & 0 & 0 & 0 & 164.3 & 46.14 & 36.99 & -2.19 \\
 $\delta\sigma(Zhh,U)$ & -10.67 & -55.34 & 64.27 & 64.27 & 0 & 82.76 & 10.98 & 153.7 & -2.19 \\
 \hline
\end{tabular}$}
\caption{Tree-level contributions of Higgs operators to the observables that we use. The numbers agree with \cite{Barklow:2017awn} and are presented in absolute values not in \%. They are understood as follow: for example, $\delta \alpha^{-1} = -1.852 c_{WW} + 3.704 c_{WB} - 1.852 c_{BB}$ + (top-RG effects in \autoref{tab:Topops1}) + (SM parameters in \autoref{tab:SMparam1}).}\label{tab:Higgsops}
\end{center}
\end{table}

\begin{table}[t]
\begin{center}
\begin{tabular}{ |c||p{1.3cm}|p{1.4cm}|p{1.3cm}|p{1.3cm}|p{1.3cm}|p{1.3cm}|p{1.3cm}|  }
 \hline
 \multicolumn{8}{|c|}{Top operators $(Q_0 =$ 1 TeV) } \\
 \hline
 Observables &  $c_{tH}$ & $c_{HQ}^{(1)}$ & $c_{HQ}^{(3)}$ & $c_{Ht}$ & $c_{Htb}$ & $c_{tW}$ & $c_{tB}$  \\
  \hline
 $\delta\alpha^{-1}$	&0	&0	&0	&0	&0	&-0.0531 &-0.0967\\
 $\delta G_F$	&0	&0	&-0.0187	&0	&0	&0	&0\\
 $\delta A_l$	&0	&0.0386	&-0.116	&0.0773	&0	&1.3	&-0.228\\
 $\delta m_W$	&0	&0	&0	&0	&0	&0.043	&0\\
 $\delta m_Z$	&0	&-0.0668	&0	&0.0639	&0.0016	&0.0164	&-0.009\\
 $\delta m_h$	&0	&0	&0.169	&0	&-0.0049	&0	&0\\
 $\delta \Gamma_l$	&0	&-0.0682	&-0.02	&0.061	&0.0016	&0.0636	&-0.0296\\
 $\delta \Gamma_W$	&0	&0	&0	&0	&0	&0.129	&0\\
 $\delta \Gamma_Z$	&0	&-0.0668	&0	&0.0639	&0.0016	&0.0621	&-0.0293\\
 \hline\hline
 $\delta g_{Z eff} (250)$	&0	&0	&0.0096	&0	&0	&0	&0\\
 $\delta \kappa_{A eff} (250)$	&0	&0.0025	&0.0074	&0.0049	&0	&0	&0\\
 $\delta g_{Z eff} (500)$	&0	&0	&0.0048	&0	&0	&0	&0\\
 $\delta \kappa_{A eff} (500)$	&0	&0.0012	&0.0037	&0.0025	&0	&0	&0\\
 \hline\hline
 $a_L(250)$	&0	&-0.189	&-0.161	&0.212	&0.0066	&0.107	&-0.0324\\
 $a_R(250)$	&0	&-0.224	&-0.0577	&0.143	&0.0066	&-0.0464	&-0.0053\\
 $b_L(250)$	&0	&0	&0	&0	&0	&0.0877	&-0.0251\\
 $b_R(250)$	&0	&0	&0	&0	&0	&-0.0456	&-0.0017\\
 $a_L(500)$	&0	&-0.119	&-0.17	&0.175	&0.0023	&0.0635	&-0.0216\\
 $a_R(500)$	&0	&-0.188	&0.0372	&0.0373	&0.0023	&-0.0134	&-0.008\\
 $b_L(500)$	&0	&0	&0	&0	&0	&0.0474	&-0.0132\\
 $b_R(500)$	&0	&0	&0	&0	&0	&-0.0269	&-0.0001\\
 \hline\hline
 $\delta \sigma (\nu \bar{\nu} h,250)$	&0	&-0.0003	&-0.329	&-0.0005	&0.0104	&0.156	&0\\
 $\delta \sigma (\nu \bar{\nu} h,500)$	&0	&-0.0001	&-0.045	&-0.0001	&0.002	&0.156	&0\\
 \hline \hline
 $\delta\Gamma(h \to WW^*)$	&0	&0	&2.64	&0	&-0.0766	&-0.259	&0\\
 $\delta\Gamma(h \to ZZ^*)$	&0	&0.308	&2.98	&-0.295	&-0.0939	&-0.104	&0.0629\\
 $\delta\Gamma(h \to \gamma\gamma)$	&-0.565* &0	&1.05	&0	&-0.0304	&28.1	&51.2\\
 $\delta\Gamma(h \to Z\gamma)$	&-0.114* &0.141	&2.26	&-0.709	&-0.0676	&16.6	&-2.79\\
 $\delta\Gamma(h \to gg)$	&2*	&0	&0.508	&0	&-0.0147	&0	&0\\
 $\delta\Gamma(h \to f\bar{f})$	&0	&0	&0.508	&0	&-0.0147	&0	&0\\
 \hline \hline
 $\delta\sigma(Zhh,L)$ & 0 & -1.06 & -0.643 & 1.14 & 0.0337 & 1.08 & -0.354 \\
 $\delta\sigma(Zhh,R)$ & 0 & -1.20 & -0.229 & 0.864 & 0.0337 & -0.291 & -0.114 \\
 $\delta\sigma(Zhh,U)$ & 0 & -1.12 & -0.466 & 1.02 & 0.0337 & 0.493 & -0.251 \\
 \hline
\end{tabular}
\caption{Contributions from top operators, written in terms of Wilson coefficients renormalized at $Q_0=$ 1 TeV. The default RG scale $Q = {\rm max}(m_t, Q_{\rm proc})$ is used.  The ${\cal O}_{tH}$ contributions to $\delta\Gamma(h \to \gamma\gamma,\, Z\gamma,\, gg)$ marked with $(*)$ are from the tree-level shift of the top Yukawa discussed in \Sec{sec:yukawa}.}
\label{tab:Topops1}
\end{center}
\end{table}

\begin{table}[t]
\begin{center}
\begin{tabular}{ |c||p{1.3cm}|p{1.4cm}|p{1.3cm}|p{1.3cm}||p{1.3cm}| }
 \hline
 \multicolumn{6}{|c|}{ SM parameters } \\
 \hline
 Observalbles & $\delta g$ & $\delta g^\prime$ & $\delta v$ & $\delta \bar{\lambda}$ & $\delta m_h$ \\
 \hline
 $\delta\alpha^{-1}$ & -0.463 & -1.54 & 0 & 0 & 0 \\
 $\delta G_F$ & 0 & 0 & -2. & 0 & 0\\
 $\delta A_l$ & 19. & -19. & 0 & 0 & 0\\
 $\delta m_W$ & 1. & 0 & 1. & 0 & 0\\
 $\delta m_Z$ & 0.768 & 0.232 & 1. & 0 & 0\\
 $\delta m_h$ &  0 & 0 & 1. & 0.5 & 1\\
 $\delta \Gamma_l$ & 2.51 & 0.485 & 1. & 0 & 0 \\
 $\delta \Gamma_{W}$ & 3. & 0 & 1. & 0 & 0 \\
 $\delta \Gamma_{Z}$ & 2.68 & 0.318 & 1. & 0 & 0 \\
 \hline
 \hline
 $\delta g_{Z\,eff} $ & 0 & 0 & 0 & 0 & 0 \\
 $\delta \kappa_{A\,eff} $ &0 & 0 & 0 & 0 & 0 \\
 $\delta \lambda_{A\,eff}$ & 0 & 0 & 0 & 0 & 0 \\
 \hline
 \hline
 $a_L(250)$ & 3.18 & -0.767 & -0.720 & -0.566 & -1.13\\
 $a_R(250)$ & 0.316 & 2.10 & -0.72 & -0.566 & -1.13\\
 $b_L(250)$ & 0 & 0 & 0 & 0 & 0 \\
 $b_R(250)$ & 0 & 0 & 0 & 0 & 0 \\
 $a_L(500)$ & 3.09 & -0.793 & 0.125 & -0.088 & -0.175\\
 $a_R(500)$ & 0.231 & 2.07 & 0.125 & -0.088 & -0.175\\
 $b_L(500)$ & 0 & 0 & 0 & 0 & 0 \\ 
 $b_R(500)$ & 0 & 0 & 0 & 0 & 0 \\
 \hline
 \hline
 $\delta \sigma (\nu \bar{\nu} h,250)$ & 4.40 & 0 & -5.30 & -1.85 & -3.70 \\ 
 $\delta \sigma (\nu \bar{\nu} h,500)$ & 5.15 & 0 & -2.05 & -0.60 & -1.20 \\
 \hline
 \hline
 $\delta\Gamma(h \to WW^*)$ & -4.52 & 0 & 2.96 & 6.80 & 13.6\\
 $\delta\Gamma(h \to ZZ^*)$ & -4.37 & -1.89 & 2.98 & 7.80 & 15.6\\
 $\delta\Gamma(h \to \gamma\gamma)$ & -0.374 & 3.07 & 0.9 & 2.10 & 4.20\\
 $\delta\Gamma(h \to Z\gamma)$ &  -5.61 & 3.11 & 1.10 & 4.80 & 9.60\\
 $\delta\Gamma(h \to gg)$ & 0 & 0 & 1 & 0.5 & 1\\
 $\delta\Gamma(h \to f\bar{f})$ & 0 & 0 & -1 & 0.5 & 1\\
 \hline \hline
 $\delta\sigma(Zhh,L)$ & 10.6 & -0.257 & 1.56 & -1.39 & -3.90\\
 $\delta\sigma(Zhh,R)$ & 4.87 & 5.47 & 1.56 & -1.39 & -3.90\\
 $\delta\sigma(Zhh,U)$ & 8.16 & 2.18 & 1.56 & -1.39 & -3.90\\
 \hline
\end{tabular}
\caption{Contributions from SM parameters $\delta g,\, \delta g^\prime,\, \delta v,\, \delta \bar{\lambda}$. And the contributions through $\delta m_h$ are also shown although they are already rewritten in terms of other parameters in \autoref{tab:SMparam1}, \ref{tab:Topops1}, and \ref{tab:Higgsops}.}
\label{tab:SMparam1}
\end{center}
\end{table}

We collect numerical expressions of observables in terms of operator coefficients in \autoref{tab:Higgsops} (Higgs operator), \autoref{tab:Topops1} (top-operators renormalized at $Q_0=$ 1 TeV), and \autoref{tab:SMparam1} (SM parameters and the Higgs mass, although $\delta m_h$ is already rewritten in terms of other parameters in the tables).

\clearpage

\section{Numerical results}
\label{app:numbers}

We tabulate numerical values for the the figures in \Sec{sec:results}: \autoref{tab:HiggsCoup} (Higgs couplings) and \autoref{tab:OpConst} (EFT coefficients).

\begin{table}[h]
\begin{center}
 \begin{tabular}{c||c|c|c||c|c|c|c}
     \toprule
     & \multicolumn{3}{c||}{ILC 250} & \multicolumn{4}{c}{ILC 250 + 500} \\
     \hline
     & w/o top & \multicolumn{2}{c||}{w/ top} & w/o top &  \multicolumn{3}{c}{w/ top}\\
     \hline
     top data  & - &  LHC Run 2 & HL-LHC S2 & - & - &HL-LHC S2 & ILC 500  \\
     \hline\hline
     \text{$hbb$}	&0.94	&1.02	&0.95	&0.55	&1.85	&0.56	&0.55\\
     \text{$hcc$}	&1.73	&1.77	&1.73	&1.10	&2.08	&1.10	&1.10\\
     \text{$h\tau\tau$}	&1.08	&1.15	&1.08	&0.71	&1.90	&0.72	&0.71\\
     \text{$hWW$}	&0.48	&0.62	&0.49	&0.35	&1.81	&0.36	&0.35\\
     \text{$hZZ$}	&0.48	&0.62	&0.49	&0.35	&1.70	&0.35	&0.35\\
     \text{$h\gamma\gamma$}	&1.08	&1.15	&1.09	&1.00	&1.97	&1.00	&1.00\\
     \text{$hgg$}	&1.54	&1.59	&1.54	&0.90	&1.98	&0.90	&0.90\\
     \text{$h\mu\mu$}	&3.95	&3.97	&3.95	&3.71	&4.09	&3.71	&3.71\\
     \text{$hZ\gamma$}	&9.04	&9.11	&9.07	&6.37	&10.40	&6.62	&6.37\\
     \hline
$\Gamma_{\text{inv}}$	&0.30	&0.30	&0.30	&0.28	&0.28	&0.28	&0.28\\
$\Gamma_{\text{other}}$	&1.51	&1.51	&1.51	&1.16	&1.21	&1.16	&1.16\\
$\Gamma_{\text{tot}}$	&2.20	&2.34	&2.21	&1.51	&3.82	&1.52	&1.51\\
     \hline
     \text{$\delta y_t$}	&0	&12.8	&3.29	&0	&396.	&3.29	&3.29\\
$\delta \bar{\lambda}$	&0.87	&1.20	&0.90	&0.66	&3.29	&0.68	&0.66\\
     \bottomrule
 \end{tabular}
\caption{The projected uncertainties of Higgs couplings in ILC 250 (\Fig{fig:250higgs}) and 500 (\Fig{fig:500higgs}) stages, in $\%$. First column shows the values without top operators. Second and third columns represent the uncertainties with direct top data of LHC Run 2 and HL-LHC S2. In ILC 250+500 fits, we present the Higgs precision evaluated without top (fourth column), ``indirect only" including top operators (fifth column), plus HL-LHC S2 data (sixth column) and adding ILC 500 direct-top in the last column.} \label{tab:HiggsCoup}
\end{center}
\end{table}
\begin{table}[h]
\begin{center}
 \begin{tabular}{c||c|c|c||c|c|c|c}
     \toprule
     & \multicolumn{3}{c||}{ILC 250} & \multicolumn{4}{c}{ILC 250 + 500} \\
     \hline
     & w/o top & \multicolumn{2}{c||}{w/ top} & w/o top &  \multicolumn{3}{c}{w/ top}\\
     \hline
     top data  & - &  LHC Run 2 & HL-LHC S2 & - &  &HL-LHC S2 & ILC 500  \\
     \hline\hline
     \text{$c_{T}$}	&0.029	&2.514	&0.666	&0.02	&3.336	&0.61	&0.027\\
     \text{$c_{HE}$}	&0.024	&0.066	&0.029	&0.007	&0.035	&0.012	&0.007\\
     \text{$c_{HL}$}	&0.025	&0.042	&0.027	&0.011	&0.027	&0.013	&0.011\\
     \text{$c_{HL}'$}	&0.035	&0.04	&0.036	&0.012	&0.05	&0.014	&0.012\\
     \text{$c_{3W}$}	&0.017	&0.017	&0.017	&0.008	&0.008	&0.008	&0.008\\
     \text{$c_{WB}$}	&0.007	&0.108	&0.079	&0.004	&0.784	&0.069	&0.004\\
     \text{$c_{BB}$}	&0.016	&0.355	&0.262	&0.011	&2.457	&0.228	&0.011\\
     \text{$c_{WW}$}	&0.016	&0.021	&0.016	&0.01	&0.178	&0.01	&0.01\\
     \text{$c_{H}$}	&0.885	&1.385	&0.916	&0.656	&2.313	&0.674	&0.657\\
     \hline
     \text{$c_{tH}$}	&0	&12.779	&3.288	&0	&396.48	&3.286	&3.286\\
     \text{$c_{Hq}^{(1)}$}	&0	&2.866	&0.448	&0	&20.67	&0.445	&0.016\\
     \text{$c_{Hq}^{(3)}$}	&0	&2.149	&0.448	&0	&9.301	&0.445	&0.016\\
     \text{$c_{Ht}$}	&0	&19.778	&5.069	&0	&19.046	&4.579	&0.03\\
     \text{$c_{Htb}$}	&0	&14.34	&8.963	&0	&692.62	&8.938	&5.372\\
     \text{$c_{tW}$}	&0	&1.237	&0.234	&0	&10.433	&0.233	&0.078\\
     \text{$c_{tB}$}	&0	&10.904	&8.043	&0	&71.313	&7.044	&0.041\\
     \bottomrule
\end{tabular}
\caption{The global 1 $\sigma$ constraints on the EFT coefficients at ILC 250 (\Fig{fig:250higgs}) and 500 (\Fig{fig:500higgs}) stages, in $\%$. Details are as in \autoref{tab:HiggsCoup}.} \label{tab:OpConst}
\end{center}
\end{table}

\clearpage

\begin{landscape}
\section{Covariance matrices}
\label{app:cov}

We present the covariance matrices of the global fits for the three benchmark scenarios. The matrices are written down in the basis,

\begin{align}
( \delta g,\, \delta g^\prime,\, \delta v,\, \delta \bar{\lambda},\, c_{T},\, c_{HE},\, c_{HL},\, c_{HL}^\prime,\, c_{3W},\, c_{WB},\,& c_{BB},\, c_{WW},\, c_H,\, c_{tH},\, c_{Hq}^{(1)},\, c_{Hq}^{(3)},\, c_{Ht},\, c_{Htb}, \\\nonumber 
&{} c_{tW},\, c_{tB},\, c_{bH},\, c_{cH},\, c_{gH},\, c_{\tau H},\, c_{\mu H},\, \delta a_{inv},\, \delta a_{other},\, C_W,\, C_Z )
\end{align}

\begin{itemize} 

\item Covariance matrix without top operators at ILC 250
    
 \begin{equation}\label{covwot500}
    \resizebox{0.85\hsize}{!}{$10^{-8} \cdot
     \left(
     \begin{array}{cccccccccccccccccccccc}
    17.	&-4.4	&4.3	&-90.	&2.5	&1.1	&3.9	&4.3	&0.18	&-0.067	&5.2	&-5.4	&-81.	&3.5	&3.9	&3.9	&8.2	&-45.	&0	&6.7	&4.4	&1.5\\
-4.4	&2.	&-1.2	&19.	&-1.1	&-0.35	&-1.1	&-1.2	&-0.055	&-0.14	&-1.7	&1.4	&17.	&-1.4	&-1.4	&-1.4	&-2.6	&-7.8	&0	&-2.3	&-1.2	&-0.22\\
4.3	&-1.2	&13.	&39.	&-1.3	&-4.1	&3.	&13.	&2.	&-1.8	&0.55	&-4.3	&64.	&-17.	&-18.	&-18.	&-5.8	&-23.	&0	&-3.2	&13.	&13.\\
-90.	&19.	&39.	&7.6\times 10^{3}	&8.9	&-64.	&72.	&39.	&-13.	&-2.8	&-18.	&13.	&7.7\times 10^{3}	&1.1\times 10^{2}	&1.4\times 10^{2}	&1.3\times 10^{2}	&-96.	&1.8\times 10^{2}	&0	&5.0\times 10^{3}	&36.	&48.\\
2.5	&-1.1	&-1.3	&8.9	&8.4	&4.5	&3.7	&-1.3	&-0.88	&1.6	&3.3	&-0.047	&6.2	&-0.31	&-0.24	&-0.23	&-1.6	&-15.	&0	&3.4	&-2.8	&-5.3\\
1.1	&-0.35	&-4.1	&-64.	&4.5	&5.9	&0.32	&-4.1	&-0.61	&1.4	&2.1	&0.75	&-72.	&6.3	&7.1	&7.1	&3.	&-3.3	&0	&3.4	&-4.2	&-6.3\\
3.9	&-1.1	&3.	&72.	&3.7	&0.32	&6.5	&3.	&-0.82	&0.43	&2.6	&-1.8	&78.	&-1.7	&-2.5	&-2.5	&0.57	&-17.	&0	&0.88	&3.3	&1.3\\
4.3	&-1.2	&13.	&39.	&-1.3	&-4.1	&3.	&13.	&2.	&-1.8	&0.55	&-4.3	&64.	&-17.	&-18.	&-18.	&-5.8	&-23.	&0	&-3.2	&13.	&13.\\
0.18	&-0.055	&2.	&-13.	&-0.88	&-0.61	&-0.82	&2.	&2.9	&-0.44	&-0.33	&-0.55	&-9.	&-3.3	&-3.3	&-3.3	&-1.3	&-1.7	&0	&-0.73	&2.	&2.5\\
-0.067	&-0.14	&-1.8	&-2.8	&1.6	&1.4	&0.43	&-1.8	&-0.44	&0.55	&0.61	&0.5	&-6.4	&2.4	&2.6	&2.6	&0.76	&-0.031	&0	&1.1	&-2.	&-2.6\\
5.2	&-1.7	&0.55	&-18.	&3.3	&2.1	&2.6	&0.55	&-0.33	&0.61	&2.6	&-1.4	&-17.	&2.3	&2.4	&2.4	&3.	&1.2	&0	&3.1	&0.46	&-1.4\\
-5.4	&1.4	&-4.3	&13.	&-0.047	&0.75	&-1.8	&-4.3	&-0.55	&0.5	&-1.4	&2.4	&4.7	&3.1	&3.3	&3.3	&-0.97	&17.	&0	&-0.84	&-4.5	&-3.9\\
-81.	&17.	&64.	&7.7\times 10^{3}	&6.2	&-72.	&78.	&64.	&-9.	&-6.4	&-17.	&4.7	&7.8\times 10^{3}	&1.7\times 10^{2}	&1.9\times 10^{2}	&1.9\times 10^{2}	&1.3\times 10^{2}	&1.1\times 10^{2}	&0	&5.0\times 10^{3}	&62.	&74.\\
3.5	&-1.4	&-17.	&1.1\times 10^{2}	&-0.31	&6.3	&-1.7	&-17.	&-3.3	&2.4	&2.3	&3.1	&1.7\times 10^{2}	&6.8\times 10^{3}	&6.2\times 10^{3}	&6.2\times 10^{3}	&6.2\times 10^{3}	&8.1\times 10^{2}	&0	&2.3\times 10^{3}	&-2.	&-17.\\
3.9	&-1.4	&-18.	&1.4\times 10^{2}	&-0.24	&7.1	&-2.5	&-18.	&-3.3	&2.6	&2.4	&3.3	&1.9\times 10^{2}	&6.2\times 10^{3}	&2.8\times 10^{4}	&6.2\times 10^{3}	&6.2\times 10^{3}	&8.1\times 10^{2}	&0	&1.7\times 10^{3}	&-3.3	&-18.\\
3.9	&-1.4	&-18.	&1.3\times 10^{2}	&-0.23	&7.1	&-2.5	&-18.	&-3.3	&2.6	&2.4	&3.3	&1.9\times 10^{2}	&6.2\times 10^{3}	&6.2\times 10^{3}	&2.2\times 10^{4}	&6.2\times 10^{3}	&8.1\times 10^{2}	&0	&3.5\times 10^{2}	&-3.3	&-18.\\
8.2	&-2.6	&-5.8	&-96.	&-1.6	&3.	&0.57	&-5.8	&-1.3	&0.76	&3.	&-0.97	&1.3\times 10^{2}	&6.2\times 10^{3}	&6.2\times 10^{3}	&6.2\times 10^{3}	&9.4\times 10^{3}	&7.8\times 10^{2}	&0	&2.6\times 10^{3}	&9.5	&-5.\\
-45.	&-7.8	&-23.	&1.8\times 10^{2}	&-15.	&-3.3	&-17.	&-23.	&-1.7	&-0.031	&1.2	&17.	&1.1\times 10^{2}	&8.1\times 10^{2}	&8.1\times 10^{2}	&8.1\times 10^{2}	&7.8\times 10^{2}	&1.5\times 10^{5}	&0	&2.5\times 10^{2}	&-21.	&-2.2\\
0	&0	&0	&0	&0	&0	&0	&0	&0	&0	&0	&0	&0	&0	&0	&0	&0	&0	&3.3\times 10^{2}	&3.3\times 10^{2}	&0	&0\\
6.7	&-2.3	&-3.2	&5.0\times 10^{3}	&3.4	&3.4	&0.88	&-3.2	&-0.73	&1.1	&3.1	&-0.84	&5.0\times 10^{3}	&2.3\times 10^{3}	&1.7\times 10^{3}	&3.5\times 10^{2}	&2.6\times 10^{3}	&2.5\times 10^{2}	&3.3\times 10^{2}	&8.0\times 10^{3}	&-3.1	&-5.9\\
4.4	&-1.2	&13.	&36.	&-2.8	&-4.2	&3.3	&13.	&2.	&-2.	&0.46	&-4.5	&62.	&-2.	&-3.3	&-3.3	&9.5	&-21.	&0	&-3.1	&37.	&14.\\
1.5	&-0.22	&13.	&48.	&-5.3	&-6.3	&1.3	&13.	&2.5	&-2.6	&-1.4	&-3.9	&74.	&-17.	&-18.	&-18.	&-5.	&-2.2	&0	&-5.9	&14.	&38.\\
     \end{array}\right)$}
 \end{equation}
 
 \clearpage
 
\item Covariance matrix for LHC run 2 + ILC 250
\begin{equation}\label{covInd}
    \resizebox{0.85\hsize}{!}{$10^{-7} \cdot
     \left(
     \begin{array}{ccccccccccccccccccccccccccccc}
    2.8	&-0.67	&0.33	&10.	&-64.	&1.7	&1.1	&0.47	&0.014	&1.3	&-4.4	&-0.66	&16.	&1.2\times 10^{2}	&-21.	&15.	&5.2\times 10^{2}	&0.69	&-10.	&1.5\times 10^{2}	&1.4	&1.3	&1.2\times 10^{2}	&1.7	&-3.6	&0	&0.36	&0.34	&0.1\\
-0.67	&0.32	&-0.078	&-4.7	&19.	&-0.51	&-0.32	&-0.12	&-0.004	&-0.4	&1.3	&0.15	&-6.4	&-77.	&5.6	&-4.5	&1.6\times 10^{2}	&-0.35	&3.1	&-45.	&-0.43	&-0.39	&77.	&-0.46	&-1.	&0	&-0.16	&-0.08	&-0.0016\\
0.33	&-0.078	&1.3	&4.9	&-3.5	&-0.3	&0.41	&1.3	&0.21	&-0.9	&2.6	&-0.48	&5.4	&-14.	&0.0055	&-6.5	&-27.	&0.11	&5.3	&-78.	&-1.9	&-2.1	&12.	&-0.71	&-2.4	&0	&-0.41	&1.4	&1.4\\
10.	&-4.7	&4.9	&1.4\times 10^{3}	&2.0\times 10^{3}	&40.	&31.	&3.8	&-1.3	&9.9	&-41.	&-4.3	&1.4\times 10^{3}	&1.1\times 10^{3}	&2.5\times 10^{2}	&1.1\times 10^{2}	&1.5\times 10^{4}	&1.5\times 10^{2}	&57.	&1.2\times 10^{3}	&6.9	&-0.28	&1.1\times 10^{3}	&4.6	&37.	&0	&5.1\times 10^{2}	&4.6	&7.6\\
-64.	&19.	&-3.5	&2.0\times 10^{3}	&6.3\times 10^{3}	&1.5\times 10^{2}	&-77.	&-8.	&-0.24	&0.98	&21.	&16.	&2.2\times 10^{3}	&1.6\times 10^{2}	&12.	&4.8\times 10^{2}	&4.9\times 10^{4}	&6.5\times 10^{2}	&36.	&7.1\times 10^{2}	&-59.	&-44.	&1.2\times 10^{2}	&-47.	&-65.	&0	&39.	&-3.8	&-9.8\\
1.7	&-0.51	&-0.3	&40.	&1.5\times 10^{2}	&4.3	&2.	&-0.11	&-0.055	&-0.1	&0.52	&-0.36	&46.	&-4.5	&-39.	&21.	&1.2\times 10^{3}	&0.89	&1.3	&-7.8	&2.	&1.8	&6.3	&1.4	&1.2	&0	&-0.63	&-0.31	&-0.39\\
1.1	&-0.32	&0.41	&31.	&-77.	&2.	&1.7	&0.53	&-0.077	&-0.64	&2.4	&-0.44	&36.	&-13.	&-22.	&12.	&6.3\times 10^{2}	&0.64	&4.9	&-64.	&0.4	&0.14	&13.	&0.55	&-0.95	&0	&-0.48	&0.44	&0.31\\
0.47	&-0.12	&1.3	&3.8	&-8.	&-0.11	&0.53	&1.6	&0.21	&-0.95	&2.8	&-0.51	&19.	&-13.	&-18.	&37.	&-81.	&0.57	&6.6	&-84.	&-1.7	&-1.8	&11.	&-0.56	&-2.2	&0	&-0.46	&1.3	&1.4\\
0.014	&-0.004	&0.21	&-1.3	&-0.24	&-0.055	&-0.077	&0.21	&0.29	&-0.077	&0.084	&-0.057	&-0.84	&-0.57	&-0.17	&0.12	&-1.4	&0.011	&0.26	&-3.6	&-0.33	&-0.34	&0.23	&-0.14	&-0.18	&0	&-0.078	&0.2	&0.25\\
1.3	&-0.4	&-0.9	&9.9	&0.98	&-0.1	&-0.64	&-0.95	&-0.077	&12.	&-38.	&-0.27	&6.3	&-13.	&4.5	&-5.3	&15.	&-0.066	&16.	&1.2\times 10^{3}	&2.4	&2.4	&15.	&1.5	&0.95	&0	&0.9	&-0.93	&-1.\\
-4.4	&1.3	&2.6	&-41.	&21.	&0.52	&2.4	&2.8	&0.084	&-38.	&1.3\times 10^{2}	&0.54	&-27.	&1.0\times 10^{2}	&-14.	&24.	&1.3\times 10^{2}	&-1.1	&-9.2	&3.9\times 10^{3}	&-7.6	&-7.5	&1.1\times 10^{2}	&-5.	&-3.5	&0	&-2.3	&2.6	&2.4\\
-0.66	&0.15	&-0.48	&-4.3	&16.	&-0.36	&-0.44	&-0.51	&-0.057	&-0.27	&0.54	&0.45	&-6.5	&33.	&6.1	&-3.5	&1.3\times 10^{2}	&0.22	&-15.	&-21.	&0.25	&0.32	&-32.	&-0.16	&1.6	&0	&0.055	&-0.5	&-0.46\\
16.	&-6.4	&5.4	&1.4\times 10^{3}	&2.2\times 10^{3}	&46.	&36.	&19.	&-0.84	&6.3	&-27.	&-6.5	&1.9\times 10^{3}	&1.1\times 10^{3}	&4.1\times 10^{2}	&1.4\times 10^{3}	&1.7\times 10^{4}	&1.9\times 10^{3}	&1.1\times 10^{2}	&8.4\times 10^{2}	&9.1	&1.2	&1.1\times 10^{3}	&6.7	&37.	&0	&5.1\times 10^{2}	&5.2	&8.\\
1.2\times 10^{2}	&-77.	&-14.	&1.1\times 10^{3}	&1.6\times 10^{2}	&-4.5	&-13.	&-13.	&-0.57	&-13.	&1.0\times 10^{2}	&33.	&1.1\times 10^{3}	&1.6\times 10^{5}	&1.1\times 10^{3}	&1.5\times 10^{2}	&-45.	&1.8\times 10^{2}	&83.	&1.4\times 10^{3}	&-56.	&-64.	&1.6\times 10^{5}	&-78.	&-56.	&0	&33.	&-14.	&-7.6\\
-21.	&5.6	&0.0055	&2.5\times 10^{2}	&12.	&-39.	&-22.	&-18.	&-0.17	&4.5	&-14.	&6.1	&4.1\times 10^{2}	&1.1\times 10^{3}	&8.2\times 10^{3}	&2.0\times 10^{3}	&8.6\times 10^{3}	&1.5\times 10^{2}	&-80.	&3.8\times 10^{2}	&-15.	&-16.	&1.1\times 10^{3}	&-16.	&-21.	&0	&7.2	&-0.096	&-0.5\\
15.	&-4.5	&-6.5	&1.1\times 10^{2}	&4.8\times 10^{2}	&21.	&12.	&37.	&0.12	&-5.3	&24.	&-3.5	&1.4\times 10^{3}	&1.5\times 10^{2}	&2.0\times 10^{3}	&4.6\times 10^{3}	&5.8\times 10^{3}	&50.	&1.4\times 10^{2}	&6.8\times 10^{2}	&24.	&22.	&1.2\times 10^{2}	&16.	&19.	&0	&-5.2	&-6.4	&-7.3\\
5.2\times 10^{2}	&1.6\times 10^{2}	&-27.	&1.5\times 10^{4}	&4.9\times 10^{4}	&1.2\times 10^{3}	&6.3\times 10^{2}	&-81.	&-1.4	&15.	&1.3\times 10^{2}	&1.3\times 10^{2}	&1.7\times 10^{4}	&-45.	&8.6\times 10^{3}	&5.8\times 10^{3}	&3.9\times 10^{5}	&3.0\times 10^{2}	&1.7\times 10^{2}	&4.7\times 10^{3}	&4.7\times 10^{2}	&3.5\times 10^{2}	&3.1\times 10^{2}	&3.8\times 10^{2}	&5.2\times 10^{2}	&0	&3.1\times 10^{2}	&-28.	&-73.\\
0.69	&-0.35	&0.11	&1.5\times 10^{2}	&6.5\times 10^{2}	&0.89	&0.64	&0.57	&0.011	&-0.066	&-1.1	&0.22	&1.9\times 10^{3}	&1.8\times 10^{2}	&1.5\times 10^{2}	&50.	&3.0\times 10^{2}	&2.1\times 10^{5}	&-39.	&39.	&-1.6	&0.38	&1.8\times 10^{2}	&0.48	&0.73	&0	&2.3	&0.1	&0.17\\
-10.	&3.1	&5.3	&57.	&36.	&1.3	&4.9	&6.6	&0.26	&16.	&-9.2	&-15.	&1.1\times 10^{2}	&83.	&-80.	&1.4\times 10^{2}	&1.7\times 10^{2}	&-39.	&1.5\times 10^{3}	&2.6\times 10^{2}	&-15.	&-16.	&-99.	&-11.	&-7.8	&0	&-3.6	&5.5	&5.4\\
1.5\times 10^{2}	&-45.	&-78.	&1.2\times 10^{3}	&7.1\times 10^{2}	&-7.8	&-64.	&-84.	&-3.6	&1.2\times 10^{3}	&3.9\times 10^{3}	&-21.	&8.4\times 10^{2}	&1.4\times 10^{3}	&3.8\times 10^{2}	&6.8\times 10^{2}	&4.7\times 10^{3}	&39.	&2.6\times 10^{2}	&1.2\times 10^{5}	&2.4\times 10^{2}	&2.4\times 10^{2}	&1.6\times 10^{3}	&1.6\times 10^{2}	&1.1\times 10^{2}	&0	&79.	&-80.	&-79.\\
1.4	&-0.43	&-1.9	&6.9	&-59.	&2.	&0.4	&-1.7	&-0.33	&2.4	&-7.6	&0.25	&9.1	&-56.	&-15.	&24.	&4.7\times 10^{2}	&-1.6	&-15.	&2.4\times 10^{2}	&6.8\times 10^{2}	&6.2\times 10^{2}	&6.8\times 10^{2}	&6.2\times 10^{2}	&82.	&0	&2.3\times 10^{2}	&-0.41	&-1.9\\
1.3	&-0.39	&-2.1	&-0.28	&-44.	&1.8	&0.14	&-1.8	&-0.34	&2.4	&-7.5	&0.32	&1.2	&-64.	&-16.	&22.	&3.5\times 10^{2}	&0.38	&-16.	&2.4\times 10^{2}	&6.2\times 10^{2}	&2.8\times 10^{3}	&6.9\times 10^{2}	&6.2\times 10^{2}	&82.	&0	&1.7\times 10^{2}	&-0.55	&-2.\\
1.2\times 10^{2}	&77.	&12.	&1.1\times 10^{3}	&1.2\times 10^{2}	&6.3	&13.	&11.	&0.23	&15.	&1.1\times 10^{2}	&-32.	&1.1\times 10^{3}	&1.6\times 10^{5}	&1.1\times 10^{3}	&1.2\times 10^{2}	&3.1\times 10^{2}	&1.8\times 10^{2}	&-99.	&1.6\times 10^{3}	&6.8\times 10^{2}	&6.9\times 10^{2}	&1.7\times 10^{5}	&7.0\times 10^{2}	&1.4\times 10^{2}	&0	&1.6	&14.	&5.5\\
1.7	&-0.46	&-0.71	&4.6	&-47.	&1.4	&0.55	&-0.56	&-0.14	&1.5	&-5.	&-0.16	&6.7	&-78.	&-16.	&16.	&3.8\times 10^{2}	&0.48	&-11.	&1.6\times 10^{2}	&6.2\times 10^{2}	&6.2\times 10^{2}	&7.0\times 10^{2}	&9.4\times 10^{2}	&79.	&0	&2.6\times 10^{2}	&0.82	&-0.59\\
-3.6	&-1.	&-2.4	&37.	&-65.	&1.2	&-0.95	&-2.2	&-0.18	&0.95	&-3.5	&1.6	&37.	&-56.	&-21.	&19.	&5.2\times 10^{2}	&0.73	&-7.8	&1.1\times 10^{2}	&82.	&82.	&1.4\times 10^{2}	&79.	&1.5\times 10^{4}	&0	&24.	&-2.1	&-0.25\\
0	&0	&0	&0	&0	&0	&0	&0	&0	&0	&0	&0	&0	&0	&0	&0	&0	&0	&0	&0	&0	&0	&0	&0	&0	&33.	&-33.	&0	&0\\
0.36	&-0.16	&-0.41	&5.1\times 10^{2}	&39.	&-0.63	&-0.48	&-0.46	&-0.078	&0.9	&-2.3	&0.055	&5.1\times 10^{2}	&33.	&7.2	&-5.2	&3.1\times 10^{2}	&2.3	&-3.6	&79.	&2.3\times 10^{2}	&1.7\times 10^{2}	&1.6	&2.6\times 10^{2}	&24.	&-33.	&8.0\times 10^{2}	&-0.41	&-0.72\\
0.34	&-0.08	&1.4	&4.6	&-3.8	&-0.31	&0.44	&1.3	&0.2	&-0.93	&2.6	&-0.5	&5.2	&-14.	&-0.096	&-6.4	&-28.	&0.1	&5.5	&-80.	&-0.41	&-0.55	&14.	&0.82	&-2.1	&0	&-0.41	&3.8	&1.5\\
0.1	&-0.0016	&1.4	&7.6	&-9.8	&-0.39	&0.31	&1.4	&0.25	&-1.	&2.4	&-0.46	&8.	&-7.6	&-0.5	&-7.3	&-73.	&0.17	&5.4	&-79.	&-1.9	&-2.	&5.5	&-0.59	&-0.25	&0	&-0.72	&1.5	&3.9\\
     \end{array}\right)$}
 \end{equation}
 
 \item Covariance matrix for HL-LHC S2 + ILC 250
 \begin{equation}\label{covHL250}
     \resizebox{0.85\hsize}{!}{$10^{-8} \cdot
     \left(
     \begin{array}{ccccccccccccccccccccccccccccc}
     19.	&-4.8	&3.8	&-72.	&-37.	&1.9	&3.9	&3.8	&0.16	&7.6	&-20.	&-5.6	&-64.	&-72.	&0.43	&-0.43	&3.0\times 10^{2}	&-9.7	&-3.9	&7.9\times 10^{2}	&5.5	&5.8	&78.	&9.6	&-44.	&0	&7.	&3.9	&1.\\
-4.8	&2.1	&-1.	&13.	&11.	&-0.58	&-1.1	&-1.	&-0.048	&-2.5	&6.2	&1.4	&11.	&-53.	&-0.17	&0.17	&91.	&3.2	&1.2	&2.4\times 10^{2}	&-2.	&-2.	&52.	&-3.	&-8.1	&0	&-2.4	&-1.	&-0.066\\
3.8	&-1.	&13.	&38.	&-6.6	&-3.9	&3.4	&13.	&2.	&-6.1	&15.	&-4.2	&62.	&-12.	&3.4	&-3.4	&-39.	&4.1	&2.5	&4.4\times 10^{2}	&-18.	&-19.	&-6.9	&-6.3	&-23.	&0	&-3.6	&13.	&14.\\
-72.	&13.	&38.	&8.2\times 10^{3}	&1.5\times 10^{3}	&-32.	&86.	&34.	&-13.	&45.	&1.8\times 10^{2}	&8.6	&8.1\times 10^{3}	&7.6\times 10^{2}	&4.0\times 10^{2}	&3.9\times 10^{2}	&1.1\times 10^{4}	&4.5\times 10^{2}	&46.	&5.0\times 10^{3}	&-91.	&1.2\times 10^{2}	&8.8\times 10^{2}	&-81.	&2.0\times 10^{2}	&0	&5.0\times 10^{3}	&35.	&47.\\
-37.	&11.	&-6.6	&1.5\times 10^{3}	&4.4\times 10^{3}	&-97.	&-50.	&0.75	&-1.	&13.	&-19.	&12.	&1.2\times 10^{3}	&-95.	&7.9\times 10^{2}	&7.9\times 10^{2}	&3.4\times 10^{4}	&3.0\times 10^{3}	&-70.	&6.7\times 10^{2}	&-32.	&-22.	&73.	&-28.	&-53.	&0	&31.	&-8.2	&-15.\\
1.9	&-0.58	&-3.9	&-32.	&-97.	&8.3	&1.6	&-4.	&-0.6	&-0.08	&6.8	&0.49	&-43.	&-1.3	&9.3	&-9.3	&7.8\times 10^{2}	&-10.	&2.2	&1.4\times 10^{2}	&6.8	&7.3	&8.7	&3.5	&-2.5	&0	&2.7	&-4.	&-6.\\
3.9	&-1.1	&3.4	&86.	&-50.	&1.6	&7.4	&3.3	&-0.81	&-3.6	&16.	&-1.9	&92.	&-10.	&4.2	&-4.2	&4.1\times 10^{2}	&-2.4	&3.	&4.1\times 10^{2}	&-2.1	&-3.	&7.5	&0.36	&-17.	&0	&0.25	&3.6	&1.7\\
3.8	&-1.	&13.	&34.	&0.75	&-4.	&3.3	&13.	&2.	&-6.1	&15.	&-4.2	&65.	&-13.	&-15.	&15.	&-1.3	&0.36	&2.2	&4.4\times 10^{2}	&-18.	&-19.	&-6.	&-6.3	&-23.	&0	&-3.5	&13.	&14.\\
0.16	&-0.048	&2.	&-13.	&-1.	&-0.6	&-0.81	&2.	&2.9	&-0.63	&0.32	&-0.55	&-9.1	&-0.54	&-0.03	&0.03	&-1.2	&0.16	&0.11	&-20.	&-3.3	&-3.4	&-2.8	&-1.3	&-1.8	&0	&-0.75	&2.	&2.5\\
7.6	&-2.5	&-6.1	&45.	&13.	&-0.08	&-3.6	&-6.1	&-0.63	&62.	&2.0\times 10^{2}	&-0.31	&32.	&40.	&3.9	&-3.9	&1.1\times 10^{2}	&-42.	&-0.6	&6.3\times 10^{3}	&15.	&15.	&-24.	&9.2	&5.6	&0	&5.7	&-6.4	&-7.\\
-20.	&6.2	&15.	&1.8\times 10^{2}	&-19.	&6.8	&16.	&15.	&0.32	&2.0\times 10^{2}	&6.9\times 10^{2}	&1.2	&1.5\times 10^{2}	&-93.	&-16.	&16.	&2.7\times 10^{2}	&1.4\times 10^{2}	&18.	&2.1\times 10^{4}	&-40.	&-40.	&53.	&-25.	&-18.	&0	&-12.	&15.	&13.\\
-5.6	&1.4	&-4.2	&8.6	&12.	&0.49	&-1.9	&-4.2	&-0.55	&-0.31	&1.2	&2.5	&0.51	&21.	&-1.3	&1.3	&91.	&2.3	&-5.6	&-79.	&2.8	&3.1	&-18.	&-1.1	&17.	&0	&-0.81	&-4.4	&-3.9\\
-64.	&11.	&62.	&8.1\times 10^{3}	&1.2\times 10^{3}	&-43.	&92.	&65.	&-9.1	&32.	&1.5\times 10^{2}	&0.51	&8.4\times 10^{3}	&7.1\times 10^{2}	&2.8\times 10^{2}	&2.8\times 10^{2}	&9.7\times 10^{3}	&7.6\times 10^{3}	&41.	&4.0\times 10^{3}	&1.5\times 10^{2}	&1.8\times 10^{2}	&8.8\times 10^{2}	&1.1\times 10^{2}	&1.3\times 10^{2}	&0	&5.0\times 10^{3}	&60.	&73.\\
-72.	&-53.	&-12.	&7.6\times 10^{2}	&-95.	&-1.3	&-10.	&-13.	&-0.54	&40.	&-93.	&21.	&7.1\times 10^{2}	&1.1\times 10^{5}	&81.	&-81.	&5.8\times 10^{2}	&3.8\times 10^{2}	&-6.3	&4.1\times 10^{3}	&-27.	&-32.	&1.1\times 10^{5}	&-44.	&-32.	&0	&25.	&-13.	&-8.\\
0.43	&-0.17	&3.4	&4.0\times 10^{2}	&7.9\times 10^{2}	&9.3	&4.2	&-15.	&-0.03	&3.9	&-16.	&-1.3	&2.8\times 10^{2}	&81.	&2.0\times 10^{3}	&2.0\times 10^{3}	&4.1\times 10^{3}	&4.0\times 10^{2}	&32.	&4.7\times 10^{2}	&-3.3	&-5.3	&-87.	&-2.	&-1.2	&0	&-3.2	&3.4	&4.5\\
-0.43	&0.17	&-3.4	&3.9\times 10^{2}	&7.9\times 10^{2}	&-9.3	&-4.2	&15.	&0.03	&-3.9	&16.	&1.3	&2.8\times 10^{2}	&-81.	&2.0\times 10^{3}	&2.0\times 10^{3}	&4.1\times 10^{3}	&4.0\times 10^{2}	&-32.	&4.7\times 10^{2}	&3.3	&5.4	&87.	&2.	&1.3	&0	&3.2	&-3.4	&-4.5\\
3.0\times 10^{2}	&91.	&-39.	&1.1\times 10^{4}	&3.4\times 10^{4}	&7.8\times 10^{2}	&4.1\times 10^{2}	&-1.3	&-1.2	&1.1\times 10^{2}	&2.7\times 10^{2}	&91.	&9.7\times 10^{3}	&5.8\times 10^{2}	&4.1\times 10^{3}	&4.1\times 10^{3}	&2.6\times 10^{5}	&3.2\times 10^{3}	&5.2\times 10^{2}	&8.1\times 10^{3}	&2.5\times 10^{2}	&1.7\times 10^{2}	&4.1\times 10^{2}	&2.1\times 10^{2}	&3.0\times 10^{2}	&0	&2.1\times 10^{2}	&-41.	&-73.\\
-9.7	&3.2	&4.1	&4.5\times 10^{2}	&3.0\times 10^{3}	&-10.	&-2.4	&0.36	&0.16	&-42.	&1.4\times 10^{2}	&2.3	&7.6\times 10^{3}	&3.8\times 10^{2}	&4.0\times 10^{2}	&4.0\times 10^{2}	&3.2\times 10^{3}	&8.0\times 10^{5}	&-88.	&4.2\times 10^{3}	&-23.	&-14.	&3.6\times 10^{2}	&-10.	&-8.8	&0	&8.8	&4.2	&4.1\\
-3.9	&1.2	&2.5	&46.	&-70.	&2.2	&3.	&2.2	&0.11	&-0.6	&18.	&-5.6	&41.	&-6.3	&32.	&-32.	&5.2\times 10^{2}	&-88.	&5.5\times 10^{2}	&5.6\times 10^{2}	&-6.5	&-7.1	&-0.75	&-4.5	&-2.9	&0	&-2.2	&2.6	&2.7\\
7.9\times 10^{2}	&2.4\times 10^{2}	&4.4\times 10^{2}	&5.0\times 10^{3}	&6.7\times 10^{2}	&1.4\times 10^{2}	&4.1\times 10^{2}	&4.4\times 10^{2}	&-20.	&6.3\times 10^{3}	&2.1\times 10^{4}	&-79.	&4.0\times 10^{3}	&4.1\times 10^{3}	&4.7\times 10^{2}	&4.7\times 10^{2}	&8.1\times 10^{3}	&4.2\times 10^{3}	&5.6\times 10^{2}	&6.5\times 10^{5}	&1.3\times 10^{3}	&1.3\times 10^{3}	&2.8\times 10^{3}	&8.7\times 10^{2}	&5.9\times 10^{2}	&0	&4.7\times 10^{2}	&4.5\times 10^{2}	&4.5\times 10^{2}\\
5.5	&-2.	&-18.	&-91.	&-32.	&6.8	&-2.1	&-18.	&-3.3	&15.	&-40.	&2.8	&1.5\times 10^{2}	&-27.	&-3.3	&3.3	&2.5\times 10^{2}	&-23.	&-6.5	&1.3\times 10^{3}	&6.8\times 10^{3}	&6.2\times 10^{3}	&6.2\times 10^{3}	&6.2\times 10^{3}	&8.1\times 10^{2}	&0	&2.3\times 10^{3}	&-2.9	&-18.\\
5.8	&-2.	&-19.	&1.2\times 10^{2}	&-22.	&7.3	&-3.	&-19.	&-3.4	&15.	&-40.	&3.1	&1.8\times 10^{2}	&-32.	&-5.3	&5.4	&1.7\times 10^{2}	&-14.	&-7.1	&1.3\times 10^{3}	&6.2\times 10^{3}	&2.8\times 10^{4}	&6.3\times 10^{3}	&6.2\times 10^{3}	&8.1\times 10^{2}	&0	&1.7\times 10^{3}	&-4.2	&-19.\\
78.	&52.	&-6.9	&8.8\times 10^{2}	&73.	&8.7	&7.5	&-6.	&-2.8	&-24.	&53.	&-18.	&8.8\times 10^{2}	&1.1\times 10^{5}	&-87.	&87.	&4.1\times 10^{2}	&3.6\times 10^{2}	&-0.75	&2.8\times 10^{3}	&6.2\times 10^{3}	&6.3\times 10^{3}	&1.3\times 10^{5}	&6.2\times 10^{3}	&8.4\times 10^{2}	&0	&3.3\times 10^{2}	&8.5	&-11.\\
9.6	&-3.	&-6.3	&-81.	&-28.	&3.5	&0.36	&-6.3	&-1.3	&9.2	&-25.	&-1.1	&1.1\times 10^{2}	&-44.	&-2.	&2.	&2.1\times 10^{2}	&-10.	&-4.5	&8.7\times 10^{2}	&6.2\times 10^{3}	&6.2\times 10^{3}	&6.2\times 10^{3}	&9.4\times 10^{3}	&7.8\times 10^{2}	&0	&2.6\times 10^{3}	&8.9	&-5.6\\
-44.	&-8.1	&-23.	&2.0\times 10^{2}	&-53.	&-2.5	&-17.	&-23.	&-1.8	&5.6	&-18.	&17.	&1.3\times 10^{2}	&-32.	&-1.2	&1.3	&3.0\times 10^{2}	&-8.8	&-2.9	&5.9\times 10^{2}	&8.1\times 10^{2}	&8.1\times 10^{2}	&8.4\times 10^{2}	&7.8\times 10^{2}	&1.5\times 10^{5}	&0	&2.5\times 10^{2}	&-21.	&-2.6\\
0	&0	&0	&0	&0	&0	&0	&0	&0	&0	&0	&0	&0	&0	&0	&0	&0	&0	&0	&0	&0	&0	&0	&0	&0	&3.3\times 10^{2}	&3.3\times 10^{2}	&0	&0\\
7.	&-2.4	&-3.6	&5.0\times 10^{3}	&31.	&2.7	&0.25	&-3.5	&-0.75	&5.7	&-12.	&-0.81	&5.0\times 10^{3}	&25.	&-3.2	&3.2	&2.1\times 10^{2}	&8.8	&-2.2	&4.7\times 10^{2}	&2.3\times 10^{3}	&1.7\times 10^{3}	&3.3\times 10^{2}	&2.6\times 10^{3}	&2.5\times 10^{2}	&3.3\times 10^{2}	&8.0\times 10^{3}	&-3.5	&-6.3\\
3.9	&-1.	&13.	&35.	&-8.2	&-4.	&3.6	&13.	&2.	&-6.4	&15.	&-4.4	&60.	&-13.	&3.4	&-3.4	&-41.	&4.2	&2.6	&4.5\times 10^{2}	&-2.9	&-4.2	&8.5	&8.9	&-21.	&0	&-3.5	&37.	&15.\\
1.	&-0.066	&14.	&47.	&-15.	&-6.	&1.7	&14.	&2.5	&-7.	&13.	&-3.9	&73.	&-8.	&4.5	&-4.5	&-73.	&4.1	&2.7	&4.5\times 10^{2}	&-18.	&-19.	&-11.	&-5.6	&-2.6	&0	&-6.3	&15.	&38.\\
     \end{array}\right)$}
 \end{equation}
  
 \clearpage
 
 \item Covariance matrix without top operators at ILC 500
    
 \begin{equation}\label{covwot500}
    \resizebox{0.85\hsize}{!}{$10^{-7} \cdot
    \left(
\begin{array}{cccccccccccccccccccccc}
 1.2	&-0.3	&0.15	&-4.9	&0.13	&0.064	&0.19	&0.15	&-0.004	&0.0039	&0.33	&-0.33	&-4.6	&0.5	&0.88	&0.61	&1.6	&-2.2	&0	&2.3	&0.14	&-0.038\\
-0.3	&0.16	&-0.04	&1.	&-0.075	&-0.028	&-0.05	&-0.04	&0.00044	&-0.017	&-0.12	&0.084	&0.93	&-0.18	&-0.25	&-0.2	&-0.4	&-1.	&0	&-0.51	&-0.04	&0.032\\
0.15	&-0.04	&0.14	&-1.4	&-0.042	&-0.0099	&-0.051	&0.14	&0.038	&-0.026	&0.019	&-0.072	&-1.1	&-0.12	&-0.1	&-0.09	&-0.062	&-0.35	&0	&-0.19	&0.14	&0.15\\
-4.9	&1.	&-1.4	&4.3\times 10^{2}	&0.82	&-0.97	&1.	&-1.4	&-0.54	&0.32	&-0.95	&1.6	&4.3\times 10^{2}	&-0.66	&-5.2	&-0.43	&-22.	&9.	&0	&3.8\times 10^{2}	&-1.7	&-1.3\\
0.13	&-0.075	&-0.042	&0.82	&0.4	&0.067	&0.078	&-0.042	&-0.02	&0.06	&0.12	&0.0015	&0.73	&-0.22	&-0.22	&-0.23	&-0.23	&-0.51	&0	&0.3	&-0.19	&-0.24\\
0.064	&-0.028	&-0.0099	&-0.97	&0.067	&0.05	&0.027	&-0.0099	&-0.0044	&0.015	&0.043	&-0.013	&-1.	&0.031	&0.086	&0.019	&0.31	&0.065	&0	&0.59	&-0.016	&-0.048\\
0.19	&-0.05	&-0.051	&1.	&0.078	&0.027	&0.12	&-0.051	&-0.034	&0.023	&0.081	&-0.035	&0.91	&0.23	&0.25	&0.25	&0.12	&-0.28	&0	&0.049	&-0.044	&-0.11\\
0.15	&-0.04	&0.14	&-1.4	&-0.042	&-0.0099	&-0.051	&0.14	&0.038	&-0.026	&0.019	&-0.072	&-1.1	&-0.12	&-0.1	&-0.09	&-0.062	&-0.35	&0	&-0.19	&0.14	&0.15\\
-0.004	&0.00044	&0.038	&-0.54	&-0.02	&-0.0044	&-0.034	&0.038	&0.067	&-0.0092	&-0.01	&-0.0083	&-0.46	&-0.07	&-0.065	&-0.069	&-0.02	&-0.01	&0	&-0.013	&0.037	&0.049\\
0.0039	&-0.017	&-0.026	&0.32	&0.06	&0.015	&0.023	&-0.026	&-0.0092	&0.015	&0.021	&0.0079	&0.26	&0.0069	&0.0032	&0.0015	&-0.0073	&-0.044	&0	&0.07	&-0.04	&-0.057\\
0.33	&-0.12	&0.019	&-0.95	&0.12	&0.043	&0.081	&0.019	&-0.01	&0.021	&0.13	&-0.084	&-0.91	&0.19	&0.28	&0.21	&0.46	&0.81	&0	&0.66	&0.004	&-0.071\\
-0.33	&0.084	&-0.072	&1.6	&0.0015	&-0.013	&-0.035	&-0.072	&-0.0083	&0.0079	&-0.084	&0.1	&1.5	&-0.13	&-0.23	&-0.17	&-0.44	&0.6	&0	&-0.52	&-0.085	&-0.04\\
-4.6	&0.93	&-1.1	&4.3\times 10^{2}	&0.73	&-1.	&0.91	&-1.1	&-0.46	&0.26	&-0.91	&1.5	&4.3\times 10^{2}	&-2.9	&-7.4	&-2.5	&-24.	&6.3	&0	&3.8\times 10^{2}	&-1.4	&-1.\\
0.5	&-0.18	&-0.12	&-0.66	&-0.22	&0.031	&0.23	&-0.12	&-0.07	&0.0069	&0.19	&-0.13	&-2.9	&1.9\times 10^{2}	&1.8\times 10^{2}	&1.8\times 10^{2}	&1.8\times 10^{2}	&51.	&0	&64.	&1.4	&0.048\\
0.88	&-0.25	&-0.1	&-5.2	&-0.22	&0.086	&0.25	&-0.1	&-0.065	&0.0032	&0.28	&-0.23	&-7.4	&1.8\times 10^{2}	&1.1\times 10^{3}	&1.8\times 10^{2}	&1.8\times 10^{2}	&51.	&0	&33.	&1.4	&0.03\\
0.61	&-0.2	&-0.09	&-0.43	&-0.23	&0.019	&0.25	&-0.09	&-0.069	&0.0015	&0.21	&-0.17	&-2.5	&1.8\times 10^{2}	&1.8\times 10^{2}	&6.9\times 10^{2}	&1.8\times 10^{2}	&51.	&0	&-5.6	&1.4	&0.073\\
1.6	&-0.4	&-0.062	&-22.	&-0.23	&0.31	&0.12	&-0.062	&-0.02	&-0.0073	&0.46	&-0.44	&-24.	&1.8\times 10^{2}	&1.8\times 10^{2}	&1.8\times 10^{2}	&3.8\times 10^{2}	&49.	&0	&69.	&1.5	&-0.0014\\
-2.2	&-1.	&-0.35	&9.	&-0.51	&0.065	&-0.28	&-0.35	&-0.01	&-0.044	&0.81	&0.6	&6.3	&51.	&51.	&51.	&49.	&1.4\times 10^{4}	&0	&10.	&0.19	&1.1\\
0	&0	&0	&0	&0	&0	&0	&0	&0	&0	&0	&0	&0	&0	&0	&0	&0	&0	&28.	&-28.	&0	&0\\
2.3	&-0.51	&-0.19	&3.8\times 10^{2}	&0.3	&0.59	&0.049	&-0.19	&-0.013	&0.07	&0.66	&-0.52	&3.8\times 10^{2}	&64.	&33.	&-5.6	&69.	&10.	&-28.	&4.7\times 10^{2}	&-0.14	&-0.59\\
0.14	&-0.04	&0.14	&-1.7	&-0.19	&-0.016	&-0.044	&0.14	&0.037	&-0.04	&0.004	&-0.085	&-1.4	&1.4	&1.4	&1.4	&1.5	&0.19	&0	&-0.14	&2.5	&0.22\\
-0.038	&0.032	&0.15	&-1.3	&-0.24	&-0.048	&-0.11	&0.15	&0.049	&-0.057	&-0.071	&-0.04	&-1.	&0.048	&0.03	&0.073	&-0.0014	&1.1	&0	&-0.59	&0.22	&2.4\\
\end{array}
\right)$}
 \end{equation}

\item Covariance matrix for ILC 250+500 (indirect only)
\begin{equation}\label{covInd}
    \resizebox{0.85\hsize}{!}{$10^{-5} \cdot
    \left(
\begin{array}{ccccccccccccccccccccccccccccc}
 0.89	&0.54	&0.048	&-9.1	&0.66	&-0.058	&-0.028	&-0.058	&-0.0012	&2.	&-6.6	&-0.46	&5.9	&1.2\times 10^{3}	&-14.	&-11.	&39.	&1.9\times 10^{3}	&21.	&1.9\times 10^{2}	&0.55	&0.57	&1.2\times 10^{3}	&0.62	&0.42	&0	&-0.11	&0.047	&-0.0069\\
0.54	&0.35	&0.032	&-5.7	&0.74	&-0.038	&-0.015	&-0.032	&-0.00073	&1.3	&-4.1	&-0.29	&3.5	&7.4\times 10^{2}	&-9.4	&-6.9	&26.	&1.2\times 10^{3}	&14.	&1.2\times 10^{2}	&0.33	&0.34	&7.4\times 10^{2}	&0.37	&0.27	&0	&-0.081	&0.032	&-0.00043\\
0.048	&0.032	&0.026	&-0.66	&0.64	&0.000078	&0.0056	&-0.013	&0.0003	&0.2	&-0.63	&-0.051	&0.062	&-67.	&-7.1	&-4.1	&2.9	&2.1\times 10^{2}	&3.	&18.	&-0.0095	&-0.0088	&67.	&0.017	&-0.000048	&0	&-0.0086	&0.026	&0.026\\
-9.1	&-5.7	&-0.66	&1.1\times 10^{2}	&-40.	&0.72	&0.18	&0.34	&0.001	&-23.	&74.	&5.3	&-57.	&1.2\times 10^{4}	&3.0\times 10^{2}	&1.1\times 10^{2}	&5.2\times 10^{2}	&2.0\times 10^{4}	&2.7\times 10^{2}	&2.1\times 10^{3}	&-5.3	&-5.4	&1.2\times 10^{4}	&-6.1	&-4.1	&0	&-2.4	&-0.67	&-0.086\\
0.66	&0.74	&0.64	&-40.	&1.1\times 10^{2}	&-0.32	&0.41	&0.62	&0.018	&12.	&-32.	&-2.8	&-3.8	&1.3\times 10^{3}	&5.6\times 10^{2}	&-2.	&3.8\times 10^{2}	&3.6\times 10^{3}	&2.3\times 10^{2}	&9.8\times 10^{2}	&-0.63	&-0.46	&1.3\times 10^{3}	&0.51	&-0.36	&0	&2.2	&0.66	&0.55\\
-0.058	&-0.038	&0.000078	&0.72	&-0.32	&0.012	&0.0045	&-0.0026	&0.00002	&-0.11	&0.37	&0.024	&-0.49	&79.	&-0.41	&-0.29	&-5.7	&1.1\times 10^{2}	&-0.92	&-10.	&-0.041	&-0.042	&-79.	&-0.042	&-0.022	&0	&0.0044	&0.000051	&0.0063\\
-0.028	&-0.015	&0.0056	&0.18	&0.41	&0.0045	&0.0071	&0.0015	&-0.00024	&-0.016	&0.079	&-0.0011	&-0.28	&33.	&-3.2	&-0.43	&-0.98	&33.	&0.61	&-2.1	&-0.029	&-0.029	&-33.	&-0.024	&-0.017	&0	&0.0014	&0.0058	&0.0083\\
-0.058	&-0.032	&-0.013	&0.34	&0.62	&-0.0026	&0.0015	&0.025	&0.00066	&-0.14	&0.47	&0.027	&-0.26	&71.	&0.82	&4.	&0.66	&2.0\times 10^{2}	&-0.83	&-14.	&-0.02	&-0.021	&-71.	&-0.033	&-0.021	&0	&0.0079	&-0.012	&-0.012\\
-0.0012	&-0.00073	&0.0003	&0.001	&0.018	&0.00002	&-0.00024	&0.00066	&0.00068	&-0.001	&0.004	&0.00011	&-0.012	&1.6	&-0.072	&0.039	&-0.0029	&2.7	&0.012	&-0.11	&-0.0014	&-0.0014	&-1.6	&-0.0011	&-0.00075	&0	&0.000072	&0.0003	&0.00047\\
2.	&1.3	&0.2	&-23.	&12.	&-0.11	&-0.016	&-0.14	&-0.001	&6.1	&-19.	&-1.4	&12.	&2.7\times 10^{3}	&-99.	&-37.	&1.1\times 10^{2}	&4.8\times 10^{3}	&76.	&5.6\times 10^{2}	&1.1	&1.1	&2.7\times 10^{3}	&1.3	&0.92	&0	&-0.039	&0.21	&0.094\\
-6.6	&-4.1	&-0.63	&74.	&-32.	&0.37	&0.079	&0.47	&0.004	&-19.	&60.	&4.3	&-39.	&8.7\times 10^{3}	&2.9\times 10^{2}	&1.2\times 10^{2}	&3.4\times 10^{2}	&1.5\times 10^{4}	&2.3\times 10^{2}	&1.8\times 10^{3}	&-3.6	&-3.7	&8.7\times 10^{3}	&-4.4	&-3.	&0	&0.2	&-0.63	&-0.27\\
-0.46	&-0.29	&-0.051	&5.3	&-2.8	&0.024	&-0.0011	&0.027	&0.00011	&-1.4	&4.3	&0.32	&-2.6	&6.2\times 10^{2}	&23.	&8.3	&-25.	&1.1\times 10^{3}	&-18.	&1.2\times 10^{2}	&-0.24	&-0.25	&6.2\times 10^{2}	&-0.3	&-0.2	&0	&0.026	&-0.052	&-0.027\\
5.9	&3.5	&0.062	&-57.	&-3.8	&-0.49	&-0.28	&-0.26	&-0.012	&12.	&-39.	&-2.6	&54.	&7.6\times 10^{3}	&-14.	&-34.	&2.7\times 10^{2}	&1.2\times 10^{4}	&1.1\times 10^{2}	&1.1\times 10^{3}	&4.	&3.8	&7.6\times 10^{3}	&3.3	&2.8	&0	&-6.2	&0.056	&-0.34\\
1.2\times 10^{3}	&7.4\times 10^{2}	&-67.	&1.2\times 10^{4}	&1.3\times 10^{3}	&79.	&33.	&71.	&1.6	&2.7\times 10^{3}	&8.7\times 10^{3}	&6.2\times 10^{2}	&7.6\times 10^{3}	&1.6\times 10^{6}	&2.0\times 10^{4}	&1.5\times 10^{4}	&5.4\times 10^{4}	&2.5\times 10^{6}	&2.9\times 10^{4}	&2.5\times 10^{5}	&7.1\times 10^{2}	&7.4\times 10^{2}	&1.6\times 10^{6}	&8.0\times 10^{2}	&6.0\times 10^{2}	&0	&1.6\times 10^{2}	&-67.	&4.\\
-14.	&-9.4	&-7.1	&3.0\times 10^{2}	&5.6\times 10^{2}	&-0.41	&-3.2	&0.82	&-0.072	&-99.	&2.9\times 10^{2}	&23.	&-14.	&2.0\times 10^{4}	&4.3\times 10^{3}	&8.5\times 10^{2}	&1.4\times 10^{3}	&6.0\times 10^{4}	&1.7\times 10^{3}	&8.7\times 10^{3}	&1.9	&1.6	&2.0\times 10^{4}	&-6.3	&-3.4	&0	&-5.2	&-7.3	&-7.4\\
-11.	&-6.9	&-4.1	&1.1\times 10^{2}	&-2.	&-0.29	&-0.43	&4.	&0.039	&-37.	&1.2\times 10^{2}	&8.3	&-34.	&1.5\times 10^{4}	&8.5\times 10^{2}	&8.7\times 10^{2}	&2.4\times 10^{2}	&4.3\times 10^{4}	&4.1\times 10^{2}	&3.4\times 10^{3}	&-1.2	&-1.3	&1.5\times 10^{4}	&-5.4	&-2.2	&0	&1.8	&-4.1	&-4.\\
39.	&26.	&2.9	&5.2\times 10^{2}	&3.8\times 10^{2}	&-5.7	&-0.98	&0.66	&-0.0029	&1.1\times 10^{2}	&3.4\times 10^{2}	&-25.	&2.7\times 10^{2}	&5.4\times 10^{4}	&1.4\times 10^{3}	&2.4\times 10^{2}	&3.6\times 10^{3}	&8.8\times 10^{4}	&1.4\times 10^{3}	&9.9\times 10^{3}	&22.	&23.	&5.4\times 10^{4}	&26.	&14.	&0	&0.27	&2.9	&-0.62\\
1.9\times 10^{3}	&1.2\times 10^{3}	&2.1\times 10^{2}	&2.0\times 10^{4}	&3.6\times 10^{3}	&1.1\times 10^{2}	&33.	&2.0\times 10^{2}	&2.7	&4.8\times 10^{3}	&1.5\times 10^{4}	&1.1\times 10^{3}	&1.2\times 10^{4}	&2.5\times 10^{6}	&6.0\times 10^{4}	&4.3\times 10^{4}	&8.8\times 10^{4}	&4.8\times 10^{6}	&5.2\times 10^{4}	&4.4\times 10^{5}	&9.9\times 10^{2}	&9.9\times 10^{2}	&2.5\times 10^{6}	&1.1\times 10^{3}	&7.9\times 10^{2}	&0	&4.4\times 10^{2}	&2.1\times 10^{2}	&1.1\times 10^{2}\\
21.	&14.	&3.	&2.7\times 10^{2}	&2.3\times 10^{2}	&-0.92	&0.61	&-0.83	&0.012	&76.	&2.3\times 10^{2}	&-18.	&1.1\times 10^{2}	&2.9\times 10^{4}	&1.7\times 10^{3}	&4.1\times 10^{2}	&1.4\times 10^{3}	&5.2\times 10^{4}	&1.1\times 10^{3}	&6.7\times 10^{3}	&9.6	&10.	&2.9\times 10^{4}	&13.	&9.2	&0	&0.26	&3.1	&2.\\
1.9\times 10^{2}	&1.2\times 10^{2}	&18.	&2.1\times 10^{3}	&9.8\times 10^{2}	&-10.	&-2.1	&-14.	&-0.11	&5.6\times 10^{2}	&1.8\times 10^{3}	&1.2\times 10^{2}	&1.1\times 10^{3}	&2.5\times 10^{5}	&8.7\times 10^{3}	&3.4\times 10^{3}	&9.9\times 10^{3}	&4.4\times 10^{5}	&6.7\times 10^{3}	&5.1\times 10^{4}	&1.0\times 10^{2}	&1.1\times 10^{2}	&2.5\times 10^{5}	&1.3\times 10^{2}	&85.	&0	&-4.2	&19.	&8.3\\
0.55	&0.33	&-0.0095	&-5.3	&-0.63	&-0.041	&-0.029	&-0.02	&-0.0014	&1.1	&-3.6	&-0.24	&4.	&7.1\times 10^{2}	&1.9	&-1.2	&22.	&9.9\times 10^{2}	&9.6	&1.0\times 10^{2}	&2.3	&2.2	&7.2\times 10^{2}	&2.2	&0.82	&0	&0.56	&0.0052	&-0.045\\
0.57	&0.34	&-0.0088	&-5.4	&-0.46	&-0.042	&-0.029	&-0.021	&-0.0014	&1.1	&-3.7	&-0.25	&3.8	&7.4\times 10^{2}	&1.6	&-1.3	&23.	&9.9\times 10^{2}	&10.	&1.1\times 10^{2}	&2.2	&11.	&7.4\times 10^{2}	&2.2	&0.84	&0	&0.31	&0.0059	&-0.045\\
1.2\times 10^{3}	&7.4\times 10^{2}	&67.	&1.2\times 10^{4}	&1.3\times 10^{3}	&-79.	&-33.	&-71.	&-1.6	&2.7\times 10^{3}	&8.7\times 10^{3}	&6.2\times 10^{2}	&7.6\times 10^{3}	&1.6\times 10^{6}	&2.0\times 10^{4}	&1.5\times 10^{4}	&5.4\times 10^{4}	&2.5\times 10^{6}	&2.9\times 10^{4}	&2.5\times 10^{5}	&7.2\times 10^{2}	&7.4\times 10^{2}	&1.6\times 10^{6}	&8.1\times 10^{2}	&6.0\times 10^{2}	&0	&1.6\times 10^{2}	&67.	&-4.\\
0.62	&0.37	&0.017	&-6.1	&0.51	&-0.042	&-0.024	&-0.033	&-0.0011	&1.3	&-4.4	&-0.3	&3.3	&8.0\times 10^{2}	&-6.3	&-5.4	&26.	&1.1\times 10^{3}	&13.	&1.3\times 10^{2}	&2.2	&2.2	&8.1\times 10^{2}	&4.2	&0.84	&0	&0.77	&0.032	&-0.019\\
0.42	&0.27	&-0.000048	&-4.1	&-0.36	&-0.022	&-0.017	&-0.021	&-0.00075	&0.92	&-3.	&-0.2	&2.8	&6.0\times 10^{2}	&-3.4	&-2.2	&14.	&7.9\times 10^{2}	&9.2	&85.	&0.82	&0.84	&6.0\times 10^{2}	&0.84	&1.4\times 10^{2}	&0	&0.092	&0.0047	&-0.0098\\
0	&0	&0	&0	&0	&0	&0	&0	&0	&0	&0	&0	&0	&0	&0	&0	&0	&0	&0	&0	&0	&0	&0	&0	&0	&0.28	&-0.28	&0	&0\\
-0.11	&-0.081	&-0.0086	&-2.4	&2.2	&0.0044	&0.0014	&0.0079	&0.000072	&-0.039	&0.2	&0.026	&-6.2	&1.6\times 10^{2}	&-5.2	&1.8	&0.27	&4.4\times 10^{2}	&0.26	&-4.2	&0.56	&0.31	&1.6\times 10^{2}	&0.77	&0.092	&-0.28	&5.2	&-0.0084	&-0.005\\
0.047	&0.032	&0.026	&-0.67	&0.66	&0.000051	&0.0058	&-0.012	&0.0003	&0.21	&-0.63	&-0.052	&0.056	&-67.	&-7.3	&-4.1	&2.9	&2.1\times 10^{2}	&3.1	&19.	&0.0052	&0.0059	&67.	&0.032	&0.0047	&0	&-0.0084	&0.05	&0.027\\
-0.0069	&-0.00043	&0.026	&-0.086	&0.55	&0.0063	&0.0083	&-0.012	&0.00047	&0.094	&-0.27	&-0.027	&-0.34	&4.	&-7.4	&-4.	&-0.62	&1.1\times 10^{2}	&2.	&8.3	&-0.045	&-0.045	&-4.	&-0.019	&-0.0098	&0	&-0.005	&0.027	&0.053\\
\end{array}
\right)$}
 \end{equation}
  
 \clearpage
 
 \item Covariance matrix for ILC 250+500 + HL-LHC S2
 \begin{equation}\label{covHL250}
     \resizebox{0.85\hsize}{!}{$10^{-7} \cdot
     \left(
\begin{array}{ccccccccccccccccccccccccccccc}
 1.4	&-0.35	&0.054	&-4.5	&-2.5	&0.099	&0.11	&0.059	&-0.0039	&0.89	&-2.7	&-0.35	&-4.1	&-5.1	&-0.47	&0.47	&-20.	&-10.	&-0.46	&92.	&0.82	&1.3	&6.1	&2.	&-2.	&0	&2.9	&0.047	&-0.12\\
-0.35	&0.18	&-0.012	&0.89	&0.59	&-0.037	&-0.027	&-0.013	&0.00038	&-0.29	&0.78	&0.09	&0.8	&-5.7	&0.14	&-0.14	&5.1	&2.2	&0.16	&-28.	&-0.27	&-0.36	&5.5	&-0.5	&-1.1	&0	&-0.71	&-0.012	&0.058\\
0.054	&-0.012	&0.19	&-1.7	&1.8	&-0.038	&-0.02	&0.18	&0.038	&-0.41	&1.3	&-0.062	&-1.5	&-0.64	&0.48	&-0.48	&14.	&0.81	&0.12	&-40.	&-0.27	&-0.27	&0.4	&-0.19	&-0.44	&0	&-0.41	&0.19	&0.19\\
-4.5	&0.89	&-1.7	&4.6\times 10^{2}	&-89.	&0.49	&1.2	&-2.	&-0.54	&-0.2	&-0.37	&1.6	&4.4\times 10^{2}	&38.	&38.	&-38.	&6.6\times 10^{2}	&2.2\times 10^{2}	&2.6	&-31.	&0.34	&-4.9	&-38.	&-23.	&11.	&0	&3.8\times 10^{2}	&-2.1	&-1.2\\
-2.5	&0.59	&1.8	&-89.	&3.7\times 10^{2}	&-6.	&-0.56	&2.5	&-0.023	&0.76	&2.7	&0.23	&-63.	&-0.89	&-75.	&75.	&2.8\times 10^{3}	&2.2\times 10^{2}	&-3.9	&-25.	&-5.8	&-2.5	&-4.3	&6.2	&-10.	&0	&27.	&1.7	&-0.28\\
0.099	&-0.037	&-0.038	&0.49	&-6.	&0.15	&0.042	&-0.046	&-0.0043	&-0.04	&0.15	&-0.017	&0.13	&0.51	&0.86	&-0.86	&-46.	&-0.8	&0.24	&-4.2	&0.11	&0.11	&-0.43	&0.18	&0.22	&0	&0.086	&-0.044	&-0.044\\
0.11	&-0.027	&-0.02	&1.2	&-0.56	&0.042	&0.16	&-0.02	&-0.034	&-0.38	&1.4	&-0.026	&1.1	&-0.72	&0.0005	&-0.0012	&-5.1	&1.1	&0.17	&-42.	&0.13	&0.095	&0.86	&-0.064	&-0.3	&0	&-0.4	&-0.014	&-0.069\\
0.059	&-0.013	&0.18	&-2.	&2.5	&-0.046	&-0.02	&0.2	&0.038	&-0.41	&1.3	&-0.062	&-1.2	&-0.72	&-1.4	&1.4	&18.	&0.51	&0.09	&-40.	&-0.26	&-0.26	&0.48	&-0.2	&-0.44	&0	&-0.42	&0.18	&0.18\\
-0.0039	&0.00038	&0.038	&-0.54	&-0.023	&-0.0043	&-0.034	&0.038	&0.067	&-0.0089	&-0.011	&-0.0084	&-0.45	&0.038	&-0.017	&0.017	&-0.039	&-0.11	&0.0039	&0.023	&-0.07	&-0.065	&-0.11	&-0.02	&-0.01	&0	&-0.014	&0.037	&0.049\\
0.89	&-0.29	&-0.41	&-0.2	&0.76	&-0.04	&-0.38	&-0.41	&-0.0089	&4.8	&-16.	&-0.12	&-0.94	&2.6	&0.15	&-0.15	&7.9	&-12.	&0.53	&4.9\times 10^{2}	&1.3	&1.8	&-1.1	&2.	&0.37	&0	&4.7	&-0.43	&-0.48\\
-2.7	&0.78	&1.3	&-0.37	&2.7	&0.15	&1.4	&1.3	&-0.011	&-16.	&52.	&0.34	&2.3	&-3.8	&-1.2	&1.2	&10.	&38.	&-0.16	&1.6\times 10^{3}	&-4.4	&-5.7	&-0.99	&-6.2	&-0.73	&0	&-14.	&1.3	&1.3\\
-0.35	&0.09	&-0.062	&1.6	&0.23	&-0.017	&-0.026	&-0.062	&-0.0084	&-0.12	&0.34	&0.11	&1.4	&1.4	&-0.03	&0.031	&1.6	&2.4	&-0.5	&-13.	&-0.17	&-0.27	&-1.6	&-0.47	&0.59	&0	&-0.6	&-0.074	&-0.03\\
-4.1	&0.8	&-1.5	&4.4\times 10^{2}	&-63.	&0.13	&1.1	&-1.2	&-0.45	&-0.94	&2.3	&1.4	&4.5\times 10^{2}	&34.	&-29.	&29.	&5.0\times 10^{2}	&5.7\times 10^{2}	&1.8	&1.1\times 10^{2}	&-1.8	&-7.3	&-35.	&-26.	&7.9	&0	&3.9\times 10^{2}	&-1.8	&-1.1\\
-5.1	&-5.7	&-0.64	&38.	&-0.89	&0.51	&-0.72	&-0.72	&0.038	&2.6	&-3.8	&1.4	&34.	&1.1\times 10^{4}	&8.2	&-8.2	&3.4	&29.	&-1.8	&2.6\times 10^{2}	&-2.6	&-5.4	&1.1\times 10^{4}	&-12.	&-11.	&0	&-11.	&-0.56	&-0.6\\
-0.47	&0.14	&0.48	&38.	&-75.	&0.86	&0.0005	&-1.4	&-0.017	&0.15	&-1.2	&-0.03	&-29.	&8.2	&2.0\times 10^{2}	&2.0\times 10^{2}	&3.7\times 10^{2}	&32.	&2.8	&28.	&-0.99	&-0.62	&-9.	&0.13	&0.19	&0	&1.6	&0.47	&0.73\\
0.47	&-0.14	&-0.48	&-38.	&75.	&-0.86	&-0.0012	&1.4	&0.017	&-0.15	&1.2	&0.031	&29.	&-8.2	&2.0\times 10^{2}	&2.0\times 10^{2}	&3.7\times 10^{2}	&-32.	&-2.8	&-28.	&0.99	&0.62	&9.	&-0.14	&-0.19	&0	&-1.6	&-0.47	&-0.74\\
-20.	&5.1	&14.	&6.6\times 10^{2}	&2.8\times 10^{3}	&-46.	&-5.1	&18.	&-0.039	&7.9	&10.	&1.6	&5.0\times 10^{2}	&3.4	&3.7\times 10^{2}	&3.7\times 10^{2}	&2.1\times 10^{4}	&3.2\times 10^{2}	&-29.	&83.	&-43.	&-18.	&-41.	&49.	&-75.	&0	&2.1\times 10^{2}	&14.	&0.19\\
-10.	&2.2	&0.81	&2.2\times 10^{2}	&2.2\times 10^{2}	&-0.8	&1.1	&0.51	&-0.11	&-12.	&38.	&2.4	&5.7\times 10^{2}	&29.	&32.	&-32.	&3.2\times 10^{2}	&8.0\times 10^{4}	&0.18	&1.2\times 10^{3}	&-7.4	&12.	&-32.	&47.	&26.	&0	&1.2\times 10^{2}	&0.55	&2.\\
-0.46	&0.16	&0.12	&2.6	&-3.9	&0.24	&0.17	&0.09	&0.0039	&0.53	&-0.16	&-0.5	&1.8	&-1.8	&2.8	&-2.8	&-29.	&0.18	&54.	&2.9	&-0.51	&-0.9	&1.3	&-1.7	&-0.62	&0	&-3.1	&0.13	&0.14\\
92.	&-28.	&-40.	&-31.	&-25.	&-4.2	&-42.	&-40.	&0.023	&4.9\times 10^{2}	&1.6\times 10^{3}	&-13.	&1.1\times 10^{2}	&2.6\times 10^{2}	&28.	&-28.	&83.	&1.2\times 10^{3}	&2.9	&5.0\times 10^{4}	&1.4\times 10^{2}	&1.8\times 10^{2}	&1.1\times 10^{2}	&2.1\times 10^{2}	&46.	&0	&4.6\times 10^{2}	&-40.	&-43.\\
0.82	&-0.27	&-0.27	&0.34	&-5.8	&0.11	&0.13	&-0.26	&-0.07	&1.3	&-4.4	&-0.17	&-1.8	&-2.6	&-0.99	&0.99	&-43.	&-7.4	&-0.51	&1.4\times 10^{2}	&1.9\times 10^{2}	&1.8\times 10^{2}	&1.8\times 10^{2}	&1.8\times 10^{2}	&52.	&0	&64.	&1.2	&-0.082\\
1.3	&-0.36	&-0.27	&-4.9	&-2.5	&0.11	&0.095	&-0.26	&-0.065	&1.8	&-5.7	&-0.27	&-7.3	&-5.4	&-0.62	&0.62	&-18.	&12.	&-0.9	&1.8\times 10^{2}	&1.8\times 10^{2}	&1.1\times 10^{3}	&1.8\times 10^{2}	&1.8\times 10^{2}	&51.	&0	&34.	&1.3	&-0.14\\
6.1	&5.5	&0.4	&-38.	&-4.3	&-0.43	&0.86	&0.48	&-0.11	&-1.1	&-0.99	&-1.6	&-35.	&1.1\times 10^{4}	&-9.	&9.	&-41.	&-32.	&1.3	&1.1\times 10^{2}	&1.8\times 10^{2}	&1.8\times 10^{2}	&1.1\times 10^{4}	&1.9\times 10^{2}	&62.	&0	&6.5	&1.8	&0.53\\
2.	&-0.5	&-0.19	&-23.	&6.2	&0.18	&-0.064	&-0.2	&-0.02	&2.	&-6.2	&-0.47	&-26.	&-12.	&0.13	&-0.14	&49.	&47.	&-1.7	&2.1\times 10^{2}	&1.8\times 10^{2}	&1.8\times 10^{2}	&1.9\times 10^{2}	&3.8\times 10^{2}	&49.	&0	&71.	&1.4	&-0.18\\
-2.	&-1.1	&-0.44	&11.	&-10.	&0.22	&-0.3	&-0.44	&-0.01	&0.37	&-0.73	&0.59	&7.9	&-11.	&0.19	&-0.19	&-75.	&26.	&-0.62	&46.	&52.	&51.	&62.	&49.	&1.4\times 10^{4}	&0	&9.8	&0.094	&1.\\
0	&0	&0	&0	&0	&0	&0	&0	&0	&0	&0	&0	&0	&0	&0	&0	&0	&0	&0	&0	&0	&0	&0	&0	&0	&28.	&-28.	&0	&0\\
2.9	&-0.71	&-0.41	&3.8\times 10^{2}	&27.	&0.086	&-0.4	&-0.42	&-0.014	&4.7	&-14.	&-0.6	&3.9\times 10^{2}	&-11.	&1.6	&-1.6	&2.1\times 10^{2}	&1.2\times 10^{2}	&-3.1	&4.6\times 10^{2}	&64.	&34.	&6.5	&71.	&9.8	&-28.	&4.8\times 10^{2}	&-0.37	&-0.98\\
0.047	&-0.012	&0.19	&-2.1	&1.7	&-0.044	&-0.014	&0.18	&0.037	&-0.43	&1.3	&-0.074	&-1.8	&-0.56	&0.47	&-0.47	&14.	&0.55	&0.13	&-40.	&1.2	&1.3	&1.8	&1.4	&0.094	&0	&-0.37	&2.6	&0.26\\
-0.12	&0.058	&0.19	&-1.2	&-0.28	&-0.044	&-0.069	&0.18	&0.049	&-0.48	&1.3	&-0.03	&-1.1	&-0.6	&0.73	&-0.74	&0.19	&2.	&0.14	&-43.	&-0.082	&-0.14	&0.53	&-0.18	&1.	&0	&-0.98	&0.26	&2.4\\
\end{array}
\right)$}
 \end{equation}
 
\item Covariance matrix for ILC 250+500 + HL-LHC S2 + ILC 500-top
\begin{equation}\label{covFull}
    \resizebox{0.83\hsize}{!}{$10^{-8} \cdot
    \left(
\begin{array}{ccccccccccccccccccccccccccccc}
 17.	&-4.4	&4.3	&-91.	&2.5	&1.1	&3.9	&4.3	&0.18	&-0.069	&5.1	&-5.4	&-82.	&-78.	&-0.093	&0.093	&-0.18	&-0.11	&-0.52	&0.26	&3.6	&3.9	&82.	&8.2	&-45.	&0	&6.7	&4.4	&1.5\\
-4.4	&2.	&-1.2	&19.	&-1.1	&-0.35	&-1.1	&-1.2	&-0.055	&-0.14	&-1.7	&1.4	&16.	&-52.	&0.028	&-0.029	&0.056	&0.059	&0.16	&-0.081	&-1.3	&-1.4	&50.	&-2.6	&-7.8	&0	&-2.3	&-1.2	&-0.22\\
4.3	&-1.2	&13.	&39.	&-1.3	&-4.1	&3.	&13.	&2.	&-1.8	&0.55	&-4.3	&64.	&-9.8	&0.047	&-0.049	&0.094	&0.47	&0.26	&-0.13	&-17.	&-18.	&-8.6	&-5.8	&-23.	&0	&-3.2	&13.	&13.\\
-91.	&19.	&39.	&7.6\times 10^{3}	&9.5	&-64.	&72.	&39.	&-13.	&-2.8	&-18.	&13.	&7.7\times 10^{3}	&6.9\times 10^{2}	&0.58	&-0.57	&0.97	&2.1\times 10^{2}	&3.	&-1.5	&1.1\times 10^{2}	&1.4\times 10^{2}	&8.3\times 10^{2}	&-96.	&1.8\times 10^{2}	&0	&5.0\times 10^{3}	&36.	&48.\\
2.5	&-1.1	&-1.3	&9.5	&12.	&4.5	&3.7	&-1.3	&-0.88	&1.6	&3.3	&-0.063	&-2.5	&-6.6	&0.14	&-0.24	&0.74	&9.5\times 10^{2}	&1.5	&-0.75	&-0.34	&-0.26	&6.4	&-1.6	&-15.	&0	&3.4	&-2.8	&-5.2\\
1.1	&-0.35	&-4.1	&-64.	&4.5	&5.9	&0.32	&-4.1	&-0.61	&1.4	&2.1	&0.75	&-72.	&-2.2	&-0.0017	&0.00059	&-0.0066	&-0.28	&-0.0098	&0.0048	&6.3	&7.1	&9.3	&3.	&-3.3	&0	&3.4	&-4.2	&-6.3\\
3.9	&-1.1	&3.	&72.	&3.7	&0.32	&6.5	&3.	&-0.82	&0.43	&2.6	&-1.8	&78.	&-8.8	&0.032	&-0.033	&0.062	&0.28	&0.18	&-0.091	&-1.7	&-2.5	&6.4	&0.57	&-17.	&0	&0.88	&3.3	&1.3\\
4.3	&-1.2	&13.	&39.	&-1.3	&-4.1	&3.	&13.	&2.	&-1.8	&0.55	&-4.3	&64.	&-9.8	&0.03	&-0.025	&0.06	&0.46	&0.16	&-0.082	&-17.	&-18.	&-8.6	&-5.7	&-23.	&0	&-3.2	&13.	&13.\\
0.18	&-0.055	&2.	&-13.	&-0.88	&-0.61	&-0.82	&2.	&2.9	&-0.44	&-0.33	&-0.55	&-9.	&-0.41	&0.002	&-0.0021	&0.0041	&0.024	&0.011	&-0.0058	&-3.3	&-3.3	&-2.9	&-1.3	&-1.7	&0	&-0.73	&2.	&2.5\\
-0.069	&-0.14	&-1.8	&-2.8	&1.6	&1.4	&0.43	&-1.8	&-0.44	&0.55	&0.61	&0.5	&-6.4	&-0.15	&0.044	&-0.044	&0.09	&-0.048	&0.25	&-0.12	&2.4	&2.6	&2.7	&0.75	&-0.032	&0	&1.1	&-2.	&-2.6\\
5.1	&-1.7	&0.55	&-18.	&3.3	&2.1	&2.6	&0.55	&-0.33	&0.61	&2.7	&-1.4	&-17.	&40.	&0.18	&-0.18	&0.34	&0.0071	&0.96	&-0.52	&2.3	&2.4	&-38.	&2.9	&1.2	&0	&3.1	&0.46	&-1.4\\
-5.4	&1.4	&-4.3	&13.	&-0.063	&0.75	&-1.8	&-4.3	&-0.55	&0.5	&-1.4	&2.5	&4.8	&22.	&-0.11	&0.11	&-0.21	&-0.13	&-0.59	&0.3	&3.1	&3.3	&-19.	&-0.97	&17.	&0	&-0.84	&-4.5	&-3.9\\
-82.	&16.	&64.	&7.7\times 10^{3}	&-2.5	&-72.	&78.	&64.	&-9.	&-6.4	&-17.	&4.8	&7.9\times 10^{3}	&6.8\times 10^{2}	&0.046	&0.23	&-0.064	&2.6\times 10^{3}	&-0.24	&0.16	&1.7\times 10^{2}	&1.9\times 10^{2}	&8.7\times 10^{2}	&1.3\times 10^{2}	&1.1\times 10^{2}	&0	&5.0\times 10^{3}	&62.	&74.\\
-78.	&-52.	&-9.8	&6.9\times 10^{2}	&-6.6	&-2.2	&-8.8	&-9.8	&-0.41	&-0.15	&40.	&22.	&6.8\times 10^{2}	&1.1\times 10^{5}	&0.34	&0.71	&0.5	&-62.	&0.77	&-0.055	&-36.	&-40.	&1.1\times 10^{5}	&-50.	&-36.	&0	&23.	&-10.	&-5.4\\
-0.093	&0.028	&0.047	&0.58	&0.14	&-0.0017	&0.032	&0.03	&0.002	&0.044	&0.18	&-0.11	&0.046	&0.34	&2.6	&-1.9	&3.6	&-0.42	&11.	&-5.6	&-0.14	&-0.15	&-0.48	&-0.1	&-0.075	&0	&-0.03	&0.049	&0.049\\
0.093	&-0.029	&-0.049	&-0.57	&-0.24	&0.00059	&-0.033	&-0.025	&-0.0021	&-0.044	&-0.18	&0.11	&0.23	&0.71	&-1.9	&2.6	&-3.6	&-1.2	&-11.	&5.6	&0.14	&0.15	&-0.56	&0.1	&0.076	&0	&0.03	&-0.05	&-0.05\\
-0.18	&0.056	&0.094	&0.97	&0.74	&-0.0066	&0.062	&0.06	&0.0041	&0.09	&0.34	&-0.21	&-0.064	&0.5	&3.6	&-3.6	&8.9	&1.8	&22.	&-11.	&-0.28	&-0.29	&-0.79	&-0.2	&-0.15	&0	&-0.057	&0.096	&0.096\\
-0.11	&0.059	&0.47	&2.1\times 10^{2}	&9.5\times 10^{2}	&-0.28	&0.28	&0.46	&0.024	&-0.048	&0.0071	&-0.13	&2.6\times 10^{3}	&-62.	&-0.42	&-1.2	&1.8	&2.9\times 10^{5}	&3.5	&-2.1	&-3.7	&-0.84	&61.	&-0.37	&-0.091	&0	&3.3	&0.48	&0.52\\
-0.52	&0.16	&0.26	&3.	&1.5	&-0.0098	&0.18	&0.16	&0.011	&0.25	&0.96	&-0.59	&-0.24	&0.77	&11.	&-11.	&22.	&3.5	&61.	&-31.	&-0.78	&-0.82	&-1.6	&-0.56	&-0.42	&0	&-0.16	&0.27	&0.27\\
0.26	&-0.081	&-0.13	&-1.5	&-0.75	&0.0048	&-0.091	&-0.082	&-0.0058	&-0.12	&-0.52	&0.3	&0.16	&-0.055	&-5.6	&5.6	&-11.	&-2.1	&-31.	&17.	&0.39	&0.42	&0.47	&0.28	&0.21	&0	&0.083	&-0.14	&-0.14\\
3.6	&-1.3	&-17.	&1.1\times 10^{2}	&-0.34	&6.3	&-1.7	&-17.	&-3.3	&2.4	&2.3	&3.1	&1.7\times 10^{2}	&-36.	&-0.14	&0.14	&-0.28	&-3.7	&-0.78	&0.39	&6.8\times 10^{3}	&6.2\times 10^{3}	&6.2\times 10^{3}	&6.2\times 10^{3}	&8.1\times 10^{2}	&0	&2.3\times 10^{3}	&-2.	&-17.\\
3.9	&-1.4	&-18.	&1.4\times 10^{2}	&-0.26	&7.1	&-2.5	&-18.	&-3.3	&2.6	&2.4	&3.3	&1.9\times 10^{2}	&-40.	&-0.15	&0.15	&-0.29	&-0.84	&-0.82	&0.42	&6.2\times 10^{3}	&2.8\times 10^{4}	&6.3\times 10^{3}	&6.2\times 10^{3}	&8.1\times 10^{2}	&0	&1.7\times 10^{3}	&-3.3	&-18.\\
82.	&50.	&-8.6	&8.3\times 10^{2}	&6.4	&9.3	&6.4	&-8.6	&-2.9	&2.7	&-38.	&-19.	&8.7\times 10^{2}	&1.1\times 10^{5}	&-0.48	&-0.56	&-0.79	&61.	&-1.6	&0.47	&6.2\times 10^{3}	&6.3\times 10^{3}	&1.3\times 10^{5}	&6.2\times 10^{3}	&8.4\times 10^{2}	&0	&3.3\times 10^{2}	&6.7	&-13.\\
8.2	&-2.6	&-5.8	&-96.	&-1.6	&3.	&0.57	&-5.7	&-1.3	&0.75	&2.9	&-0.97	&1.3\times 10^{2}	&-50.	&-0.1	&0.1	&-0.2	&-0.37	&-0.56	&0.28	&6.2\times 10^{3}	&6.2\times 10^{3}	&6.2\times 10^{3}	&9.4\times 10^{3}	&7.8\times 10^{2}	&0	&2.6\times 10^{3}	&9.5	&-5.\\
-45.	&-7.8	&-23.	&1.8\times 10^{2}	&-15.	&-3.3	&-17.	&-23.	&-1.7	&-0.032	&1.2	&17.	&1.1\times 10^{2}	&-36.	&-0.075	&0.076	&-0.15	&-0.091	&-0.42	&0.21	&8.1\times 10^{2}	&8.1\times 10^{2}	&8.4\times 10^{2}	&7.8\times 10^{2}	&1.5\times 10^{5}	&0	&2.5\times 10^{2}	&-21.	&-2.2\\
0	&0	&0	&0	&0	&0	&0	&0	&0	&0	&0	&0	&0	&0	&0	&0	&0	&0	&0	&0	&0	&0	&0	&0	&0	&3.3\times 10^{2}	&3.3\times 10^{2}	&0	&0\\
6.7	&-2.3	&-3.2	&5.0\times 10^{3}	&3.4	&3.4	&0.88	&-3.2	&-0.73	&1.1	&3.1	&-0.84	&5.0\times 10^{3}	&23.	&-0.03	&0.03	&-0.057	&3.3	&-0.16	&0.083	&2.3\times 10^{3}	&1.7\times 10^{3}	&3.3\times 10^{2}	&2.6\times 10^{3}	&2.5\times 10^{2}	&3.3\times 10^{2}	&8.0\times 10^{3}	&-3.1	&-5.9\\
4.4	&-1.2	&13.	&36.	&-2.8	&-4.2	&3.3	&13.	&2.	&-2.	&0.46	&-4.5	&62.	&-10.	&0.049	&-0.05	&0.096	&0.48	&0.27	&-0.14	&-2.	&-3.3	&6.7	&9.5	&-21.	&0	&-3.1	&37.	&14.\\
1.5	&-0.22	&13.	&48.	&-5.2	&-6.3	&1.3	&13.	&2.5	&-2.6	&-1.4	&-3.9	&74.	&-5.4	&0.049	&-0.05	&0.096	&0.52	&0.27	&-0.14	&-17.	&-18.	&-13.	&-5.	&-2.2	&0	&-5.9	&14.	&38.\\
\end{array}
\right)$}
\end{equation}
\end{itemize}
\end{landscape}


\bibliographystyle{JHEP_2}
\bibliography{biblio.bib}

\end{document}